\documentclass[twocolumn,epjc3]{svjour3}
\pdfoutput=1 % if your are submitting a pdflatex (i.e. if you have
\usepackage{amsmath}
\usepackage{graphicx}
\usepackage{amssymb}
\usepackage{subfigure}
\usepackage{cancel}
\usepackage{xspace}
\usepackage{relsize}
\usepackage{bbold}
\usepackage{lineno}
\usepackage{braket}
\usepackage{slashed}
\usepackage{multirow}
\usepackage{placeins}
\usepackage{rotating}
\usepackage{url}
\usepackage{color}
\usepackage{hyperref}
\usepackage{cite}
\usepackage{multicol}
\usepackage{lscape}
\usepackage[utf8]{inputenc} 
\usepackage{float}
\usepackage[usenames,dvipsnames,svgnames,table]{xcolor}
\usepackage{colortbl}
\usepackage[all]{hypcap}
\RequirePackage{fix-cm}
\usepackage{soul}

\definecolor{oucrimsonred}{rgb}{0.6, 0.0, 0.0}
\definecolor{persianblue}{rgb}{0.11, 0.22, 0.73}
\definecolor{forestgreen}{rgb}{0.13,0.35,0.13}
\hypersetup{colorlinks, citecolor=oucrimsonred, linkcolor=persianblue, urlcolor=oucrimsonred}

\newcommand{\beq}{\begin{equation}} 
\newcommand{\eeq}{\end{equation}}
\newcommand{\bea}{\begin{eqnarray}}  
\newcommand{\eea}{\end{eqnarray}}
\newcommand{\nnl}{\nonumber \\}  
\newcommand{\GeV}{\,{\rm GeV}}

\definecolor{Gray}{gray}{0.9}
\newcommand{\cg}{\cellcolor{Gray}}

\usepackage[T1]{fontenc} % if needed

\smartqed  % flush right qed marks, e.g. at end of proof
\RequirePackage{graphicx}
\journalname{Eur. Phys. J. C}
\begin{document}
\title{New Physics in {\boldmath$b \to s \ell^+ \ell^-$} confronts new data on Lepton Universality}
\author{Marco Ciuchini\thanksref{em1,addr1}
\and
Ant\'onio M. Coutinho\thanksref{em2,addr2}
\and
Marco Fedele\thanksref{em3,addr3}
\and
Enrico Franco\thanksref{em4,addr4}
\and
Ayan Paul\thanksref{em5,addr5,addr6}
\and
Luca Silvestrini\thanksref{em6,addr4,addr7}
\and
Mauro Valli\thanksref{em7,addr8}
}
\thankstext{em1}{marco.ciuchini@roma3.infn.it}
\thankstext{em2}{antonio.coutinho@roma3.infn.it}
\thankstext{em3}{marco.fedele@icc.ub.edu}
\thankstext{em4}{enrico.franco@roma1.infn.it}
\thankstext{em5}{ayan.paul@roma1.infn.it,}
\thankstext{em6}{luca.silvestrini@roma1.infn.it}
\thankstext{em7}{mvalli@uci.edu}
\institute{INFN, Sezione di Roma Tre, Via della Vasca Navale 84,
  I-00146 Roma, Italy \label{addr1}
\and
  CFTP, Instituto Superior T\'{e}cnico, Universidade de Lisboa, Avenida Rovisco Pais 1, 1049-001 Lisboa, Portugal\label{addr2}
\and
  Dept.~de F\'{\i}sica Qu\`antica i Astrof\'{\i}sica, Institut de Ci\`encies del Cosmos (ICCUB), 
Universitat de Barcelona, Mart\'i Franqu\`es 1, E-08028 Barcelona, Spain  \label{addr3}
\and
  INFN, Sezione di Roma, P.le A. Moro 2,
  I-00185 Roma, Italy \label{addr4}
\and 
DESY, Notkestra{\ss}e 85, D-22607 Hamburg, Germany \label{addr5}
\and
Institut f\"ur Physik, Humboldt-Universit\"at zu Berlin, D-12489 Berlin, Germany  \label{addr6}
\and
Theoretical Physics Department, CERN, Geneva, Switzerland \label{addr7}
\and
Department of Physics and Astronomy, University of California, Irvine, CA 92697-4575 USA  \label{addr8}
}
\date{Received: date / Accepted: date}
\maketitle
\begin{abstract}{In light of the very recent updates on the $R_K$ and $R_{K^*}$ measurements from the LHCb and Belle collaborations, we systematically explore here imprints of New Physics in $b \to s \ell^+ \ell^- $ transitions using the language of effective field theories. We focus on effects that violate Lepton Flavour Universality both in the Weak Effective Theory and in the Standard Model Effective Field Theory. 
In the Weak Effective Theory we find a preference for scenarios with the simultaneous presence of two operators, a left-handed quark current with vector muon coupling and a right-handed quark current with axial muon coupling, irrespective of the treatment of hadronic uncertainties. In the Standard Model Effective Field Theory we select different scenarios according to the treatment of hadronic effects: while an aggressive estimate of hadronic uncertainties points to the simultaneous presence of two operators, one with left-handed quark and muon couplings and one with left-handed quark and right-handed muon couplings, a more conservative treatment of hadronic matrix elements leaves room for a broader set of scenarios, including the one involving only the purely left-handed operator with muon coupling.}
  \end{abstract}
%\flushbottom

%%%%%%%%%%%%%%%%%%%%%%%%%%%%%%%%%%%%%%%%%%%%%%%%%%%%%%%%

\section{Introduction}
\label{sec:intro}

%%%%%%%%%%%%%%%%%%%%%%%%%%%%%%%%%%%%%%%%%%%%%%%%%%%%%%%%

The past few years have brought us a thriving debate on the possible hints of New Physics (NP) from measurements of semileptonic $B$ decays. In particular, Flavour Changing Neutral Current (FCNC) decay modes into multi-body final states, e.g. $B\to K^{(*)}\ell^+\ell^-$ and $B_s \to \phi \, \ell^+ \ell^-$, bring forth a large number of experimental handles, see e.g.~\cite{Gratrex:2015hna}, that are extremely useful for NP investigations while also allowing to probe the Standard Model (SM) itself in detail \cite{Hiller:2003js,Bobeth:2007dw,Bobeth:2008ij,Egede:2008uy,Matias:2012xw}. The inference of what pattern is being revealed by the experimental observations is the crux of the debate.

Two distinct classes of observables characterize these semileptonic decays. The first is the class of angular observables arising from the kinematic distribution of the differential decay widths that have been measured at LHCb~\cite{Aaij:2013qta,Aaij:2014pli,Aaij:2014tfa,Aaij:2015oid,Aaij:2015dea,Aaij:2015esa,Aaij:2016flj}, Belle~\cite{Wehle:2016yoi}, ATLAS~\cite{Aaboud:2018krd} and CMS~\cite{Khachatryan:2015isa,Sirunyan:2017dhj,Sirunyan:2018jll}. These observables, mostly related to the muonic decay channel, while being sensitive to NP~\cite{Bobeth:2011nj,Matias:2012xw,DescotesGenon:2012zf,Descotes-Genon:2013vna,Matias:2014jua} are besieged by hadronic uncertainties~\cite{Khodjamirian:2010vf,Khodjamirian:2012rm,Lyon:2014hpa,Capdevila:2017ert,Blake:2017fyh,Bobeth:2017vxj}. The latter, associated with QCD long-distance effects -- hard to estimate from first principles~\cite{Jager:2012uw,Jager:2014rwa} -- can saturate the measurements so as to be interpreted as possibly arising from the SM or can obfuscate the gleaning of NP from SM contributions~\cite{Ciuchini:2015qxb,Ciuchini:2016weo,Ciuchini:2017gva}. Therefore, in the absence of a complete and reliable calculation of the hadronic long-distance contributions, a clear resolution of this debate based solely on the present set of angular measurements is hard to achieve. Improved experimental information in the near future~\cite{Kou:2018nap} concerning, in particular, the electron modes is a subject of current cross-talk between the theoretical and experimental communities~\cite{Capdevila:2016ivx,Serra:2016ivr}, and may shed new light on this matter~\cite{Albrecht:2017odf,Mauri:2018vbg,Ciuchini:2018anp}.
{\sloppy
The second class of observables then becomes crucial to this debate. These are the Lepton Flavour Universality Violating (LFUV) ratios that hold the potential to conclusively disentangle NP contributions from SM hadronic effects. The latter are indeed lepton flavour universal~\cite{Hiller:2003js,Bordone:2016gaq}. Several hints in favour of LFUV have surfaced in the past few years in experimental searches at LHCb~\cite{Aaij:2014ora,Aaij:2017vbb} and Belle~\cite{Wehle:2016yoi}. These have led to a plethora of theoretical investigations~\cite{Alonso:2014csa,
Hiller:2014yaa,Ghosh:2014awa,Glashow:2014iga,Hiller:2014ula,Gripaios:2014tna,Sahoo:2015wya,Crivellin:2015lwa,Crivellin:2015era,Celis:2015ara,Alonso:2015sja,Greljo:2015mma,Calibbi:2015kma,Falkowski:2015zwa,Carmona:2015ena,Allanach:2015gkd,Chiang:2016qov,Becirevic:2016zri,Feruglio:2016gvd,Megias:2016bde,Becirevic:2016oho,Arnan:2016cpy,Altmannshofer:2016jzy,Sahoo:2016pet,Alonso:2016onw,Hiller:2016kry,Galon:2016bka,Crivellin:2016ejn,GarciaGarcia:2016nvr,Cox:2016epl,Jager:2017gal,Megias:2017ove,Crivellin:2017zlb,
Celis:2017doq,Becirevic:2017jtw,Cai:2017wry,Kamenik:2017tnu,Sala:2017ihs,DiChiara:2017cjq,Ghosh:2017ber,Alonso:2017bff,Greljo:2017vvb,
Bonilla:2017lsq,Feruglio:2017rjo,Ellis:2017nrp,Bishara:2017pje,Alonso:2017uky,Tang:2017gkz,Datta:2017ezo,Das:2017kfo,Dinh:2017smk,Bardhan:2017xcc,DiLuzio:2017chi,Chiang:2017hlj,Chauhan:2017ndd,King:2017anf,Chivukula:2017qsi,Dorsner:2017ufx,Buttazzo:2017ixm,Choudhury:2017qyt,Cline:2017ihf,Crivellin:2017dsk,
Guo:2017gxp,Chen:2017usq,Baek:2017sew,Bian:2017rpg,Megias:2017vdg,
Lee:2017fin,Assad:2017iib,DiLuzio:2017vat,
Calibbi:2017qbu,Cline:2017aed,Romao:2017qnu,Descotes-Genon:2017ptp,Altmannshofer:2017bsz,Bian:2017xzg,Cline:2017qqu,Botella:2017caf,DiLuzio:2017fdq,Barbieri:2017tuq,Sannino:2017utc,DAmbrosio:2017wis,Raby:2017igl,Blanke:2018sro,Falkowski:2018dsl,Arcadi:2018tly,CarcamoHernandez:2018aon,Marzocca:2018wcf,
Camargo-Molina:2018cwu,Bordone:2018nbg,Earl:2018snx,Matsuzaki:2018jui,
Becirevic:2018afm,Baek:2018aru,King:2018fcg,Kumar:2018kmr,Hati:2018fzc,Guadagnoli:2018ojc,deMedeirosVarzielas:2018bcy,Li:2018rax,Alonso:2018bcg,Azatov:2018kzb,DiLuzio:2018zxy,Duan:2018akc,Heeck:2018ntp,Angelescu:2018tyl,
Grinstein:2018fgb,Singirala:2018mio,Balaji:2018zna,Rocha-Moran:2018jzu,Kamada:2018kmi,Fornal:2018dqn,Geng:2018xzd,Kumar:2019qbv,
Baek:2019qte,Marzo:2019ldg,Cerdeno:2019vpd,Bhattacharya:2019eji,Ko:2019tts}, all oriented towards physics Beyond the Standard Model (BSM) able to accommodate such LFUV signals, mainly involving $Z'$ or leptoquark mediators at scales typically larger than a few TeV and with some peculiar flavour structure needed to avoid clashing with the stringent bounds from meson-antimeson mixing and from other observables. Despite possible model-building challenges, the primary message here is  clear: a statistically significant measurement of LFUV effects in FCNCs such as $b \to s \ell^+ \ell^-$ decays would herald the discovery of NP unambiguously~\cite{DAmico:2017mtc,Geng:2017svp,Capdevila:2017bsm,Altmannshofer:2017yso,Ciuchini:2017mik,Hiller:2017bzc}. 
}

In this work we focus on the progress of this debate with the new measurements of $R_{K}$ and $R_{K^*}$ recently presented by the LHCb~\cite{Aaij:2019wad} and Belle collaborations~\cite{Abdesselam:2019wac}:
\begin{align}
    \label{eq:RK_measurement}
    R_{K}\,[1.1,6] 
 &\equiv \frac{Br \left(B^{+} \to K^{+} \mu^{+}\mu^{-}\right)}{Br\left(B^{+} \to K^{+} e^{+}e^{-}\right)} \nonumber\\
 &= 0.846\phantom{}^{+0.060}_{-0.054}\phantom{}^{+0.016}_{-0.014} \ \textrm{(LHCb)}, \\ 
    R_{K^*}\,[0.045,1.1] 
 &\equiv \frac{Br \left(B \to K^{*} \mu^{+}\mu^{-}\right)}{Br\left(B \to K^{*} e^{+}e^{-}\right)} \nonumber \\
 &= 0.52^{+0.36}_{-0.26} \pm 0.05 \ \textrm{(Belle)} , \label{eq:RKst_measurement_1} \\
    R_{K^*}\,[1.1,6] &= 0.96^{+0.45}_{-0.29} \pm 0.11 \ \textrm{(Belle)} .  \label{eq:RKst_measurement_2} 
\end{align}
The LHCb result combines the re-analysis of the 2014 measurement together with more recent data, partially including the experimental information from Run~II, and covers an invariant dilepton mass $q^{2}$ ranging in [1.1,6]~GeV$^2$. The preliminary Belle measurement also covers larger values of the dilepton invariant mass, which however are not used in our analysis, as detailed below. While the central value of the measurement in  eq.~\eqref{eq:RK_measurement} shifts towards the SM prediction~\cite{Hiller:2003js,Bordone:2016gaq}, the statistical significance of the corresponding \textit{$R_{K}$ anomaly} remains interesting, at the level of 2.5$\sigma$. On the other hand, the result in eq.~\eqref{eq:RKst_measurement_2} slightly weakens the significance of the $R_{K^*}$ anomaly in this range of dilepton invariant mass.

In an attempt to better disentangle SM hadronic uncertainties and to zoom in on the importance of NP contributions, here we present a state-of-the-art analysis of $b \to s \ell^+ \ell^-$ transitions where:
\begin{itemize}
    \item We revisit our approach to QCD power corrections streamlined for efficiently capturing long-distance effects, which are of utmost relevance in the interpretation of the current experimental information on the $B \to K^{*} \mu^+ \mu^-$ channel. We discuss several novelties about our new parameterization of hadronic contributions, recently introduced in \cite{Ciuchini:2018anp};
    \item  We make use of two distinct Effective Field Theory (EFT) frameworks, namely the $\Delta B = 1$ Weak Effective Hamiltonian and the Standard Model Effective Field Theory (SMEFT). The former EFT allows us to obtain a better insight on the dynamics at the decay scale, while the latter can offer a deeper link with BSM interpretations.
\end{itemize}
We review our theoretical framework in section~\ref{sec:framework}, where we also present a fresh look at the anatomy of $R_{K}$ in light of the new LHCb measurement. We then provide a thorough description of our analysis procedure in section~\ref{sec:inputs}. Finally, we collect and discuss all our main results in section~\ref{sec:results}, supported also by~\ref{app:A}~and~\ref{app:B}. We present our conclusions in section~\ref{sec:conclusion}.

\section{Theoretical framework}
\label{sec:framework}
{\sloppy
As an introduction to the basic ingredients of our analysis we start by reviewing the standard EFT for $\Delta B = 1$ transitions, highlighting the distinction between short-distance and hadronic contributions. We then move on to LFUV effects in terms of SM gauge-invariant dimension-six operators, completing the EFT dictionary useful for understanding the results we present in section~\ref{sec:results}.
}

\subsection{Short distance vs long distance}\label{sec:SDvsLS}
The anatomy of 
$B \to K^{(*)} \ell^+ \ell^-$, $B \to K^{*} \gamma$, $B_s \to \phi \ell^+ \ell^-$ and $B_s \to \phi \gamma$ decays can be inspected with an effective field theory of weak interactions for $\Delta B = 1$ processes \cite{Chetyrkin:1996vx, Buras:1994dj}.
%The corresponding effective Hamiltonian can be split in two parts:
The corresponding effective Hamiltonian at the scale $\mu_b \sim m_b$ can be split in two parts:
\beq \label{eq:HdB1}
{\cal H}_{\rm eff}^{\Delta B = 1} = {\cal H}_{\rm eff}^{\rm had} + {\cal H}_{\rm eff}^{\rm sl+\gamma}\,,
\eeq
where the first ``hadronic'' term contains only nonleptonic operators:
\beq\label{eq:H_had}
{\cal H}_{\rm eff}^{\rm had} = \frac{4G_F}{\sqrt{2}} \sum_{p=u,c}\lambda_p\bigg[ C_1Q_1^p + C_2Q_2^p + \sum_{i=3,\dots,6} C_iP_i + C_{8}Q_{8g} \bigg]\,,
\eeq
involving the following set of relevant operators up to dimension six:
\begin{align}
\label{eq:quark_O}
  Q^p_1 &= (\bar{s}_L\gamma_{\mu}T^a p_L)(\bar{p}_L\gamma^{\mu}T^ab_L)\,,\nnl
  Q^p_2 &= (\bar{s}_L\gamma_{\mu} p_L)(\bar{p}_L\gamma^{\mu}b_L)\,, \nnl
  P_3 &= (\bar{s}_L\gamma_{\mu}b_L)\sum\phantom{} _q(\bar{q}\gamma^{\mu}q)\,, \nnl
  P_4 &= (\bar{s}_L\gamma_{\mu}T^ab_L)\sum\phantom{}_q(\bar{q}\gamma^{\mu}T^aq) \,,\nnl 
  P_5 &= (\bar{s}_L\gamma_{\mu1}\gamma_{\mu2}\gamma_{\mu3}b_L)\sum\phantom{}_q
  (\bar{q}\gamma^{\mu1}\gamma^{\mu2}\gamma^{\mu3}q) \,, \nnl
  P_6 &= (\bar{s}_L\gamma_{\mu1}\gamma_{\mu2}\gamma_{\mu3}T^ab_L)\sum\phantom{}_q
  (\bar{q}\gamma^{\mu1}\gamma^{\mu2}\gamma^{\mu3}T^aq) \,, \nnl 
  Q_{8g} &= \sqrt{\frac{\alpha_{s}}{64 \pi^3}}
  m_b \bar{s}_L\sigma_{\mu\nu}G^{\mu\nu}b_R \,.
\end{align} 
The second term features four-fermion operators constructed with leptonic and quark bilinears, together with the electromagnetic dipole operators,
\begin{align}
\label{eq:H_sl}
{\cal H}_{\rm eff}^{\rm sl+\gamma} = -\frac{4G_F}{\sqrt{2}} \lambda_t\bigg[ &C_7^{(\prime)}Q_{7\gamma}^{(\prime)} + C_9^{(\prime)}Q_{9V}^{(\prime)} + C_{10}^{(\prime)}Q_{10A}^{(\prime)} \nonumber\\
&+ C_{S}^{(\prime)}Q_{S}^{(\prime)} + C_{P}^{(\prime)}Q_{P}^{(\prime)} \bigg]\,,
\end{align}
including, up to dimension six, the operators:
\begin{align}
\label{eq:WET_operators}
Q_{7\gamma} &= \sqrt{\frac{\alpha_{e}}{64 \pi^3}} m_b \bar{s}_L\sigma_{\mu\nu}F^{\mu\nu}b_R\,, \nnl
Q_{9V} &= \frac{\alpha_{e}}{4\pi}(\bar{s}_L\gamma_{\mu}b_L)(\bar{\ell}\gamma^{\mu}\ell)\,, \nnl
Q_{10A} &= \frac{\alpha_{e}}{4\pi}(\bar{s}_L\gamma_{\mu}b_L)(\bar{\ell}\gamma^{\mu}\gamma^5\ell) \,, \nnl
Q_{S} &= \frac{\alpha_{e}}{4\pi}(\bar{s}_L b_R)(\bar{\ell}\ell) \,, \nnl
Q_{P} &= \frac{\alpha_{e}}{4\pi}(\bar{s}_L b_R)(\bar{\ell}\gamma^5\ell) \,.
\end{align}
Note that in eq.~\eqref{eq:WET_operators} we have omitted tensorial semileptonic structures under the reasonable assumption that NP exhibits a mass gap above the electroweak (EW) scale~\cite{Alonso:2014csa}. We have also omitted other hadronic operators that may arise beyond the SM, since we are focusing on LFUV. 
The primed operators $Q_i'$ are obtained from eq.~\eqref{eq:WET_operators} substituting
$P_{R,L} \to P_{L,R}$ in the corresponding quark bilinears. Throughout the paper, CKM factors are defined as \mbox{$\lambda_i= V_{is}^{}V_{ib}^*=V_{i2}^{}V_{i3}^*$}, with $i=\{u,c,t\}=\{1,2,3\}$. 

The short-distance physics in eqs.~\eqref{eq:HdB1}-\eqref{eq:WET_operators} is, in general, captured
by the Wilson coefficients (WCs), denoted as effective couplings $C^{(\prime)}$. Within the SM, at the dimension-six level, semileptonic chirality-flipped and (pseudo)scalar operators can be neglected, however they are potentially relevant for the study of NP effects. In our analysis, we evaluate SM WCs at the scale $\mu_b = 4.8 \GeV$ using state-of-the-art QCD and QED perturbative corrections, both in the
matching~\cite{Bobeth:1999mk,Gambino:2001au,Misiak:2004ew} and in the  anomalous dimension of the operators involved~\cite{Misiak:2004ew,Bobeth:2003at,Gambino:2003zm,Huber:2005ig}.\footnote{The scale $\mu_b$ is here set by the scale at which form factors have been computed~\cite{Straub:2015ica}.} We note that the remaining theoretical uncertainty on the SM WCs, at the level of few percent, can be neglected in this work.

In the absence of a unique UV complete model that can potentially be responsible for the measured hints of anomalies in the $b \to s$ transitions, the formalism of the effective Hamiltonian is extremely powerful. It allows to study the effects of BSM physics in a model-independent manner, where the presence of NP effects manifests itself as (lepton-flavour dependent) shifts 
of the WCs with respect to the SM values\footnote{In the present work, while we treat the SM short distance with all available quantum corrections included, for the NP WCs we neglect the running induced by gauge couplings. Consequently, they stay constant from the scale they have been generated, with the notable exception of the SMEFT contributions arsing only at one-loop via RGE, see section~\ref{sec:SMEFT} and~\ref{app:B}.}. On the basis
of previous global analyses which allow for LFUV effects~\cite{Altmannshofer:2014rta,Descotes-Genon:2015uva,Chobanova:2017ghn,Capdevila:2017bsm,Altmannshofer:2017yso,DAmico:2017mtc,Hiller:2017bzc,Geng:2017svp,Ciuchini:2017mik,Alok:2017sui,Hurth:2017hxg}, 
%in this paper we allow for NP effects in the  $C_{9,10,S,P}^{(\prime) \, \ell=e,\mu}$. 
in this paper we allow for NP effects in the WCs of the operators $Q_{9,10,S,P}^{(\prime) \, \ell=e,\mu}$. 
We do not consider the case of NP effects in dipole operator coefficients $C_7^{(\prime)}$, since such a possibility is severely constrained by the inclusive radiative branching fraction of $B \to X_s \gamma$ among other measurements~\cite{Paul:2016urs} and it is anyway irrelevant for LFUV. Moreover, in the following we also set aside the possibility of NP effects entering in eq.~\eqref{eq:H_had}, a case considered in the study by the authors of~\cite{Jager:2017gal}. Our choice is, once more, primarily driven by our focus on LFUV effects. On more general grounds, one should stress that decoding LFU-conserving NP effects in current $b \to s $ data -- a possibility recently considered in~\cite{Alguero:2018nvb} -- may be a challenging task~\cite{Alguero:2019pjc}, especially in light of unknown hadronic contributions.

Let us consider the $\bar{B} \to \bar{K}^{*} \ell^+ \ell^-$ transition as an example for setting up our notation. From the Hamiltonian defined in eq.~\eqref{eq:HdB1}, it is possible to write down seven independent helicity amplitudes that, in full generality, describe a (pseudo)scalar particle decaying into a vector state and a dilepton pair. These helicity amplitudes can be combined together to define the decay branching ratio and the largest independent set of angular observables. In the basis defined in~\cite{Jager:2012uw}, these structures within the SM can be schematically written as:\footnote{Here we do not include the negligible contributions to $C^{(')}_{S,P}$ from the SM for clarity.}
{\allowdisplaybreaks
\begin{align}
\label{eq:helamp}
H_V^{\lambda} &\propto \left\{C_9^{\rm SM}\widetilde{V}_{L\lambda} + \frac{m_B^2}{q^2} \left[\frac{2m_b}{m_B}C_7^{\rm SM}\widetilde{T}_{L\lambda}  - 16\pi^2h_{\lambda} \right]\right\}\,,\nonumber\\
H_A^{\lambda} &\propto C_{10}^{\rm SM}\widetilde{V}_{L\lambda} \ , \quad
H_P \propto \frac{m_{\ell} \, m_{b}}{q^2} \,  C_{10}^{\rm SM} \left( \widetilde{S}_{L} - \frac{m_s}{m_b}\widetilde{S}_{R} \right) \ ,
\end{align}}
with $\lambda=0,\pm$. The factorizable part of these amplitudes, i.e. the one involving matrix elements of semileptonic local operators, is described by means of seven independent form 
factors, $\widetilde{V}_{0,\pm}$, $\widetilde{T}_{0,\pm}$ and $\widetilde{S}$ which are smooth functions of $q^{2}$. In eq.~\eqref{eq:helamp} these are defined following the convention described in appendix~A of ref.~\cite{Ciuchini:2015qxb}. In addition to form factors, at first order in $\alpha_{e}$, non-local contributions arise from the insertion of a quark current with each of the operators appearing in eq.~\eqref{eq:H_had} \cite{Khodjamirian:2010vf,Jager:2012uw}. As a result, non-factorizable QCD power corrections appear in $H_V^{\lambda}$ according to the hadronic correlator~\cite{Jager:2014rwa,Ciuchini:2015qxb,Chobanova:2017ghn,Arbey:2018ics}:
\beq \label{eq:hlambda}
 h_\lambda(q^2) = \frac{\epsilon^*_\mu(\lambda)}{m_B^2} \int d^4x\ e^{iqx} \langle \bar K^* \vert T\{j^{\mu}_\mathrm{em} (x) 
 \mathcal{H}_\mathrm{eff}^\mathrm{had} (0)\} \vert \bar B \rangle \,.
\eeq

For the factorizable part, in the large-recoil region (i.e. low dilepton invariant mass $q^2$) two light-cone sum rule (LCSR) computations are currently available~\cite{Straub:2015ica,Gubernari:2018wyi}. These 
results are in reasonable agreement with the extrapolation of the form factors computed in lattice QCD at low recoil~\cite{Horgan:2013hoa}. % Moreover, the information on the same form factors is enriched by a correlation matrix that keeps track of heavy-quark symmetry relations.
The information on the same form factors is enriched by a correlation matrix that keeps track also of heavy-quark symmetry relations as discussed in section~2.2 of ref.~\cite{Straub:2015ica}.

On the contrary, the theoretical estimate of non-factorizable terms -- denoted here generically by $h_{\lambda}$ -- is not so well under control. For the processes of interest, the largest contribution arises from current-current operators involving charm quarks, specifically $Q_{2}^{c}$~\cite{Khodjamirian:2010vf,Khodjamirian:2012rm}, not parametrically suppressed by CKM factors or small WCs. This charm-loop effect stemming out of eq.~\eqref{eq:hlambda} is therefore a genuine long-distance contribution: it implies potentially sizable non-perturbative effects involving the charm quark pair with strong phases that are very difficult to estimate.
While at low $q^2$ hard-gluon exchanges in the charm-loop amplitude can be addressed in the framework of QCD factorization (QCDF)~\cite{Beneke:2001at}, the evaluation of soft-gluon exchange effects remains, in this context, the toughest theoretical task~\cite{Kozachuk:2018yxf}. A detailed  analysis of soft-gluon exchanges in $B \to K$ transitions has been performed in ref.~\cite{Khodjamirian:2012rm}. There these 
contributions turned out to be sub-dominant in comparison with the QCDF estimate of the hard-gluon contributions, supporting previous results presented in~\cite{Khodjamirian:2010vf}. 

For $B \to K^*$, the only estimate of $h_{\lambda}$ currently available is the one carried out in ref.~\cite{Khodjamirian:2010vf} using LCSR techniques
in the single soft-gluon approximation, valid for $q^2 \ll 4m_c^2$. The regime of validity of the result is then extended to the whole large-recoil region by means of a phenomenological model based on dispersion relations. While an estimate of the error budget is attempted in ref.~\cite{Khodjamirian:2010vf}, there are potentially large systematic effects, related for instance to the lack of control over strong phases, that are difficult to quantify reliably, in particular when approaching the $c\bar{c}$ threshold at  $q^2 \sim 4m_c^2$~\cite{Ciuchini:2015qxb}, close to the $J/\psi$ resonance where quark-hadron duality is questionable even in the heavy quark limit ~\cite{Beneke:2009az}. Note that the same considerations also apply to the case of $B_{s} \to \phi \ell^+ \ell^-$, for which a similar LCSR evaluation of the charm-loop effect is still pending, leaving room also for appreciable $SU(3)_{F}$ breaking effects~\cite{Capdevila:2017ert}.

Recently, renewed attempts to obtain a better grasp of the non-factorizable terms have appeared in the literature~\cite{Blake:2017fyh,Bobeth:2017vxj}. 
Both works turn out to be in agreement with the results from ref.~\cite{Khodjamirian:2010vf}. However, in ref.~\cite{Blake:2017fyh} -- where $h_\lambda$ is assumed to be well-described 
as a sum of relativistic Breit-Wigner functions -- the authors found a very similar result to the one in~\cite{Khodjamirian:2010vf} only in the case of vanishing strong phases, while quite different outcomes may be obtained for different assumptions on the same phases.
In turn, the authors of ref.~\cite{Bobeth:2017vxj}
exploited the analytic properties of the amplitudes in order to perform a conformal expansion of $h_\lambda$, isolating physical poles and ensuring unitarity. They use resonant data and additional theoretical information at negative $q^2$ to fix the coefficients of the expansion, including estimates of strong phases due to the presence of a second branch cut in the amplitude (generated for instance by intermediate states with two charmed mesons), which represents a challenge for the formalism as well as for the numerical estimate.
Despite the quite good agreement with the numerical result presented in~\cite{Khodjamirian:2010vf}, the coefficients obtained in~\cite{Bobeth:2017vxj} for the $z$-expansion of $h_\lambda$ point to a poor convergence of the series, casting doubts on the actual $q^2$ shape of the $h_\lambda$ functions if more terms were to be included in the expansion.

In this work, we are therefore well-motivated to consider the available LCSR estimates on $h_\lambda$ from
ref.~\cite{Khodjamirian:2010vf} with a certain degree of caution. 
To this end, we  have already proposed in refs.~\cite{Ciuchini:2015qxb,Ciuchini:2016weo,Ciuchini:2017mik} 
a phenomenological expansion of $h_\lambda$ in powers of $q^2$ in the large-recoil region, inspired by ref.~\cite{Jager:2014rwa}.
We use $B\to K^*\mu^+\mu^-$ and $B\to K^*\gamma$ measurements in order to constrain the coefficients of the expansion, and enforce the results from ref.~\cite{Khodjamirian:2010vf} under two different scenarios: 
\begin{itemize}
\item A \textit{phenomenological model driven} (PMD) approach, employing LCSR results extrapolated by means of dispersion relations in the whole low-$q^2$ region for the decay; 
\item A \textit{phenomenological data driven} (PDD) approach, taking into account LCSR results only far from the c$\bar{c}$ threshold and exploiting the results in ref.~\cite{Khodjamirian:2010vf} for $q^2=0,1$ GeV$^2$, with their phases and $q^2$ dependence inferred from experimental data.
\end{itemize}
It is important to note that the PDD approach entails a loss of constraining power in the NP analysis, as some of the hadronic coefficients can mimic LFU NP effects, and therefore should be considered as the most conservative approach towards the assessment of NP effects. Eventually, we also highlight that -- differently from what done in ref.~\cite{Bobeth:2017xry} -- the extraction of the hadronic coefficients in the PDD approach relies essentially on the experimental information stemming from the same $q^2$ region where our phenomenological expansion is used.
To better investigate the interplay between hadronic contributions and possible NP ones, in this work we use a recent improvement of our parameterization for $h_\lambda$ with the expansion presented in~\cite{Ciuchini:2018anp}:
\begin{align}
 \label{eq:newexp}
h_-(q^2) &= -\frac{m_b}{8\pi^2 m_B} \widetilde T_{L -}(q^2) h_-^{(0)} -\frac{\widetilde V_{L -}(q^2)}{16\pi^2 m_B^2} h_-^{(1)} q^2 \nonumber \\
&\;\;\;\;+ h_-^{(2)} q^4+{\cal O}(q^6)\,, \nonumber\\
h_+(q^2) &=  -\frac{m_b}{8\pi^2 m_B} \widetilde T_{L +}(q^2) h_-^{(0)} -\frac{\widetilde V_{L +}(q^2)}{16\pi^2 m_B^2}  h_-^{(1)} q^2 \nonumber \\
&\;\;\;\;+ h_+^{(0)} + h_+^{(1)}q^2 + h_+^{(2)} q^4+{\cal O}(q^6)\,, \nnl
h_0(q^2) &= -\frac{m_b}{8\pi^2 m_B} \widetilde T_{L 0}(q^2) h_-^{(0)} -\frac{\widetilde V_{L 0}(q^2)}{16\pi^2 m_B^2}  h_-^{(1)} q^2 \nonumber \\
&\;\;\;\;+ h_0^{(0)}\sqrt{q^2} + h_0^{(1)}(q^2)^\frac{3}{2}  +{\cal O}((q^2)^\frac{5}{2})\,.
\end{align}
This choice allows us to write the helicity amplitudes $H_V^\lambda$ in eq.~\eqref{eq:helamp} as
\begin{align}
\label{eq:hv}
H_V^{-} \propto &\bigg\{\left(C_9^{\rm SM} + h_-^{(1)}\right) \widetilde V_{L -} \nnl 
& + \frac{m_B^2}{q^2} \left[ \frac{2m_b}{m_B}\left(C_7^{\rm SM} + h_-^{(0)} \right) \widetilde T_{L -} - 16\pi^2 h_-^{(2)}\, q^4 \right]\bigg\}\,, \nnl
H_V^{+} \propto &\bigg\{\left(C_9^{\rm SM} + h_-^{(1)}\right)\widetilde V_{L +} + \frac{m_B^2}{q^2}  \bigg[\frac{2m_b}{m_B}\left(C_7^{\rm SM} + h_-^{(0)} \right) \widetilde T_{L +} \nnl
& \quad - 16\pi^2\left(h_{+}^{(0)} + h_{+}^{(1)}\, q^2 +  h_{+}^{(2)}\, q^4\right) \bigg]\bigg\}\,,\nnl
H_V^{0} \propto &\bigg\{\left(C_9^{\rm SM} + h_-^{(1)}\right) \widetilde{V}_{L0} + \frac{m_B^2}{q^2}  \bigg[\frac{2m_b}{m_B} \left(C_7^{\rm SM} + h_-^{(0)} \right) \widetilde{T}_{L0}  \nnl
& \quad  - 16\pi^2\sqrt{q^2}\left({h}_{0}^{(0)} + {h}_{0}^{(1)}\, q^2\right)\bigg]\bigg\}\,.
\end{align}

With this definition for  the $h_\lambda$-coefficients, it is manifest that $h_-^{(0)}$ and $h_-^{(1)}$ can be
considered as constant shifts to the WCs $C_{7,9}^{\rm SM}$, hence indistinguishable 
from NP contributions to $Q_{7\gamma,9V}$. Consequently, it is not possible to extract $h_-^{(0)}$ and
$h_-^{(1)}$ directly from data unless one assumes the absence of NP effects. On the other hand, it is also not possible to ascertain the presence of
NP without a theory input for these hadronic effects. The advantage of the parameterization in eqs.~\eqref{eq:newexp}-\eqref{eq:hv} becomes clear when any of the remaining $h_\lambda$-coefficients turns out to be non-vanishing, since they likely spot purely hadronic contributions.\footnote{Note that this statement is accurate as long as NP effects do not feed any of the WCs in eq.~\eqref{eq:H_had}.} In section~\ref{sec:inputs} we report the details of the implementation of our PMD and PDD approaches for non-factorizable contributions in the present numerical analysis.

%%%%%%%%%%%%%%%%%%%%%%%%%%%%%%%%%%%%%%%%%%%%%%%%%%%%%%%%

\subsection{The SMEFT perspective}\label{sec:SMEFT}
Previous model-independent analyses of $b \to s \ell^+ \ell^- $ anomalies have essentially pointed to $\mathcal{O}(10~{\rm TeV})$ NP for $\mathcal{O}(1)$ effective couplings in order to produce a $\sim \, 25$\% shift of the SM WC values of the semileptonic operators $Q_{9V,10A}$. The UV dynamics underlying these NP effects is then expected to exhibit a reasonable mass gap with the SM theory. Hence, a quite natural choice for deeper BSM insights is the gauge-invariant framework of the SMEFT~\cite{Buchmuller:1985jz,Grzadkowski:2010es}. 

NP imprints in $b \to s$ transitions in the context of SM gauge-invariant operators have been extensively investigated in~\cite{Alonso:2014csa,Aebischer:2015fzz,Alonso:2015sja} and a systematic study of flavour physics constraints from $\Delta F=2$ processes in the SMEFT has been recently performed in \cite{Silvestrini:2018dos}. For $b \to s \ell^+ \ell^-$ anomalies, a dedicated analysis with SMEFT operators was already carried out in ref.~\cite{Celis:2017doq}. In what follows we proceed along the lines outlined in these works.\footnote{We are not going to take into account the SMEFT contributions to the CKM parameters recently worked out in~\cite{Descotes-Genon:2018foz}, since they cannot accommodate for LFUV effects.} The set of $SU(2)_{L} \times U(1)_{Y}$ invariant four-fermion operators in which we are mainly interested in this study reads:
\begin{align}
 \label{eq:SMEFT_op_tree}
O^{LQ^{(1)}}_{ijkl} &= (\bar{L}_i\gamma_\mu L_j)(\bar{Q}_k\gamma^\mu Q_l)\,, \nonumber \\
O^{LQ^{(3)}}_{ijkl} &= (\bar{L}_i\gamma_\mu \tau^{A} L_j)(\bar{Q}_k\gamma^\mu\tau^{A} Q_l)\,,\nonumber \\
O^{Qe}_{ijkl} &= (\bar{Q}_{i}\gamma_\mu Q_{j})(\bar{e}_{k} \gamma^\mu e_{l})\,,\nonumber \\
O^{Ld}_{ijkl} &= (\bar{L}_i\gamma_\mu L_j)(\bar{d}_k\gamma^\mu d_l)\,,\nonumber \\
O^{ed}_{ijkl} &= (\bar{e}_i\gamma_\mu e_j)(\bar{d}_k\gamma^\mu d_l)\,,\nonumber \\
O^{LedQ}_{ijkl} &= (\bar{L}_i e_j)(\bar{d}_k Q_l) \,, 
\end{align}
where $i,j,k,l=1,2,3$ are generation indices, $\tau^{A=1,2,3}$ are Pauli matrices (a sum over $A$ in the equations above is understood), weak doublets are in upper case and $SU(2)_{L}$ singlets are in lower case. 

The operators appearing in eq.~\eqref{eq:SMEFT_op_tree} correspond to the set that matches at tree level on the semileptonic operators in eq.~\eqref{eq:WET_operators} in an operator product expansion truncated at dimension six.
The SMEFT tree-level matching is naturally performed at the scale $\mu_{\textrm{\tiny{EW}}} \sim \mathcal{O}(M_{W})$. For BSM dynamics that distinguishes the lepton flavour $\ell=\{e,\mu,\tau\} = \{1, 2, 3\}$ in $b \to s$ transitions, 
%the relations connecting $C_{9,10,S,P}^{(\prime)}$ with the SMEFT basis are:
the relations connecting the WCs of $Q_{9,10,S,P}^{(\prime)}$ to the ones in the SMEFT-operator basis are:\footnote{Dimension-six operators made of Higgs doublets and quark bilinears should also appear~\cite{Alonso:2014csa}, but yield a lepton flavour universal contribution. They are severely constrained both by EW and Higgs data, see~\cite{deBlas:2016ojx,Ellis:2018gqa}, and by $\Delta F= 2$~measurements~\cite{Silvestrini:2018dos}. They will not be further considered here.}
{\allowdisplaybreaks
\begin{align}
 \label{eq:SMEFT_matching}
C_{9,\ell}^{\rm NP}&= \mathcal{N}_{\Lambda} \, \left(C^{LQ^{(1)}}_{\ell\ell 2 3} + C^{LQ^{(3)}}_{\ell\ell 2 3} + C^{Qe}_{2 3 \ell \ell} \right)\,,\nonumber \\
C_{10,\ell}^{\rm NP}&= \mathcal{N}_{\Lambda} \, \left(C^{Qe}_{2 3 \ell \ell} - C^{LQ^{(1)}}_{\ell\ell 2 3} - C^{LQ^{(3)}}_{\ell\ell 2 3}  \right)\,,\nonumber \\
C^{\prime, \rm{NP}}_{9,\ell}&=\mathcal{N}_{\Lambda}\,\left(C^{ed}_{\ell \ell 2 3} + C^{Ld}_{\ell\ell 2 3} \right)\,,\nonumber \\
C^{\prime, \rm{NP}}_{10,\ell}&=\mathcal{N}_{\Lambda}\, \left(C^{ed}_{\ell \ell 2 3} - C^{Ld}_{\ell \ell 2 3} \right)\,,\nonumber \\
C_{S,\ell}^{\rm NP}&=-C_{P,\ell}^{\rm NP}=\mathcal{N}_{\Lambda}\, C^{LedQ}_{\ell\ell 2 3}\,,\nonumber \\
C^{\prime, \rm{NP}}_{S,\ell}&=C^{\prime, \rm{NP}}_{P,\ell}=\mathcal{N}_{\Lambda}\,\left(C^{LedQ}_{\ell \ell 3 2}\right)^{*}\,,
\end{align}
}
where we introduced the complex factor $$\mathcal{N}_{\Lambda} \equiv (\pi v^2)/(\alpha_{e}\lambda_{t}\Lambda^2),$$ with $v^{2}/2 = \langle H^{\dagger} H \rangle$, $H$ being the SM Higgs doublet. For a NP scale $\Lambda$ of 30~TeV one has $|\mathcal{N}_{\Lambda}| \simeq \,0.7$. 
Eq.~\eqref{eq:SMEFT_matching} is valid in the basis where charged lepton and down-type quark Yukawa couplings are diagonal. This choice simplifies the analysis by avoiding lepton flavour violation and constraints from quark flavour transitions other than $b\to s$.

Note that even under this assumption, the operators in eq.~\eqref{eq:SMEFT_op_tree} may be, in principle, testable in other interesting processes other than $b \to s \ell^+ \ell^-$ transitions. The most notable opportunity may be offered by the channel $B \to K^{(*)} \nu \bar{\nu}$~\cite{Buras:2014fpa}, sensitive to the operators composed of weak doublets in both lepton and quark currents. At the present experimental sensitivity~\cite{Lees:2013kla}, this channel turns out to have a relatively mild interplay with $b \to s \ell^+ \ell^-$ measurements~\cite{Calibbi:2017qbu,Feruglio:2017rjo}. Interestingly, with the advent of more data~\cite{Kou:2018nap} one may hope to distinguish NP effects of $O^{LQ^{(3)}}$ from the ones of $O^{LQ^{(1)}}$ due to an accurately  measured light-lepton LFUV ratio in semileptonic $b \to c$ transitions~\cite{Greljo:2015mma,Feruglio:2017rjo}. Still, for the purposes of our model-independent study, $O^{LQ^{(1,3)}}_{ii23}$, $i =\{1,2\}$, are indistinguishable, as in ref.~\cite{Celis:2017doq}. Without loss of generality, the set of operators in eq.~\eqref{eq:SMEFT_op_tree} remains indeed the one primarily sensitive to the measurements considered in this work.

Going beyond eq.~\eqref{eq:SMEFT_op_tree}, one may extend the discussion to the operators induced at one-loop level that are a genuine product of the renormalization group evolution (RGE) in the SMEFT~\cite{Jenkins:2013zja,Jenkins:2013wua}. Equipped with the aforementioned assumption on the SMEFT flavour structure, in the leading-log approximation and leading expansion in the top Yukawa coupling $y_{t}$, the matching conditions induced by one-loop RGE read as:
\begin{align}
    \label{eq:SMEFT_matching_1loop}
C_{9,\ell}^{\rm NP}&= \mathcal{N}_{\Lambda} \,\lambda_{t} \left(\frac{y_{t} }{4 \pi}\right)^2 \log \left( \frac{\Lambda}{\mu_{\textrm{\tiny{EW}}}} \right)   \, \Big(C^{HL^{(3)}}_{\ell \ell} - C^{HL^{(1)}}_{\ell \ell}\nonumber\\
&\;\;\;-C^{He}_{\ell \ell} + C^{Lu}_{\ell \ell 3 3} + C^{eu}_{\ell \ell 3 3} \Big)\,,\nonumber \\
C_{10,\ell}^{\rm NP}&= \mathcal{N}_{\Lambda} \,\lambda_{t} \left(\frac{y_{t} }{4 \pi}\right)^2 \log \left( \frac{\Lambda}{\mu_{\textrm{\tiny{EW}}}} \right)   \, \Big(C^{HL^{(1)}}_{\ell \ell} - C^{HL^{(3)}}_{\ell \ell} \nonumber\\
&\;\;\;-C^{He}_{\ell \ell} - C^{Lu}_{\ell \ell 3 3} + C^{eu}_{\ell \ell 3 3} \Big)\, ,
\end{align} 
where we have reported again only contributions that matter for the discussion of LFUV effects coming from dimension-six operators with Higgs doublet and lepton bilinears,
\begin{align}
\label{eq:SMEFT_op_loop_H}
O^{HL^{(1)}}_{\ell \ell} &= ( H^{\dagger} i \overset{\leftrightarrow}{D}_{\mu}H ) (\bar{L}_{\ell} \gamma^{\mu}  L_{\ell} )\,, \nonumber \\
O^{HL^{(3)}}_{\ell \ell} &= ( H^{\dagger} i \overset{\, \leftrightarrow_A}{D_{\mu}}H ) (\bar{L}_{\ell} \gamma^{\mu} \tau^{A} L_{\ell} )\,, \nonumber \\
O^{He}_{\ell \ell} &= ( H^{\dagger} i \overset{\leftrightarrow}{D}_{\mu}H ) (\bar{e}_{\ell} \gamma^{\mu}  e_{\ell} )\,,
\end{align} 
together with semileptonic operators involving a right-handed top-quark current,
\begin{align}  \label{eq:SMEFT_op_loop_lu}
O^{Lu}_{\ell \ell 3 3} &= (\bar{L}_{\ell} \gamma_\mu L_{\ell})(\bar{u}_{3}\gamma^\mu u_{3})\,, \nonumber \\
O^{eu}_{\ell \ell 3 3} &= (\bar{e}_{\ell} \gamma_\mu e_{\ell})(\bar{u}_{3}\gamma^\mu u_{3})\,.
\end{align} 

The expression in eq.~\eqref{eq:SMEFT_matching_1loop} needs to be added to the tree-level matching already given in eq.~\eqref{eq:SMEFT_matching}. Next-to-leading order SMEFT matching conditions, invoked to reduce the matching-scale dependence on the overall result (and involving renormalization scheme-dependent finite parts of one-loop diagrams), have been computed  in refs.~\cite{Aebischer:2015fzz,Hurth:2019ula}. Within the leading-log approximation undertaken in this work, these corrections should be considered as sub-leading to the RGE-induced contributions given in eq.~\eqref{eq:SMEFT_matching_1loop}. Obviously, the same $\sim \, 25$\% shift of the SM WC $C_{9,10}$ needed for a qualitative explanation of $b \to s \ell^+ \ell^-$ anomalies, if obtained through the RG mixing in eq.~\eqref{eq:SMEFT_matching_1loop}, requires NP scales of $\mathcal{O}(\textrm{TeV})$ for $\mathcal{O}(1)$ couplings. Therefore, in these particular scenarios the underlying BSM physics should be much closer to collider reach compared to cases in which the operators in eq.~\eqref{eq:SMEFT_matching} are directly generated by NP \cite{DiLuzio:2017chi}. As already commented in ref.~\cite{Celis:2017doq}, operators appearing in eq.~\eqref{eq:SMEFT_op_loop_H} are particularly well constrained by EW precision observables~\cite{deBlas:2016ojx,Ellis:2018gqa}, making them irrelevant in the present context. Our analysis on SMEFT RGE-induced contributions will then focus only on the operators listed in eq.~\eqref{eq:SMEFT_op_loop_lu} above. Note, however, that these operators can also be constrained at the loop level by EW data, see refs.~\cite{Camargo-Molina:2018cwu,Aebischer:2018iyb}. We dedicate our~\ref{app:B} to the inspection of this set of operators, where we highlight a non-trivial interplay between the assumption made on hadronic contributions in analysis of $b \to s \ell^+ \ell^-$ data and the information coming from EW precision measurements relevant in this context~\cite{Efrati:2015eaa}.

\subsection{New Physics effects in \texorpdfstring{$R_{K}$}{New Physics effects in RK}}\label{sec:NPinRK}
We wish to end this section reviewing the relevance of $R_{K}$ for NP and its complementarity with other present and possibly forthcoming LFUV measurements, as $R_{K^{*}}$ and $R_{\phi}$. This completes the stage setup for the presentation and discussion of the results collected in section~\ref{sec:results}.

The ratio reported in eq.~\eqref{eq:RK_measurement} can be reasonably approximated in terms of a simple phenomenological formula. Since the minimum $q^{2}$-value probed in the bin of interest is much greater than light lepton masses and it is far from the light-cone region, one may neglect effects proportional to $m_{\ell}^{2}$ and the contribution coming from the electromagnetic dipole operator. Furthermore, one may also opt to neglect non-factorizable hadronic contributions present in $B \to K \ell^+ \ell^-$, retaining them as sub-leading effects, possibly supported by the estimates illustrated in ref~\cite{Khodjamirian:2012rm}. Then, up to percent level QED corrections discussed in ref.~\cite{Bordone:2017lsy}, similarly to refs.~\cite{DAmico:2017mtc,Geng:2017svp,Hiller:2017bzc,Celis:2017doq} we can express $R_{K}$ in terms of NP WCs simply as:
\begin{align}
\label{eq:RK_analytic}
R_{K}&[1.1,6] \simeq \nonumber\\
&\left\{ 1+0.23(C_{9,\mu}^{\rm NP}+C_{9,\mu}^{\prime, \rm{NP}})-0.25(C_{10,\mu}^{\rm NP}+C_{10,\mu}^{\prime, \rm{NP}})  \right. \nonumber \\
&+  \left. 0.057 (C_{9,\mu}^{\rm NP} C_{9,\mu}^{\prime, \rm{NP}}+C_{10,\mu}^{\rm NP} C_{10,\mu}^{\prime, \rm{NP}}) + 0.029 \left[ (C_{9,\mu}^{\rm NP})^2  \right. \right.  \nonumber \\
&+ (C_{9,\mu}^{\prime, \rm{NP}})^2 + \left. \left.  (C_{10,\mu}^{\rm NP})^2+ (C_{10,\mu}^{\prime, \rm{NP}})^2\right]\right\}\Big/\left\{1+ \mu \to e \right\}
\end{align}
or in terms of the WCs for the gauge-invariant combinations in eq.~\eqref{eq:SMEFT_op_tree}:
\begin{align}
\label{eq:RK_analytic_SMEFT}
R_{K}&[1.1,6] \simeq \nonumber\\
&\left\{ 1 + r_{\Lambda} \left[ 0.16 (C^{ed}_{2223}+C^{Qe}_{2322})  - 5.1 (C^{Ld}_{2223}+C^{LQ}_{2223})
\right. \right.  \nonumber \\
 & +  \left. \left. 13 r_{\Lambda} \left(C^{ed}_{2223} C^{Qe}_{2322}  + C^{Ld}_{2223} C^{LQ}_{2223}\right) \right. \right. \nonumber \\
 &+ \left. \left. 6.6 r_{\Lambda} \left( (C^{ed}_{2223})^2+(C^{Qe}_{2322})^2+ (C^{Ld}_{2223})^2\right.\right.\right.\nonumber\\
 &+\left.\left.\left.(C^{LQ}_{2223})^2\right)\right] \right\}\Big/\Big\{1+r_{\Lambda}\left[ \textrm{22 $\to$ 11} \right]\Big\} \ ,
\end{align}
where in the last expression $C^{LQ}_{\ell\ell23} \equiv C^{LQ^{(1)}}_{\ell\ell23} + C^{LQ^{(3)}}_{\ell\ell23} $, $r_{\Lambda} \equiv 10^{3}(v/\Lambda)^2$ and $\lambda_{t}$ is approximated as real. In both eqs.~\eqref{eq:RK_analytic}~-~\eqref{eq:RK_analytic_SMEFT},  we assume real NP coefficients. Notice that, in the cases at hand, quadratic terms are suppressed as $(\frac{\pi} {\alpha_e}\frac{v^2}{\Lambda^2})^2$, while linear terms with dimension-eight operators are suppressed as $\frac{\pi} {\alpha_e}\frac{v^4}{\Lambda^4}$. Therefore, we can meaningfully retain the quadratic terms. Moreover, we have neglected the NP contribution of (pseudo)scalar operators: while being constrained by $B \to \ell^+ \ell^-$ measurements, these operators cannot address at the same time other $b \to s \ell^+ \ell^-$ anomalies as the one(s) related to $R_{K^{*}}$.\footnote{Within the SMEFT, they cannot simultaneously explain the $R_{K}$ anomaly as well.} Finally, note that if one would like to consider also tensor structures~\cite{Hiller:2014yaa}, a combined explanation of $R_{K}$ and $R_{K^{*}}$ would not be possible \cite{Bardhan:2017xcc}, and embedding in UV models would be challenging~\cite{Bobeth:2007dw,Alonso:2014csa}.

\begin{figure*}[!t]
  \begin{center}
  \includegraphics[width=0.7\textwidth]{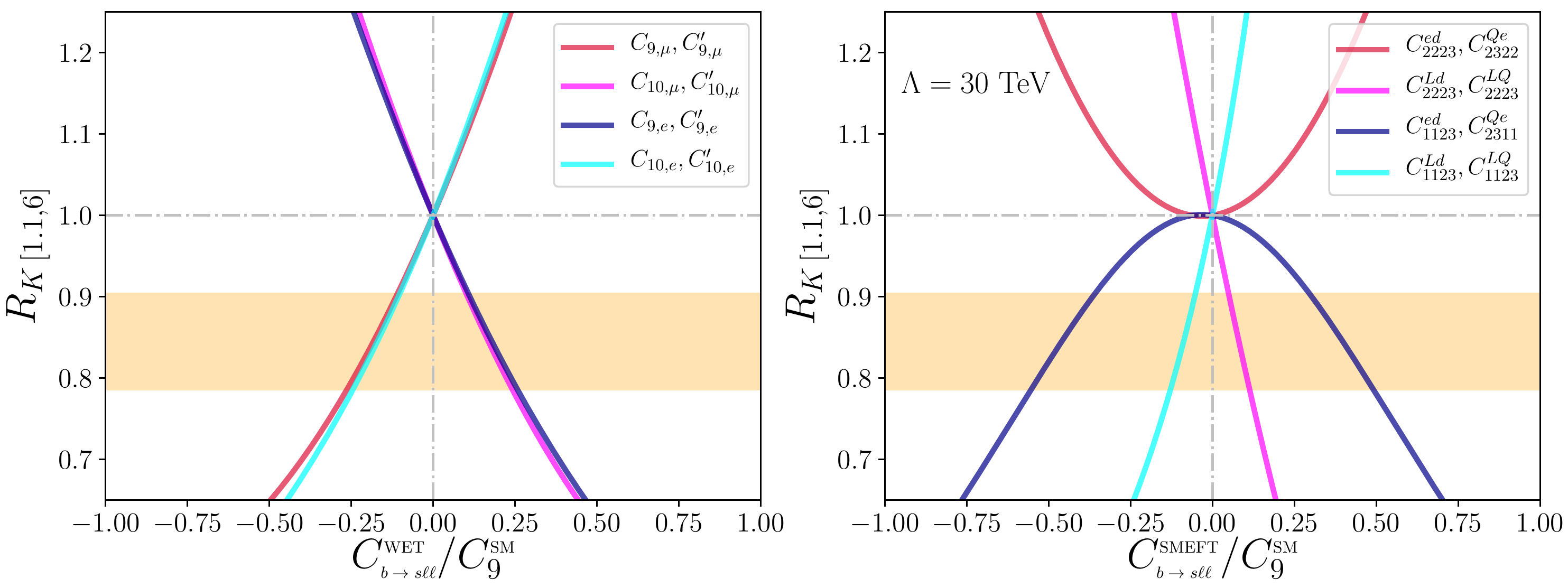}
  \end{center}
   \vspace{-0.02\textwidth}
  \caption{\emph{The impact on $R_{K}$ of each of the NP operators in the two EFT frameworks considered. The orange band highlights the new $R_{K}$ measurement, while the dashed-dot silver lines mark the SM value. In both panels the range on the $x$-axis covers up to $\mathcal{O}(1)$ effects with respect to SM short distance relatively to the low-energy scale $\mu_{b} \sim m_{b}$. SMEFT contributions are assumed to be generated at the NP scale of 30~TeV. }}
  \label{fig:RK_EFT}
\end{figure*}

In fig.~\ref{fig:RK_EFT}  we show the impact on $R_{K}$ in the bin discussed so far of each of the operators considered here in the WET (left panel), see  eq.~\eqref{eq:RK_analytic}, and in the SMEFT (right panel), see eq.~\eqref{eq:RK_analytic_SMEFT}. The range on the $x$-axis in fig.~\ref{fig:RK_EFT} covers $\mathcal{O}(1)$ effects relative to the short-distance SM contributions. The SM limit is emphasized by the silver dot-dashed lines, and the new $R_{K}$ measurement is represented by the horizontal orange band, drawn according to experimental central value and standard deviation, see eq.~\eqref{eq:RK_measurement}. 

It is clear from what is depicted for the WET that operators featuring both left-handed and right-handed $b \to s$ currents are eligible for a satisfactory explanation of the measured value of $R_{K}$. In particular, one cannot distinguish effective couplings related to left-handed or right-handed $b \to s$ currents within operators that have the same leptonic structure, since they constructively interfere in eq.~\eqref{eq:RK_analytic}. Moreover, as highlighted in the plot, the NP contribution required to explain the present $R_{K}$ measurement is now about one fifth of the SM one. Therefore, the linearized limit of eq.~\eqref{eq:RK_analytic} may be a good approximation in order to appreciate how LFUV effects actually probe the $\mu - e$ combination of the leptonic current. This fact is captured in the plot by the mirror-like behaviour of red-blue and magenta-cyan line pairs with respect to the SM limit. In the same panel, axial and vectorial leptonic effective couplings turn out also to be mirror-like as a reflection of the SM result: $C_{9}^{\textrm{\tiny SM}} \sim -C_{10}^{\textrm{\tiny SM}}$.

Similar considerations apply to the case of SMEFT operators with leptonic weak doublets, requiring only about $15 \%$ of the SM WC value for $Q_{9V,10A}$ to accommodate $R_{K}$ within a NP scale of $\Lambda = 30$ TeV, yielding $|C^{LQ,Ld}_{\ell\ell23}| \sim 0.8$. However, the correlations induced by the $SU(2)_{L} \times U(1)_{Y}$ gauge symmetry no longer allow a full set of 8 different viable solutions for the $R_{K}$ anomaly. From the right panel of fig.~\ref{fig:RK_EFT}, NP effects from SMEFT operators featuring exclusively right-handed muonic currents are ruled out, while the electronic counterparts are still available at the expense of larger NP effects, $\gtrsim 35 \%$ of the SM short-distance physics for the same $\Lambda = 30$~TeV. Interestingly, among the RGE-induced set of operators reported in eq.~\eqref{eq:SMEFT_op_loop_lu} we can then exclude $O^{eu}_{2233}$.

The bottom line drawn from fig.~\ref{fig:RK_EFT} refers merely to the inspection of one single operator at a time contributing to $R_{K}$. However, in the broader picture offered by the whole set of available $b \to s \ell^+ \ell^-$ measurements, we may end up with observable quantities that provide information on NP orthogonal to what outlined from the $R_{K}$ anatomy. Of particular significance, the LFUV ratio $R_{K^*}$  has been originally recognized in ref.~\cite{Hiller:2014ula} to be a complementary probe of NP with respect to $R_{K}$. 
Indeed, while being very sensitive to BSM physics, in the limit where the longitudinal polarization fraction in the $B \to K^* \ell^+ \ell^-$ channel were exactly equal to unity, $R_{K^*}$ would be fully sensitive to destructive interference between left-handed and right-handed $b \to s$ effective couplings, and hence complementary to what is depicted in eq.~\eqref{eq:RK_analytic} for $R_{K}$. In the same spirit of ref.~\cite{Hiller:2017bzc}, one may then look at the ratio of measured LFUV ratios, i.e. the $R_{K^*}$ experimental value in the bin [1.1,6]~GeV$^2$ from~\cite{Aaij:2017vbb,Abdesselam:2019wac} over the new measurement of $R_{K}$ from~\cite{Aaij:2019wad},
\begin{equation} \label{eq:RKst_over_RK}
    R_{K^*}[1.1,6]/R_{K}[1.1,6] \simeq 0.86 \pm 0.13 ,
\end{equation}
discovering a hint for non-zero effective couplings for the operators $Q_{9V,10A}^{\prime}$, part of eq.~\eqref{eq:H_sl}. Moreover, going beyond LFUV observables, one may supplement the information of eq.~\eqref{eq:RKst_over_RK} with the measurements of $B \to K^{(*)} \mu^+ \mu^-$ branching fractions and, most importantly, with the related angular analyses. In particular, with the inclusion of the angular observable $P_{5}^{\prime}$ -- particularly sensitive to NP effects in the operator $Q_{9V}$~\cite{Descotes-Genon:2013wba,Descotes-Genon:2015uva} -- one may end up concluding that the new experimental value of $R_K$ currently points to effects in both left-handed and right-handed $b \to s$ currents of dimension-six operators built up with the muonic vectorial current. As such, previously claimed minimal solutions for $b \to s \ell^+ \ell^-$ anomalies -- involving only $Q_{9V}$ or $O^{LQ}$ -- would now seem to be more disfavoured in view of the need for NP effects also in right-handed currents.

Unfortunately, the above qualitative considerations remain subject to several uncertainties. First of all, the longitudinal polarization fraction of $B \to K^* \ell^+ \ell^-$ in the bin of interest is not equal to unity~\cite{Hiller:2013cza}: this fact already makes the $R_{K^*}$ observable less orthogonal to $R_{K}$ in the study of NP~\cite{Geng:2017svp}. Moreover, longitudinal and transverse polarization fractions are sensitive to $\Lambda_{\textrm{\tiny QCD}}/m_{b}$ power corrections not fully under control~\cite{Hiller:2014ula,Geng:2017svp}. This also suggests an experimental information that would be important to handle in the future: the measurement of $R_{K^*}^{\textrm{\tiny T,L}}[1.1,6]$, i.e. the ratio of longitudinal and transverse parts of the $B \to K^* \ell^+ \ell^-$ amplitude in the $q^2$-bin [1.1,6] GeV$^2$. These quantities would be less sensitive to unknown hadronic effects, and distinctively sensitive to NP effects in $C^{}_{9,10} \pm C_{9,10}^{\prime}$ combinations~\cite{Geng:2017svp,Ciuchini:2017mik}. Similar information could be extracted from $B_{s} \to \phi \ell^+ \ell^-$ as well. 

Secondly, as already noted at the beginning of section~\ref{sec:framework}, the same angular observables measured in $B \to K^{*} \mu^+ \mu^-$ are also affected by non-factorizable QCD effects. Only corresponding LFUV combinations as the one proposed in ref.~\cite{Capdevila:2016ivx,Serra:2016ivr} and recently reanalyzed in~\cite{Alguero:2019pjc} may help to disentangle genuine NP effects from hadronic contributions theoretically not well-understood. At present, the only available measurement of this sort is given by Belle~\cite{Wehle:2016yoi} and it is (unfortunately) of limited statistical significance, but more will certainly come in the next years~\cite{Kou:2018nap}.

{\sloppy
In the end, a careful study of $b \to s \ell^+ \ell^-$ anomalies calls for a global analysis that can go well beyond the qualitative picture highlighted above, taking care of all the aforementioned subtleties in a framework where a non-trivial interplay between genuine NP effects and hadronic contributions is allowed. The analysis performed in this study, presented in section~\ref{sec:results}, is precisely dedicated to make interpretations of the underlying NP scenarios behind current $b \to s \ell^+ \ell^- $ anomalies as robust as possible.
}
%%%%%%%%%%%%%%%%%%%%%%%%%%%%%%%%%%%%%%%%%%%%%%%%%%%%%%%%

\section{Experimental and theoretical input}
\label{sec:inputs}

In this section we plan to review the baseline of our analysis, the experimental dataset included, and the assumptions made throughout this work. In the present study we perform a global analysis on a comprehensive set of $b \to s \ell^+ \ell^-$ data with state-of-the-art theoretical computations, within a Bayesian framework. 

We adopt for this matter the public
\texttt{HEPfit} package~\cite{HEPfit}, whose Markov Chain Monte Carlo (MCMC) analysis framework employs the Bayesian Analysis Toolkit (BAT)~\cite{Caldwell:2008fw}. In our MCMC analysis we vary from a minimum of 60 to a maximum of 80 parameters on a case by case basis. Within the Metropolis-Hastings algorithm implemented in BAT, we set up, for the scenarios presented in section~\ref{sec:results}, MCMC runs involving 240 chains with a total of 2.4 million events per run, collected after an equivalent number of pre-run iterations.

We perform a Bayesian model comparison between different scenarios evaluating for each of them an \textit{Information Criterion} (IC). This quantity offers an approximation of the predictive accuracy of the model \cite{2013arXiv1307.5928G}, and it is characterized by the mean and the variance of the posterior probability density function (\emph{p.d.f.}) of the log-likelihood $\log \mathcal{L}$, see ref.~\cite{IC}, 
\begin{equation}\label{eq:IC}
   IC \equiv -2 \overline{\log \mathcal{L}} \, + \, 4 \sigma^{2}_{\log \mathcal{L}} \ ,
\end{equation}
where the first term gives an estimate of the predictive accuracy (actually, an overestimate since the same data have already been used in the fit), and the second term corrects for the overestimate by adding a penalty factor which counts the effective number of fitted parameters.
Model selection between two scenarios proceeds according to the smallest IC value reported and the extent to which a model should be preferred over another one follows the canonical scale of evidence of ref.~\cite{BayesFactors}, related in this context to (positive) IC differences. In the following section~\ref{sec:results}, for convenience we are going to present a discussion based on $\Delta IC \equiv IC_{SM} - IC_{NP}$.\footnote{It is interesting to perform a SM global fit in order to have reference values for the $\emph{IC}$ to compare with. The fits yield an $\emph{IC}$ of 193 for the PDD approach, and 215 for the PMD one. Recalling that models with smaller values for the \textit{IC} are preferred, the PDD approach provides a better SM fit compared to the PMD one, since anomalies in the angular analysis of $B \to K^* \mu^+ \mu^-$ can be accommodated through larger long-distance contributions, see ref.~\cite{Ciuchini:2015qxb}.} In particular, we quote in tables~\ref{tab:WC_WET} and~\ref{tab:WC_SMEFT} for each NP scenario the  $\Delta IC$ value. We wish to stress that a larger value of $\Delta IC$ corresponds to a better improvement of the model compared to the SM.

Regarding the experimental dataset considered in this study, we include all the most recent measurements related to $b \to s \ell^+ \ell^-$ transitions that can have a valuable impact in our global fit. We briefly list them below with some additional comments: \begin{itemize}
    \item All the angular observables and branching ratio information on $B \to K^* \mu^+\mu^-$ from the experimental results obtained by LHCb~\cite{Aaij:2015oid,Aaij:2016flj}, Belle~\cite{Wehle:2016yoi}, ATLAS~\cite{Aaboud:2018krd} and CMS~\cite{Khachatryan:2015isa,Sirunyan:2017dhj} collaborations. When available, we always take into account experimental correlations between the measurements performed in the same bin. Note that we restrict here only to the large-recoil region, i.e. $q^{2}$ values below the $J/\psi$ resonance, excluding measurements in the (theoretically challenging) broad-charmonium region.
    \item $B \to K^* e^+e^-$ angular observables from LHCb in the available $q^2$ bin, $[0.002,1.12]$ GeV$^2$~\cite{Aaij:2015dea}. 
    \item Angular observables and branching ratio of $B_{s} \to \phi \mu^+\mu^-$ provided by LHCb~\cite{Aaij:2015esa}.
    \item Branching ratio of $B_s\to \mu^+\mu^-$ measured by LHCb \cite{Aaij:2017vad}, CMS~\cite{Chatrchyan:2013bka}, and most recently by ATLAS~\cite{Aaboud:2018mst}. Note that we also employ the upper limit on $B_s\to e^+e^-$ decay reported by HFLAV~\cite{Amhis:2016xyh}, useful for the study of NP coupled to electrons~\cite{Alonso:2017bff}.
    \item Branching ratios for $B^{(+)} \to K^{(+)} \mu^+\mu^-$ decays in the large-recoil region by LHCb~\cite{Aaij:2014pli}.
    \item Branching ratios for the radiative decay $B \to K^* \gamma$, from HFLAV~\cite{Amhis:2016xyh}, and for $B_{s} \to \phi \gamma$ as measured by LHCb~\cite{Aaij:2012ita}. While we are not going to consider NP effects in dipole operators, these measurements are relevant in our PDD approach.
    \item LFUV ratios including the very recent updates: $R_{K^*}$ in both $q^2$ bins, $[0.045,1.1]$ GeV$^{2}$ and $[1.1,6]$ GeV$^{2}$ \cite{Aaij:2017vbb,Abdesselam:2019wac}, and the $R_K$ measurement~\cite{Aaij:2019wad}.
\end{itemize}

Concerning the inputs from the theory side, our analysis is characterized in particular by the set of parameters defining form factors and non-factorizable hadronic contributions. For the former we rely on the computation presented in~\cite{Straub:2015ica} for $B \to K^*$ and $B_s \to \phi$ amplitudes, as we take into account experimental information coming from both channels;\footnote{The latest LCSR update from ref.~\cite{Gubernari:2018wyi}, while providing an important independent cross-check of several results present in~\cite{Straub:2015ica}, does not include the estimate of $B_{s} \to \phi$ matrix elements.} for the $B \to K$ channel, we adopt lattice QCD results extrapolated from the zero-recoil region to low-$q^{2}$ values as provided in ref.~\cite{Bailey:2015dka}. For all the form-factor parameters adopted in this study we adopt multi-variate Gaussian distribution priors in order to include correlation matrices reported in the relevant literature. 

Regarding the non-factorizable part of the amplitudes, we include hard-gluon contributions following what already outlined in detail in our previous work~\cite{Ciuchini:2017mik}, while we proceed here differently for what regards our treatment of soft-gluon exchanges. 

In the PMD approach, we do not expand eq.~\eqref{eq:hlambda} in powers of $q^{2}$, but we directly express it in terms of the phenomenological expression given by eq.~(7.14) of ref.~\cite{Khodjamirian:2010vf},\footnote{In \cite{Ciuchini:2017mik} we were power-expanding $h_\lambda$ correlators and enforcing the numerical results obtained from ref.~\cite{Khodjamirian:2010vf} in the whole large-recoil region as theory weights in the likelihood. Our new procedure for the PMD approach allows now to adopt the outcome of ref.~\cite{Khodjamirian:2010vf} genuinely as a set of flat priors.}  and we flatly distribute all the involved parameters according to the ranges reported in table~2 of the same reference. In order to allow for imaginary parts as well, each of the three charm-loop amplitudes in ref.~\cite{Khodjamirian:2010vf} is multiplied by a complex phase, flatly varying each angle within [0, 2$\pi$), yielding a total of 12 parameters to describe the non-perturbative hadronic contributions within this approach.

In the PDD approach, corresponding here to the parameterization in  eq.~\eqref{eq:newexp}, we allow for flat priors on the absolute values of $h_\lambda^{(i)}$ coefficients and enforce as a theory weight in the likelihood the results obtained at $q^2=0,1$ GeV$^2$ within the LCSR estimate of ref.~\cite{Khodjamirian:2010vf}. The following prior ranges are chosen in order to well determine the \emph{p.d.f.} of each parameter:
\bea \label{eq:h_lambda_range}
|h_-^{(0)}| \;\;&\in&\;\; [0,0.1]\,,
\ \ \ \ \, |h_-^{(1)}|  \in  [0,4]\,,
\ \ \ \ \ \ \ |h_-^{(2)}|  \in  [0,0.0001]\,, \nonumber \\
|h_+^{(0)}| \;\;&\in&\;\; [0,0.0001]\,,
|h_+^{(1)}|  \in  [0,0.0005]\,,
|h_+^{(2)}|  \in  [0,0.0001]\,, \nonumber \\
|h_0^{(0)}| \;\;&\in&\;\; [0,0.002]\,, 
\ \, |h_0^{(1)}|  \in  [0,0.0004]\,,
\eea
i.e. a larger range for the above priors would not alter our results.
Most importantly, each of the coefficients related to the absolute values in eq.~\eqref{eq:h_lambda_range} has a corresponding complex free phase. Therefore, our PDD approach is defined by a a total of 16 parameters.
We used the same set of parameters in eq.~\eqref{eq:h_lambda_range} to also describe the soft-gluon contributions in the case of $B_s \to \phi$, leaving possibly interesting $SU(3)_{F}$-breaking effects to a future investigation. Eventually, for $B \to K$ transitions we only include non-factorizable hadronic effects coming from hard-gluon exchanges, motivated by the results of ref.~\cite{Khodjamirian:2012rm}.
\footnote{Nevertheless, we have tested explicitly for the case of the scenario involving $C_{9, \mu}^{NP}$ that introducing the equivalent of the PDD approach also for the $B \to K$ channel does not have a relevant impact on the results of our fit.}

We conclude this section mentioning that the rest of the SM parameters varied in our analysis can be found in table~1 of ref.~\cite{Ciuchini:2017mik}, while for NP WCs, we adopt in general flat priors in the range [-10, 10], assuming they are real. Note that some of the NP scenarios here considered showed multi-modal \emph{p.d.f.}s. In such cases we focused on the NP solution closer to the SM limit, identified by $C_i^{\rm NP} = 0$. Finally, we point out that all our findings for the study of the SMEFT in section~\ref{sec:results} assume a NP scale set to 30~TeV. In order to read out SMEFT WCs at a different NP scale $\Lambda$, one needs to re-scale the results given in section~\ref{sec:results} appropriately.

%%%%%%%%%%%%%%%%%%%%%%%%%%%%%%%%%%%%%%%%%%%%%%%%%%%%%%%%

\section{EFT results from the new \texorpdfstring{$R_{K}$}{RK} measurement}
\label{sec:results}

In this section we present our results. We perform several fits to the experimental measurements listed in section~\ref{sec:inputs}, differentiated by the set of NP WC(s) considered. We employ the PDD approach in all the scenarios examined, while exploring the PMD approach only when it can provide a satisfactory fit to current data, i.e. when NP effects built up from left-handed $b \to s$ currents coupled to vector-like (purely left-handed) muonic currents are involved in the WET (SMEFT) formalism. The goodness of the fit is here evaluated by means of the \textit{IC}, defined in eq.~\eqref{eq:IC}, while the details of the PMD and PDD approaches have been presented in section~\ref{sec:SDvsLS}.

The primary goal of this analysis consists in the study of the interplay between the new $R_K$ measurement and NP. In particular, we investigate whether the update of $R_K$ combined with the current $R_{K^*}$ measurement can actually have an impact on the viable solutions to the anomalies in $b \to s$ transitions allowed by the previous $R_{K}$ from Run~I of LHC. To this end, we report in tables~\ref{tab:WC_WET} and~\ref{tab:WC_SMEFT} the results for the fitted values of the WCs in each of the models scrutinized here, employing the WET and the SMEFT formalism respectively. $\Delta IC$ values are also reported in the same table, marking the improvement with respect to the SM, see the discussion following eq.~\eqref{eq:IC} in section~\ref{sec:inputs}. Finally, results for what we retain as key observables for our study are also reported in tables~\ref{tab:OBS_WET} and~\ref{tab:OBS_SMEFT}, differentiating once again scenarios in the WET or in the SMEFT, respectively. 

\begin{figure*}[!t]
  \centering
  \includegraphics[width=0.8\textwidth]{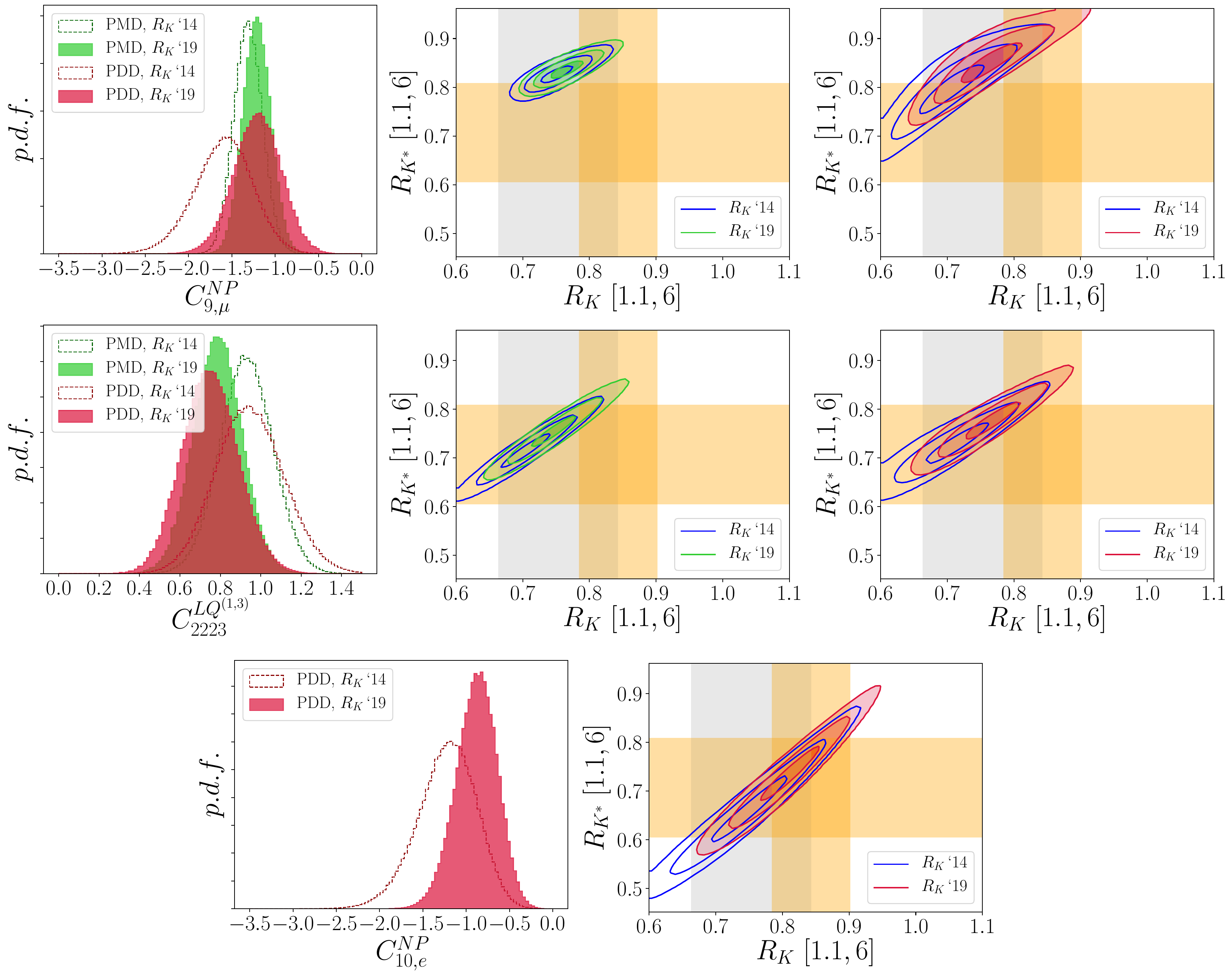}
  \vspace{-0.02\textwidth}
  \caption{\emph{First row: probability density function ({p.d.f.}) 
  for the WC $C_{9,\mu}^{\rm NP}$, where the green-filled {p.d.f.} shows
  the posterior obtained in the PMD approach after the inclusion
  of the updated measurement for $R_K$, while the red-filled
  {p.d.f.} is the analogous posterior within the PDD approach 
  (the dashed posteriors are the ones obtained employing the 2014
  $R_K$ measurement); the following panels report the combined
  2D {p.d.f.} of the corresponding results for $R_K$ and $R_{K^*}$, where the colour 
  scheme follows the one employed in the first panel. The horizontal band
  corresponds to the 1$\sigma$ experimental region for $R_{K^*}$ from \cite{Aaij:2017vbb}, while the
  two vertical bands corresponds to the previous and the current 
  1$\sigma$ experimental regions for $R_K$.
  Second row: analogous to the first row, but relative to the WC $C_{2223}^{LQ}$.
  Third row: analogous to the first row, but relative to the WC $C_{10,e}^{\rm NP}$.}}
  \label{fig:1D_mu_e}
\end{figure*}

\begin{figure*}[!ht]
  \centering
  \includegraphics[width=\textwidth]{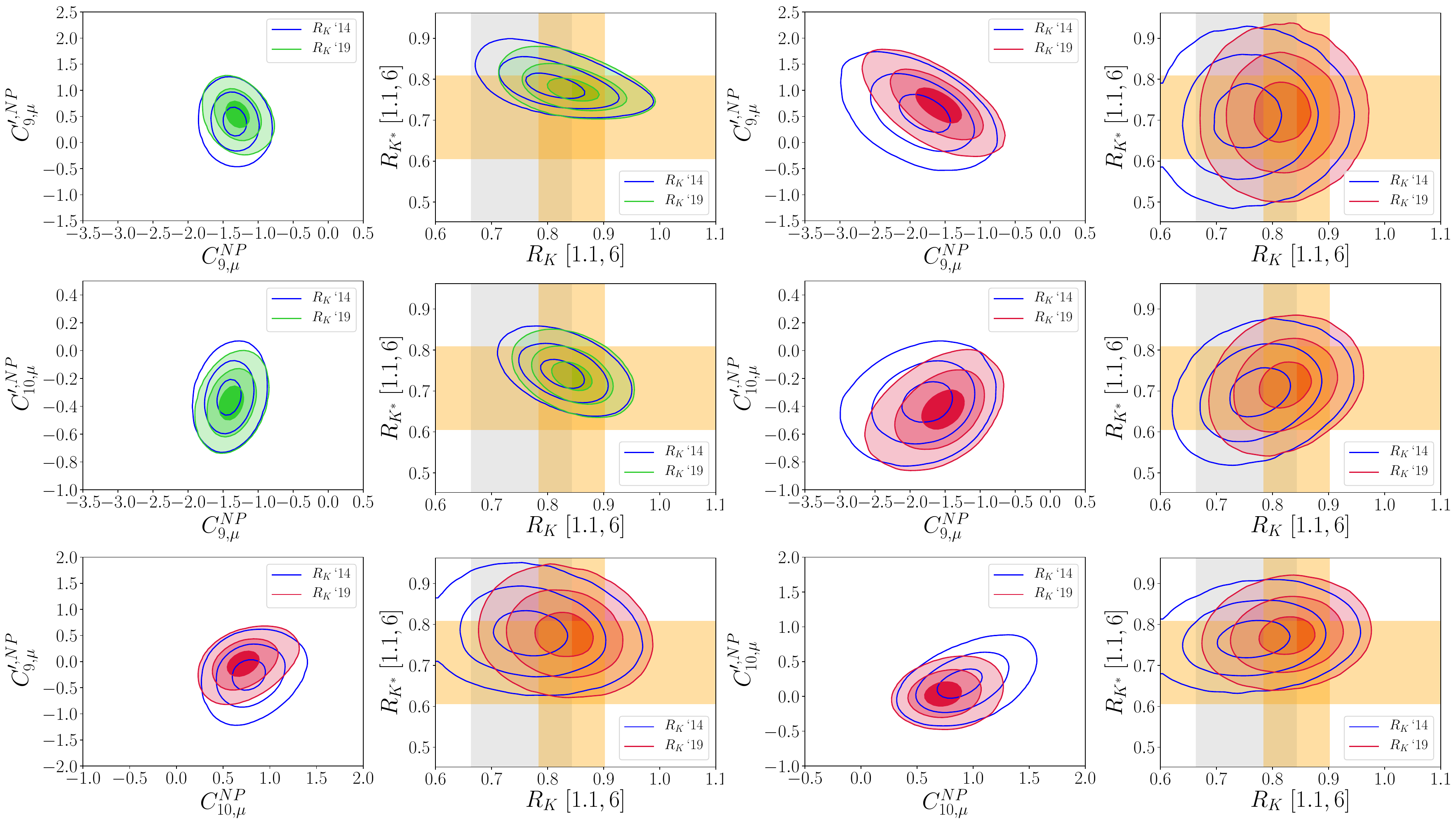}
  \vspace{-0.02\textwidth}
  \caption{\emph{First row: 2D {p.d.f.}
  for the scenario with WCs $(C_{9,\mu}^{\rm NP}, C_{9,\mu}^{\prime, \rm{NP}})$
  and combined
  2D {p.d.f.} of the corresponding results for $R_K$ and $R_{K^*}$ in both the 
  PMD and the PDD approaches, following the colour scheme defined below fig.~\ref{fig:1D_mu_e}.
  Second row: analogous to the first row, but relative to the WCs $(C_{9,\mu}^{\rm NP}, C_{10,\mu}^{\prime, \rm{NP}})$.
  Third row: analogous to the first row, but in the PDD approach only and relative to the 
  WCs $(C_{10,\mu}^{\rm NP}, C_{9,\mu}^{\prime, \rm{NP}})$ and $(C_{10,\mu}^{\rm NP}, C_{10,\mu}^{\prime, \rm{NP}})$.}}
  \label{fig:2D_WET_mu}
\end{figure*}

Our main results are illustrated here as follows. The posterior \emph{p.d.f.}s obtained for NP coefficients are shown in figures~\ref{fig:1D_mu_e}~-~\ref{fig:SMEFT_all}. Fig.~\ref{fig:1D_mu_e} refers to scenarios where a single WC is taken into account. Figs.~\ref{fig:C9_22_C9_11}~-~\ref{fig:CLQ1_CQe} involve fits with two operators at the same time and correspond to two popular benchmarks previously studied in literature. Figs.~\ref{fig:2D_WET_mu}~-~\ref{fig:2D_SMEFT_e} correspond to 2D scenarios where NP effects in the form of $b \to s$ right-handed currents are present. Finally, in fig.~\ref{fig:SMEFT_all} the result for the largest set of SM gauge-invariant operators probed by current experimental data is presented.

For each considered scenario, we show both the posterior \emph{p.d.f.}(s) of the NP WC(s) obtained employing the previous measurement of $R_K$~\cite{Aaij:2014ora}, and the new one from ref.~\cite{Aaij:2019wad}. This allows one to easily compare the impact of the new $R_K$ measurement in our analysis. Moreover, in order to have a further insight on the role of LFUV observables as $R_K$ and $R_{K^*}$, we also provide in the same figures the joint probability distribution of these ratios extracted from our fits. We give these results employing again either the~2014 measurement of $R_K$ or its 2019~update. Our attempt is to investigate whether scenarios previously capable of addressing the anomalies in both the LFUV ratios remain viable after the $R_K$ value recently presented in~\cite{Aaij:2019wad}.

\begin{table}[!t]
\centering
\renewcommand{\arraystretch}{1.5}
{\footnotesize
\begin{tabular}{|c|cc|}
\hline
& mean(rms) & $\Delta IC$ \\
\hline
\multirow{2}{*}{$ C_{9,\mu}^{\rm NP} $}
& -1.20(27) & 14 \\
&\cellcolor{Gray}  -1.21(16) &\cellcolor{Gray}  50 \\
\hline
$ C_{10,e}^{\rm NP} $
& -0.87(24) & 15 \\
\hline
\hline
\multirow{2}{*}{$ (C_{9,\mu}^{\rm NP},C_{9,e}^{\rm NP}) $}
& (-1.61(48), -0.56(53)) & 13 \\
&\cellcolor{Gray}  (-1.28(18), -0.27(34)) &\cellcolor{Gray}  48\\
\hline
\multirow{2}{*}{$ (C_{9,\mu}^{\rm NP},C_{9,\mu}^{\prime, \rm{NP}}) $}
& (-1.61(33), 0.72(34)) & 17 \\
&\cellcolor{Gray}  (-1.30(15), 0.53(24)) &\cellcolor{Gray}  54 \\
\hline
\multirow{2}{*}{$ (C_{9,\mu}^{\rm NP},C_{10,\mu}^{\prime, \rm{NP}}) $}
& (-1.55(32), -0.44(14)) & 24 \\
&\cellcolor{Gray}  (-1.38(16), -0.37(12)) &\cellcolor{Gray}  61 \\
\hline
$ (C_{10,\mu}^{\rm NP},C_{9,\mu}^{\prime, \rm{NP}}) $
& (0.73(17), -0.04(24)) & 17 \\
\hline
$ (C_{10,\mu}^{\rm NP},C_{10,\mu}^{\prime, \rm{NP}}) $
& (0.75(16), 0.04(17)) & 16 \\
\hline
$ (C_{9,e}^{\rm NP},C_{9,e}^{\prime, \rm{NP}}) $
& (1.51(38), -0.81(37)) & 10 \\
\hline
$ (C_{9,e}^{\rm NP},C_{10,e}^{\prime, \rm{NP}}) $
& (1.36(32), 0.87(40)) & 11 \\
\hline
$ (C_{10,e}^{\rm NP},C_{9,e}^{\prime, \rm{NP}}) $
& (-1.06(54), -0.46(46)) & 12 \\
\hline
$ (C_{10,e}^{\rm NP}, C_{10,e}^{\prime, \rm{NP}}) $
& (-1.01(28), 0.29(29)) & 12 \\
\hline
\end{tabular}
}
\caption{\em Values of the WET WCs fit from data in all the considered scenarios along with relative $\Delta IC$. The gray rows highlight the PMD results when this approach can be used to address the experimental data in a particular scenario. The PDD results are presented for all cases. For the definition of the two approaches, see section~\ref{sec:SDvsLS}.
\label{tab:WC_WET}}
\end{table}

\begin{table}[!t]
\centering
\renewcommand{\arraystretch}{1.5}
{\footnotesize
\begin{tabular}{|c|cc|}
\hline
& mean(rms) & $\Delta IC$ \\
\hline
\multirow{2}{*}{$ C^{LQ}_{2223} $}
& 0.75(14) & 23 \\
&\cellcolor{Gray} 0.79(12) & \cellcolor{Gray} 37 \\
\hline
\hline
\multirow{2}{*}{$ (C^{LQ}_{2223},C^{Qe}_{2322}) $}
& (0.78(18), 0.06(32)) & 21 \\
&\cellcolor{Gray} (0.94(12), 0.67(17)) &\cellcolor{Gray} 50 \\
\hline
$ (C^{LQ}_{1123},C^{Qe}_{2311}) $
& (-0.51(29), 0.96(70)) & 12 \\
\hline
\multirow{2}{*}{$ (C^{LQ}_{2223},C^{ed}_{2223}) $}
& (0.74(15), 0.16(33)) & 21 \\
&\cellcolor{Gray} (0.81(12), -0.19(29)) &\cellcolor{Gray} 35 \\
\hline
\multirow{2}{*}{$ (C^{LQ}_{2223},C^{Ld}_{2223}) $}
& (0.81(15), -0.20(15)) &  22 \\
&\cg (0.80(12), -0.11(12)) &\cg 36 \\
\hline
$ (C^{LQ}_{1123},C^{ed}_{1123}) $
& (-0.08(73), -2.1(14)) & 12 \\
\hline
$ (C^{LQ}_{1123},C^{Ld}_{1123}) $
& (-0.93(27), 0.39(27)) & 12 \\
\hline
$ (C^{Qe}_{2311},C^{ed}_{1123}) $
& (-0.2(18), -1.3(18)) & 8 \\
\hline
$ (C^{Qe}_{2311},C^{Ld}_{1123}) $
& (1.77(47), -0.18(24)) & 9 \\
\hline
\end{tabular}
}
\caption{\em Values of the SMEFT WCs fit from data in all the considered scenarios along with  relative $\Delta IC$. The gray rows highlight the PMD results when it can address the experimental data in a particular scenario. The PDD results are presented for all cases. For the definition of the two approaches, see section~\ref{sec:SDvsLS}. 
\label{tab:WC_SMEFT}}
\end{table}

\subsection{New Physics in \texorpdfstring{$b \to s$}{btos} left-handed currents}
Let us start our discussion examining the simple situation where the underlying BSM dynamics can be encoded in a single operator. We focus here on three different benchmarks, namely we assume NP effects feed into a left-handed $b \to s $ current coupled to: 
\begin{itemize}
    \item[\textit{i)}] a vectorial muonic current, i.e. $C_{9,\mu}^{\rm NP}\,$;
    \item[\textit{ii)}] a purely left-handed  muonic current, i.e. $C_{2223}^{LQ}\,$;
    \item[\textit{iii)}] an axial electronic current, i.e. $C_{10,e}^{\rm NP}\,$.
\end{itemize}
Leaving aside for a moment the role of LFUV ratios, one should note that within the PMD approach: \textit{i)} can provide an optimal outcome for the $B \to K^* \mu^+ \mu^- $ angular analysis; \textit{ii)} can provide a satisfactory description of the angular dataset (but worse than \textit{i)}, being at the same time sensitive to observables as the forward-backward asymmetry measured for $B \to K^{(*)} \mu^+ \mu^-$ and the branching fraction of $B \to \mu^+ \mu^-$); \textit{iii)} badly fails to describe such a complex experimental dataset as long as one does not allow for large QCD power corrections as in the PDD approach~\cite{Ciuchini:2017mik}\footnote{Electron LFUV couplings arising from a $Z'$ mediator may be also probed by atomic-physics data~\cite{DAmbrosio:2019tph}.}. 

From fig.~\ref{fig:1D_mu_e} we can supplement this picture with the measurement of LFUV ratios. We note how the impact of the $R_{K}$ measurement can be particularly relevant for the final outcome. Concerning case \textit{i)}, we see that the interplay of $R_{K}$ and $R_{K^*}$ does not favour this scenario any longer within the $1\sigma$ regions highlighted by the orange bands in the plot. This is in contrast to the previous situation given by the 2014 measurement of $R_{K}$ and represented in fig.~\ref{fig:1D_mu_e} by the vertical gray band. Most importantly, the tension arising in this NP scenario when accounting for current LFUV ratio measurements is also evident in the case of the PDD approach (right panel in the first row of the figure). 

A different outcome arises from the inspection of the same fig.~\ref{fig:1D_mu_e} together with the help of the $\Delta IC$ in tables~\ref{tab:WC_WET}~-~\ref{tab:WC_SMEFT} for the NP scenario \textit{ii)}. In this case, the description of LFUV ratios $R_{K}$ and $R_{K^*}$ turned out to be optimal before the advent of the present $R_{K}$ update. From the $\Delta IC$ value in table~\ref{tab:WC_SMEFT} and the comparison with the one given in table~\ref{tab:WC_WET} for the scenario \textit{i)}, we can conclude that in the PMD approach the operator $O^{LQ}_{2223}$ is not so well supported by $b \to s \ell^+ \ell^-$ data. In particular, the new $R_{K}$ value is not addressed within the 1$\sigma$ experimental uncertainty. This fact adds to the global information arising from the rest of the observables in the fit: as a consequence, in the PMD framework NP effects in $O^{LQ}_{2223}$ are now disfavoured with respect to contributions present in $Q_{9V,\mu}$. Interestingly, in the PDD approach the comparison between these two scenarios is completely reversed: in particular, an inspection of the corresponding $\Delta IC$ shows how allowing for larger QCD power corrections makes \textit{ii)} one of the scenarios favoured by data within the PDD framework. Indeed, the set of angular observables and the branching fraction of $ B \to \mu^+ \mu^-$ can now perfectly coexist in this NP scenario; the only tension remaining in the fit of \textit{ii)} is then related to this new update for $R_{K}$, shown in the right panel of central row in fig.~\ref{fig:1D_mu_e}, which as of now turns out to be a very mild one. 

\begin{figure*}[t!]
  \centering
  \includegraphics[width=0.7\textwidth]{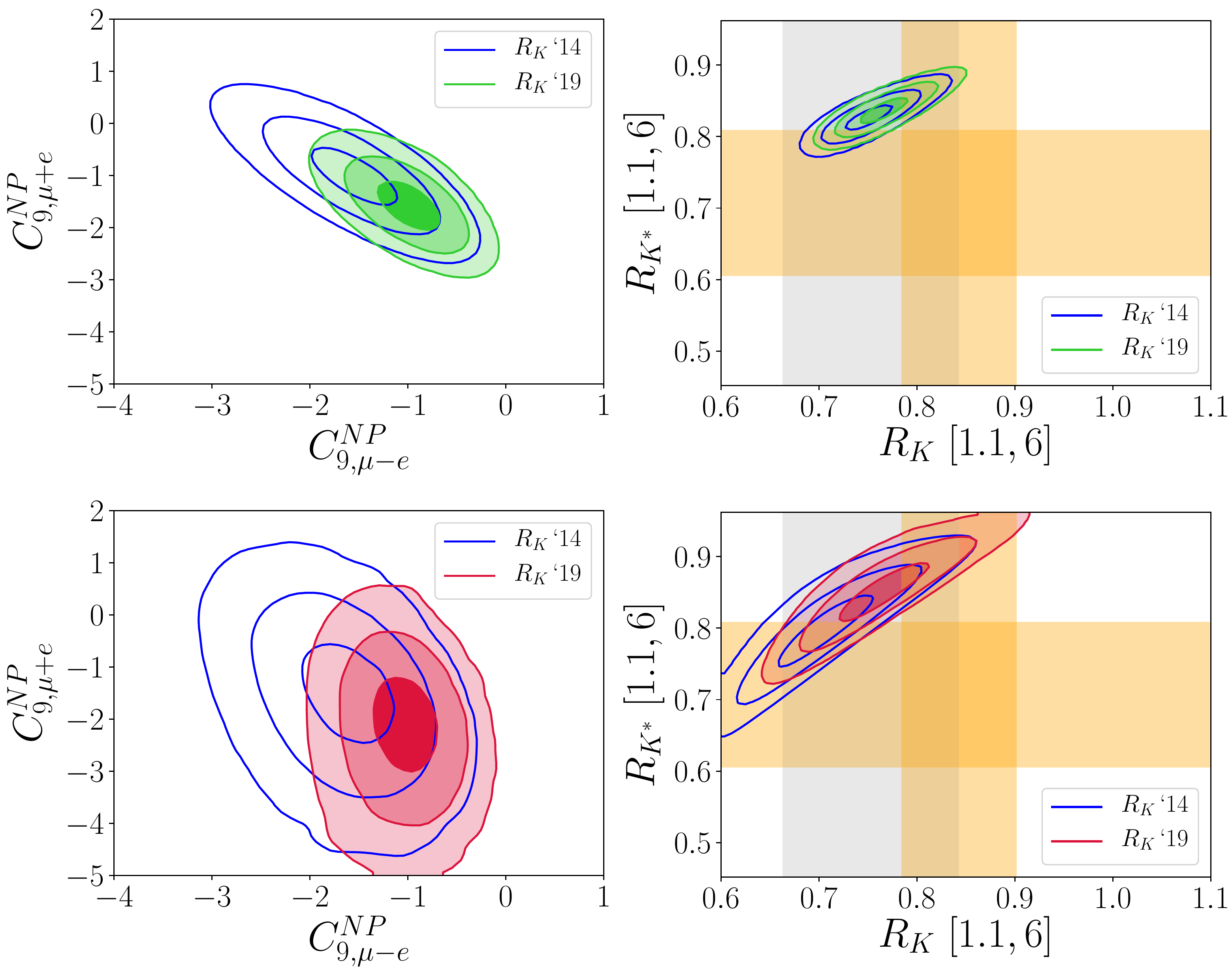}
  \vspace{-0.02\textwidth}
  \caption{\emph{2D {p.d.f.}
  for the scenario with WCs $(C_{9,\mu}^{\rm NP}, C_{9,e}^{\rm NP})$
  and combined 2D {p.d.f.} of the corresponding results 
  for $R_K$ and $R_{K^*}$ in both the PMD and the PDD approaches, following the colour scheme defined below fig.~\ref{fig:1D_mu_e}. In order to highlight the role of LFUV observables in this scenario we show in the left panels the WCs in the $\mu \pm e$ combination.}}
  \label{fig:C9_22_C9_11}
\end{figure*}

Therefore, we wish to note that -- beyond the importance of the present $R_{K}$ update -- the assumptions made in the size of the hadronic contributions when comparing NP scenarios turn out to be crucial. The most evident case of this sort is certainly \textit{iii)}. Within a conservative approach to QCD power corrections in the $B \to K^* \ell^+ \ell^-$ amplitude, this scenario offers a perfectly viable fit to $b \to s \ell^+ \ell^-$ data. In particular, \textit{iii)} provides an optimal description of LFUV ratios according to what depicted in the last row of fig.~\ref{fig:1D_mu_e}. However, in terms of model comparison, it remains globally disfavoured with respect to \textit{ii)} in virtue of the information arising from the angular measurements of $B \to K^* \mu^+ \mu^-$. Indeed, while NP effects associated to $O^{LQ}_{2223}$ can actually ameliorate $b \to s \ell^+\ell^-$ anomalies as the ones related to the so-called $P_{5}'$ observable~\cite{Descotes-Genon:2015uva}, the phenomenological viability of NP effects encoded in the effective operator $Q_{10A,e}$ necessarily relies on large hadronic contributions~\cite{Ciuchini:2017mik}, making \textit{iii)} a less economic alternative to \textit{ii)}. This is reflected by the reported $\Delta IC$: in the PDD approach the improvement of the SM fit provided by NP effects as in \textit{ii)} is several units of $IC$ larger than the one provided by \textit{iii)}, making \textit{ii)} much more favoured by the current experimental dataset.

As a bottom line for the inspection of NP effects in one single operator, in the PMD approach the $B \to K^* \mu^+ \mu^-$ angular analysis still greatly favours the presence of NP effects from $Q_{9V,\mu}$, while more NP scenarios are viable with a more conservative approach to QCD power corrections, and a particularly favoured one turns out to be $O^{LQ}_{2223}$. Finally, we observe how the three scenarios discussed so far may be distinguished with a future measurement of transverse and longitudinal ratios in the $q^2$-bin $[1.1,6]$, quite robust against hadronic uncertainties, see tables~\ref{tab:OBS_WET}~-~\ref{tab:OBS_SMEFT}. Among \textit{i), ii), iii)} $R_{K^{*},\phi}^{\textrm{\tiny T}}[1.1,6]\simeq 1.0$ would favour NP effects from $Q_{9V,\mu}$, while $R_{K^{*},\phi}^{\textrm{\tiny T}}[1.1,6]\simeq 0.8$ would point to BSM dynamics in $O^{LQ}_{2223}$, and a measurement of $R_{K^{*},\phi}^{\textrm{\tiny T}}[1.1,6]\simeq 0.7$ would hint at NP in $Q_{10A,e}$.

\begin{figure*}[!ht]
  \centering
  \includegraphics[width=\textwidth]{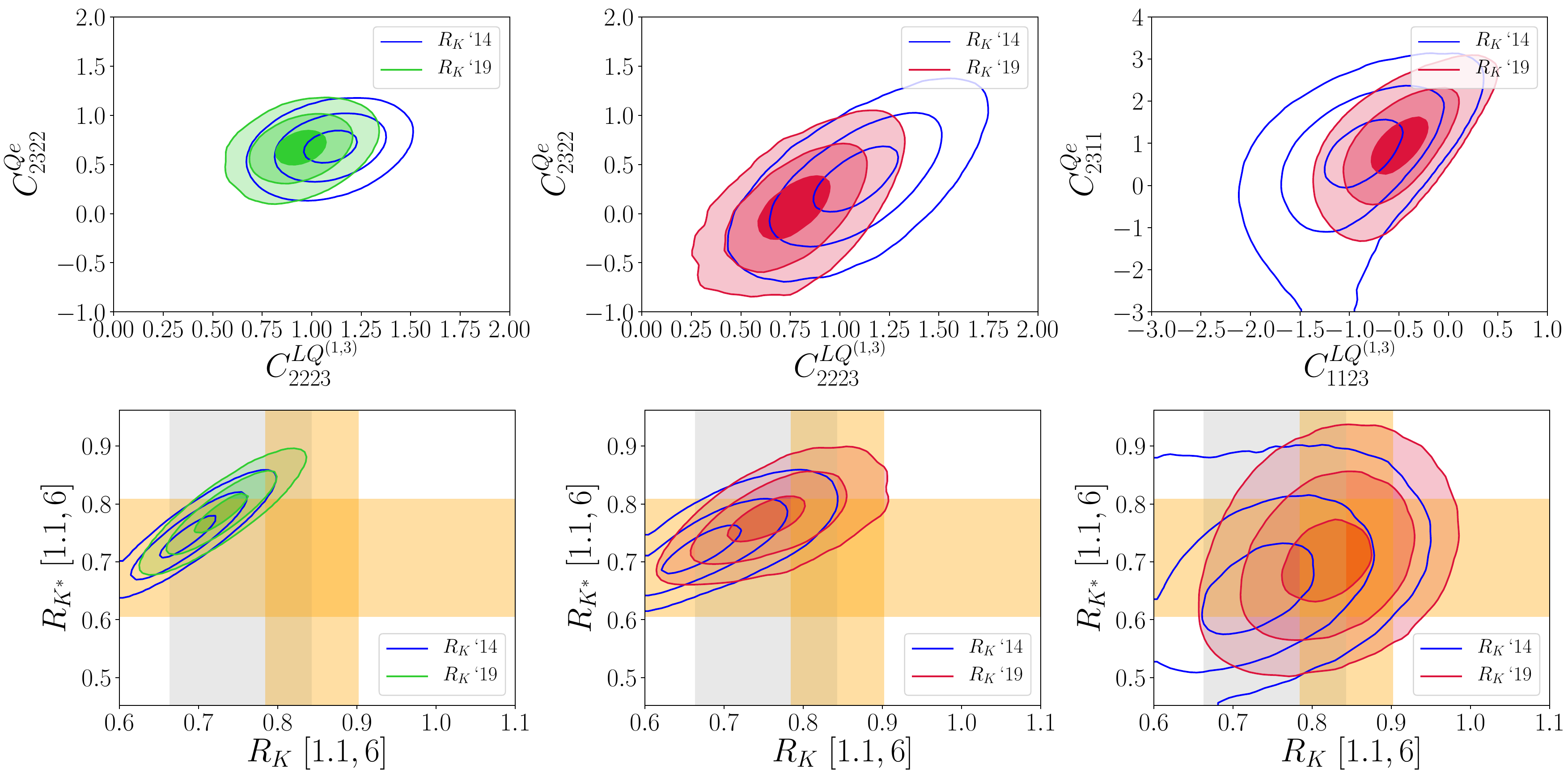}
  \vspace{-0.02\textwidth}
  \caption{\emph{2D {p.d.f.}
  for the scenario with WCs $(C_{9}^{NP}, C_{10}^{NP})$
  and combined 2D {p.d.f.} of the corresponding results 
  for $R_K$ and $R_{K^*}$ in both the PMD and the PDD approaches, following the colour scheme defined below fig.~\ref{fig:1D_mu_e}. We show the result in the SM gauge-invariant language, for both the muonic (left and central column) and electronic (right column) case.} }
  \label{fig:CLQ1_CQe}
\end{figure*}

Let us now turn to the investigation of more complex cases, where BSM dynamics is actually described by a pair of effective operators rather than just a single one. We start focussing on the scenario where the effective couplings of interest turn out to be $(C_{9,\mu}^{\rm NP}, C_{9,e}^{\rm NP})$. Note from table~\ref{tab:WC_WET} that the addition of a NP contribution coming from the electron operator $Q_{9V,e}$ does not strongly improve the fit obtained with $Q_{9V,\mu}$: in terms of $\Delta IC$, both PMD and PDD approaches slightly penalize this scenario, underlying a marginal improvement in the description of current data in correspondence to the addition of $C_{9,e}^{\rm NP}$. This is also captured by the LFUV ratios in the right panels of fig.~\ref{fig:C9_22_C9_11}, where an improvement is only seen in the value of $R_K$. Moreover, the prediction for longitudinal and transverse components of $R_{K^{*},\phi}$ remain essentially the same for the two scenarios~\ref{tab:OBS_WET}. 

Nevertheless, this NP benchmark is particularly illustrative of a study case where a robust estimate of NP effects -- i.e. as much orthogonal to hadronic effects as possible -- is actually feasible. It is indeed instructive to recast this case in the basis where NP effective couplings come into the linear combinations $(C_{9,\mu-e}^{NP}, C_{9,\mu+e}^{NP})$. As already discussed in sections~\ref{sec:SDvsLS}~-~\ref{sec:NPinRK} and highlighted e.g. in refs.~\cite{Chrzaszcz:2018yza,Ciuchini:2018anp}, such a choice is naturally driven by the presence of LFUV observables in the fit, that are maximally sensitive to $\mu-e$ combination at the linear level in the NP WCs for $Q_{9V}$. At the same time, the $\mu-e$ combination is by definition free from hadronic uncertainties of any sort and the determination of this effective coupling signals unambiguously the presence of NP, regardless of the approach chosen for the inclusion of hadronic contributions in the analysis. The independence from the approach taken for QCD power corrections is evident from the comparison of the two panels on the left column of fig.~\ref{fig:C9_22_C9_11}: going from the PMD to the PDD approach, NP in the $\mu + e$ direction gets diluted by hadronic effects, while the determination of the $\mu - e$ WC consistently differs from~0 at more than $3\sigma$.

We then move to the inspection of cases where heavy new degrees of freedom can generically couple the left-handed $b \to s$ current to both vectorial and axial leptonic structures or, from the BSM perspective drawn in the SMEFT, to both  left-handed and right-handed leptonic currents. These NP scenarios generalize the specific benchmarks \textit{i)}, \textit{ii)}, \textit{iii)} discussed at the beginning of the section. Left and central columns in figure~\ref{fig:CLQ1_CQe} report the result for the PMD and PDD approach in the case where NP effects lie in the muonic mode only, while the PDD approach for the case of the electronic mode is given in the right column. Comparing with what already illustrated for \textit{i)}, \textit{ii)}, \textit{iii)}, with the help of tab.~\ref{tab:WC_SMEFT} and the second row of fig.~\ref{fig:CLQ1_CQe} we can easily conclude that NP contributions from $O^{LQ}_{2223}$~-~$O^{Qe}_{2322}$ are still favoured by data, slightly improving the description of $R_K$ with respect to the minimal case \textit{i)} in the PMD approach, and the minimal case \textit{ii)} in the PDD approach. Moreover, the case where NP effects arise from the pair $O^{LQ}_{1123}$~-~$O^{Qe}_{2311}$ is not favoured over the simpler axial electronic proposal denoted here as \textit{iii)}. Interestingly, from table~\ref{tab:OBS_SMEFT} we can also observe that a measurement of the transverse component of the ratios $R_{K^{*},\phi}$ for these scenarios would be quite indicative. Indeed, the prediction of these LFUV observables from NP effects in $O^{LQ}_{2223}$~-~$O^{Qe}_{2322}$ points to $ R^{\textrm{\tiny T}}_{K^{*},\phi}[1.1,6]\simeq 0.95$ and $R^{\textrm{\tiny T}}_{K^{*},\phi}[1.1,6] \simeq 0.85$ in the PMD and PDD approach respectively, compatible among each other only at the 1$\sigma$ level, and different from the ones obtained for \textit{i)} and \textit{ii)}. On the contrary, the corresponding LFUV prediction from the electronic pair considered here would not be distinguishable from what assessed already in \textit{iii)}. Finally, from the same table~\ref{tab:OBS_SMEFT} we also highlight that the study of NP effects from the full set of four operators $O^{LQ}_{\ell\ell23}$~-~$O^{Qe}_{23\ell\ell}$ with  $\ell = \{1,2\}$, would not change the important phenomenological interplay found for the pair $O^{LQ}_{2223}$~-~$O^{Qe}_{2322}$, but would quite distinctively predict transverse ratios $ R^{\textrm{\tiny T}}_{K^{*},\phi}[1.1,6] \simeq 0.75$, independently of the hadronic approach considered. We postpone a thorough discussion on the analysis of these four operators all together to~\ref{app:B}, where we study them in the context of the loop-generated effects reported in eq.~\eqref{eq:SMEFT_matching_1loop}, and where we also emphasize the possible connections of $b \to s \ell^+ \ell^-$ anomalies with EW precision physics~\cite{Efrati:2015eaa,deBlas:2016ojx,Ellis:2018gqa}.

\subsection{New Physics in both \texorpdfstring{$b \to s$}{btos} left- and right-handed currents}

We continue our discussion of 2D scenarios reaching one of the highlights of this study in relation to the new $R_{K}$ measurement and what outlined in section~\ref{sec:NPinRK}: the investigation of NP effects entering both $b \to s$ left-handed and right-handed currents in dimension-six semileptonic operators. Indeed, from the discussion following eq.~\eqref{eq:RKst_over_RK} we recall that as long as $R_{K^*}$ can be retained to have a role quite complementary to the one of $R_{K}$ as a probe of NP, the new measurement appearing in eq.~\eqref{eq:RK_measurement} -- supplemented by the current one for $R_{K^*}$ in the same bin of $q^2$ -- hints at new heavy degrees of freedom coupled to $b \to s$ right-handed currents. As we show in what follows, this conclusion remains subject to the taming of non-factorizable hadronic contributions.

\begin{figure*}[!t]
  \centering
  \includegraphics[width=\textwidth]{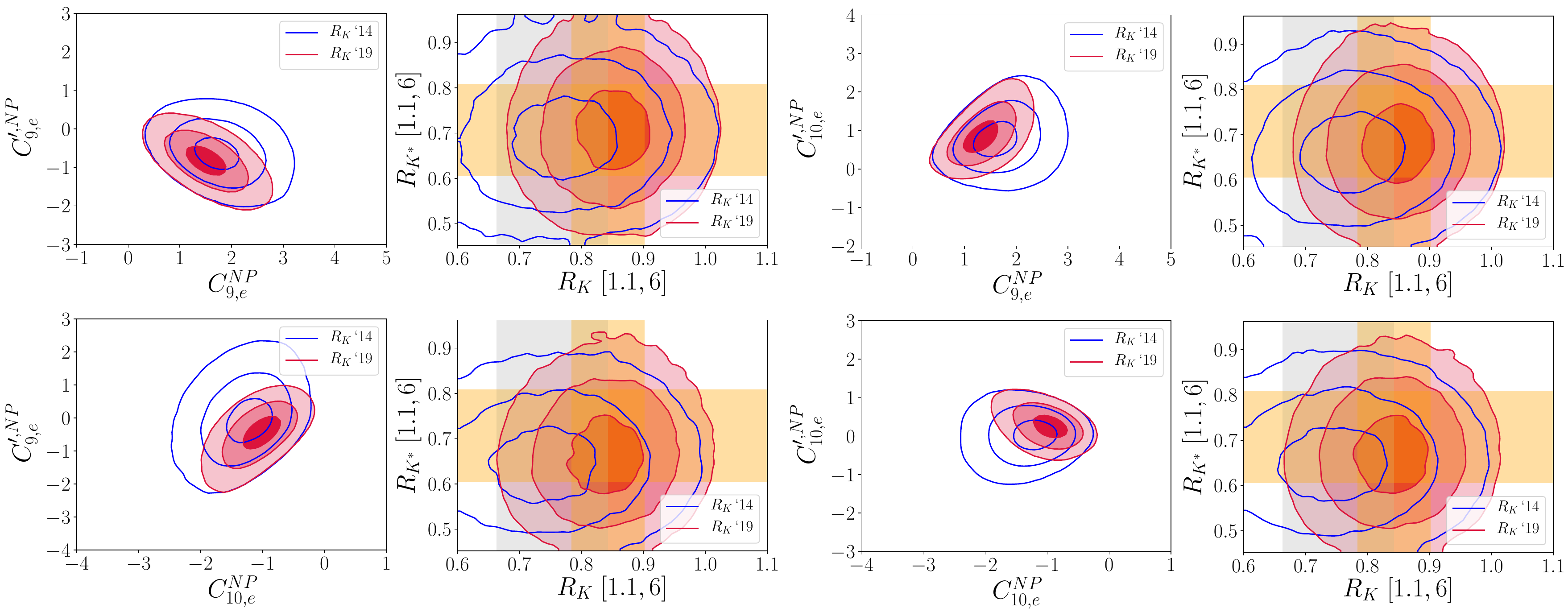}
  \vspace{-0.02\textwidth}
  \caption{\emph{First row: the first two panels show 2D {p.d.f.}
  for the scenario with WCs $(C_{9,e}^{\rm NP}, C_{9,e}^{\prime, \rm{NP}})$
  and the combined
  2D {p.d.f.} of the corresponding results for $R_K$ and $R_{K^*}$ in the 
  PDD approaches, following the colour scheme defined below fig.~\ref{fig:1D_mu_e}, while the last two panels show 
  the same for the scenario with WCs $(C_{9,e}^{\rm NP}, C_{10,e}^{\prime, \rm{NP}})$.
  Second row: analogous to the first row, but relative to the 
  WCs $(C_{10,e}^{\rm NP}, C_{9,e}^{\prime, \rm{NP}})$ and $(C_{10,e}^{\rm NP}, C_{10,e}^{\prime, \rm{NP}})$.}}
  \label{fig:2D_WET_e}
\end{figure*}

We start by considering NP effects in vectorial and axial muonic currents and described by means of the WET formalism, namely the pairs of NP WCs: $(C_{9,\mu}^{\rm NP}$, $ C_{9,\mu}^{\prime, \rm{NP}})$, $(C_{9,\mu}^{\rm NP}, C_{10,\mu}^{\prime, \rm{NP}})$, $(C_{10,\mu}^{\rm NP}, C_{9,\mu}^{\prime, \rm{NP}})$ and $(C_{10,\mu}^{\rm NP}, C_{10,\mu}^{\prime, \rm{NP}})$. The two former scenarios are generalizations of the study case \textit{i)}, and are allowed both in the PMD and PDD approaches, while the latter two can satisfactorily address $b \to s \ell^+ \ell^-$ anomalies only within the PDD approach. Results for all these scenarios can be found in fig.~\ref{fig:2D_WET_mu}. As first highlighted by the trend in the reported $\Delta IC$ and further depicted by $R_{K}$~-~$R_{K^*}$ plots, the inclusion of right-handed $b \to s$ effective couplings allows for an overall better description of data. In particular, from the inspection of the central row in figure~\ref{fig:2D_WET_mu} the scenario involving the operators  $Q_{9V,\mu}$ and $Q_{10A,\mu}^{\prime}$ provides the best match here to the newly measured $R_{K}$ together with $R_{K^*}$ in the $q^2$-bin [1.1,6] GeV$^2$. 
Moreover, it yields an optimal description of $B_s \to \mu^+ \mu^-$ and of the whole angular analysis at the same time -- independently of the hadronic approach undertaken -- and hence stands out in table~\ref{tab:WC_WET} as the study case with the highest $\Delta IC$ in both PMD and PDD approaches. 
This result comes together with the prediction for $ R^{\textrm{\tiny T}}_{K^{*},\phi}[1.1,6] \simeq 1$ in the scenarios with the pairs $Q_{9V,\mu}$~-~$Q_{9V(10A),\mu}^{\prime}$. We also note that the prediction of the longitudinal ratio unfortunately does not allow to single out within errors the NP case of $Q_{9V,\mu}$~-~$Q_{9V,\mu}^{\prime}$ with respect to $Q_{9V,\mu}$~-~$Q_{10,\mu}^{\prime}$.

A different prediction for the transverse and longitudinal LFUV ratios is instead obtained for the pairs $Q_{10A,\mu}$~-~$Q_{9V(10A),\mu}^{\prime}$, approximately giving $ R^{\textrm{\tiny T}}_{K^{*},\phi}[1.1,6] \simeq R^{\textrm{\tiny L}}_{K^{*},\phi}[1.1,6]  \simeq 0.75$. We note that the non-trivial interplay between hadronic physics -- addressing here the $B \to K^* \mu^+ \mu^-$ angular analysis -- and the experimental weights of the measured LFUV ratios and of $Br(B_s \to \mu^+ \mu^-)$ lead overall to a lower $\Delta IC$ value for these two scenarios with respect to the case of $Q_{9V,\mu}$ and $Q_{10A,\mu}^{\prime}$ (see table \ref{tab:WC_WET}).

A similar very good description of measured LFUV ratios is also obtained in the 2D scenarios with NP effects in the electron channel only, described by means of the WET formalism, namely $(C_{9,e}^{\rm NP}, C_{9,e}^{\prime, \rm{NP}})$, $(C_{9,e}^{\rm NP}, C_{10,e}^{\prime, \rm{NP}})$, $(C_{10,e}^{\rm NP}, C_{9,e}^{\prime, \rm{NP}})$ and $(C_{10,e}^{\rm NP}, C_{10,e}^{\prime, \rm{NP}})$. In these scenarios, NP cannot provide a satisfactory explanation of the angular dataset for the $B \to K^*\mu^+ \mu^-$ decay: therefore, they are viable only within the PDD approach. Results for these scenarios are reported in fig.~\ref{fig:2D_WET_e}, that capture indeed the very good description of $R_{K^*}$~-~$R_{K}$ in the $q^2$-bin [1.1,6]~GeV$^2$ in all the four cases at hand. However, comparing the $\Delta IC$ in table~\ref{tab:WC_WET}, none of these models turns out to be competitive with NP effects coming from $Q_{9V,\mu}$~-~$Q_{10A,\mu}^{\prime}$ operators. Looking again at table~\ref{tab:OBS_WET}, one can still find a particular footprint of these scenarios via the prediction of longitudinal and transverse ratios.
In particular, the two cases where $C_{9,e}^{\rm NP}$ is involved predict a quite large transverse ratio, $ R^{\textrm{\tiny T}}_{K^{*},\phi}[1.1,6] \simeq 0.95$, while the two scenarios where $C_{10,e}^{\rm NP}$ is present point to $ R^{\textrm{\tiny T}}_{K^{*},\phi}[1.1,6] \simeq 0.7$. The four scenarios here discussed qualitatively go along with the same picture drawn for the pairs $Q_{10A,\mu}$~-~$Q_{9V(10A),\mu}^{\prime}$: they turn out to be less competitive than the case of $Q_{9V,\mu}$ and $Q_{10A,\mu}^{\prime}$. 

\begin{figure*}[!t]
  \centering
  \includegraphics[width=\textwidth]{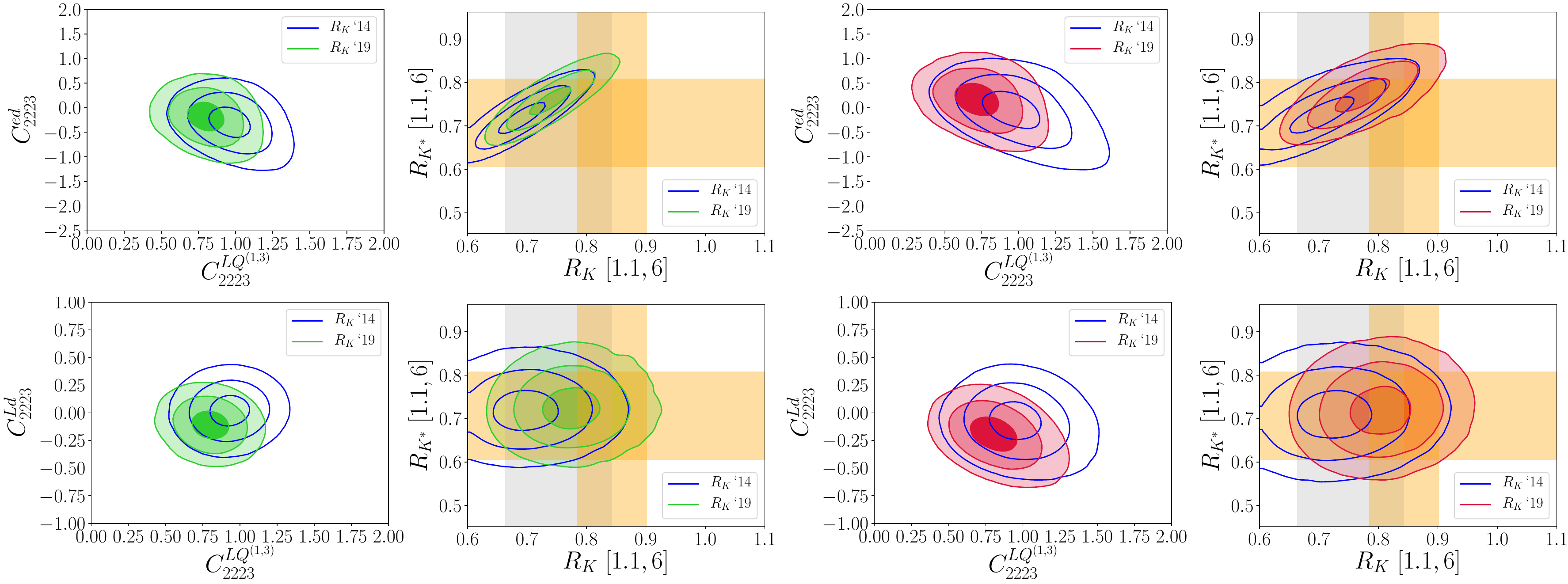}
  \vspace{-0.02\textwidth}
  \caption{\emph{First row: 2D {p.d.f.}
  for the scenario with WCs $(C_{2223}^{LQ}, C_{2223}^{ed})$
  and combined
  2D {p.d.f.} of the corresponding results for $R_K$ and $R_{K^*}$ in both the 
  PMD and the PDD approaches, following the colour scheme defined below fig.~\ref{fig:1D_mu_e}.
  Second row: analogous to the first row, but relative to the WCs $(C_{2223}^{LQ}, C_{2223}^{Ld})$.}}
  \label{fig:2D_SMEFT_mu}
\end{figure*}

\begin{figure*}[!t]
  \centering
  \includegraphics[width=\textwidth]{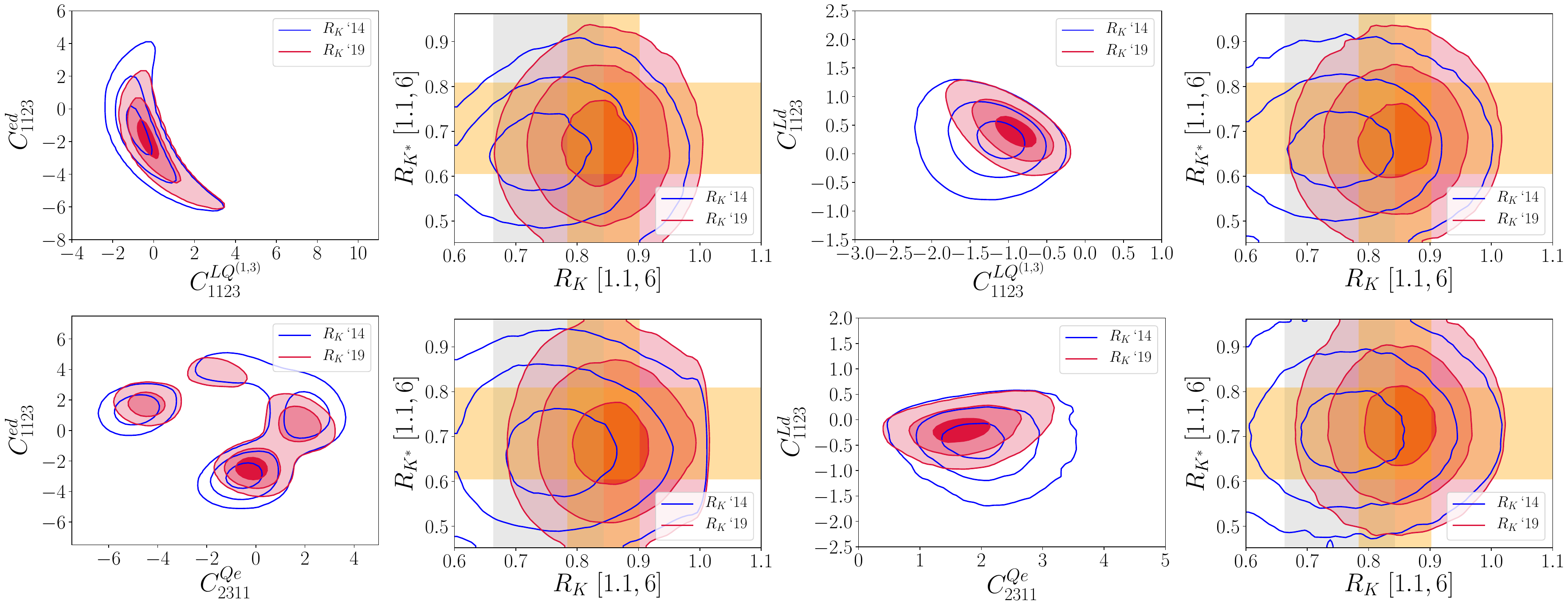}
  \vspace{-0.02\textwidth}
  \caption{\emph{First row: the first two panels show 2D {p.d.f.}
  for the scenario with WCs $(C_{1123}^{LQ^{(1,3)}}, C_{1123}^{ed})$
  and the combined
  2D {p.d.f.} of the corresponding results for $R_K$ and $R_{K^*}$ in the 
  PDD approaches, following the colour scheme defined below fig.~\ref{fig:1D_mu_e}, while the last two panels show 
  the same for the scenario with WCs  $(C_{1123}^{LQ^{(1,3)}}, C_{1123}^{Ld})$.
  Second row: analogous to the first row, but relative to the 
  WCs $(C_{2311}^{Qe}, C_{1123}^{ed})$ and $(C_{2311}^{Qe}, C_{1123}^{Ld})$.}}
  \label{fig:2D_SMEFT_e}
\end{figure*}

We now proceed considering NP effects in left-handed and right-handed muonic currents employing the gauge-invariant language of the SMEFT. In particular, we first focus on the scenarios $(C_{2223}^{LQ}, C_{2223}^{ed})$ and $(C_{2223}^{LQ}, C_{2223}^{Ld})$, that are generalizations of the study case \textit{ii)}, therefore viable both in the PMD and in the PDD approach. Their results are shown in fig.~\ref{fig:2D_SMEFT_mu}. Similarly to what found above for the pairs $Q_{10A,\mu}$~-~$Q_{9V(10A),\mu}^{\prime}$ and $Q_{9V(10A),e}$~-~$Q_{9V(10A),e}^{\prime}$, in these scenarios -- in spite of the $R_{K}$ update -- the presence of right-handed currents has an overall marginal phenomenological impact. 
These conclusions are corroborated by the values of $\Delta IC$, slightly penalizing these scenarios in comparison with the study case \textit{ii)}: a marginal improvement in the description of data is indeed obtained at the cost of model complexity in the fit. We eventually point out that the prediction for the longitudinal and transverse LFUV ratios are quite similar within these NP cases, yielding in particular $R^{\textrm{\tiny T}}_{K^{*},\phi}[1.1,6] \simeq 0.8$.

In the spirit of studying the interplay between left-handed and right-handed currents in the SMEFT framework, one may investigate also the viability of the above scenarios replacing the role carried out by $O_{2223}^{LQ}$ with the one of $O_{2322}^{Qe}$. However, eq.~\eqref{eq:RK_analytic_SMEFT} implies that the 2D scenario $(C_{2322}^{Qe}, C_{2223}^{ed})$ cannot explain the measured value of $R_K$, since both coefficients contribute to the ratio with upward shifts, in contrast with what is required to account for the experimental data. On the other hand, considering $C_{2223}^{Ld}$ as the NP term responsible of effects stemming from right-handed currents, positive solutions for this coefficient produce downward shifts in $R_K$, potentially making the $(C_{2322}^{Qe}, C_{2223}^{Ld})$ scenario a viable solution for this LFUV ratio anomaly, see right panel in fig.~\ref{fig:RK_EFT}. However, as shown e.g. in ref.~\cite{DAmico:2017mtc}, downward shifts in $R_K$ induced by $C_{2223}^{Ld}$ correspond to upward shifts in $R_{K^*}$: therefore, since $C_{2322}^{Qe}$ always contributes positively to this second ratio as well, also this second scenario cannot be considered viable in order to simultaneously address the anomalies in the two LFUV ratios.

Similar results are obtained in the last set of 2D scenarios, involving NP effects in electron channel described by means of the SMEFT formalism, namely $(C_{1123}^{LQ}, C_{1123}^{ed})$, $(C_{1123}^{LQ}, C_{1123}^{Ld})$, $(C_{2311}^{Qe}, C_{1123}^{ed})$ and $(C_{2311}^{Qe}$, $C_{1123}^{Ld})$. It is interesting to note that, contrarily to what observed for the corresponding muonic case, both scenarios involving the operator $O_{2311}^{Qe}$ are here allowed, due to the opposite direction of the contribution induced by such operator in the electron sector as shown in eq.~\eqref{eq:RK_analytic_SMEFT}. Once again, addressing the information stemming from the angular dataset for the $B \to K^*\mu^+\mu^-$ decay requires these scenarios to be considered only in the PDD approach. The results for these fits, reported in fig.~\ref{fig:2D_SMEFT_e}, show a good description of $R_K$ and $R_{K^*}$ in all the considered cases. However, once again the $\Delta IC$ values reported in table~\ref{tab:WC_SMEFT} imply that none of these models is favoured in comparison with the scenarios featuring NP effects in $O_{2223}^{LQ}$.

We conclude this section briefly discussing the case where all the SMEFT operators are inspected all together. Indeed, the experimental dataset at hand allows us to perform a fit for NP effects present in all the 12 tree-level SMEFT operators, switching on simultaneously the following effective couplings: $C_{\ell\ell23}^{LQ}$, $C_{23\ell\ell}^{Qe}$, $C_{\ell\ell23}^{Ld}$, $C_{\ell\ell23}^{ed}$, $C_{\ell\ell23}^{LedQ}$ and  $C_{\ell\ell23}^{\prime LedQ}$, with $\ell=\{1,2\}$. For the sake of completeness, in this scenario we also include scalar operators, particularly constrained by the available experimental information on $B_{s} \to \ell^{+} \ell^{-}$. We report the results of our fit in the PMD and PDD approaches in fig.~\ref{fig:SMEFT_all}. Most importantly, we observe that in both approaches $C_{2223}^{LQ}$ is found to be different from 0: at the $\sim 6\,\sigma$ level in the PMD approach, at more than $3\,\sigma$ in the PDD one. For the PMD framework we also note that NP effects in  $O_{2322}^{Qe}$ are singled out at the $\sim 5\sigma$ level. These findings pretty much reflect what already outlined from table~\ref{tab:WC_SMEFT}, where the preferred scenario in the PDD approach is indeed the one featuring only $C_{2223}^{LQ}$, while in the case of a more aggressive approach to QCD power corrections one needs to require also the presence of $C_{2322}^{Qe}$ in order to accomplish an overall good description of data within the SMEFT. It is finally worth pointing out that the results of key observables as longitudinal and transverse LFUV ratio reported in table~\ref{tab:OBS_SMEFT} are here compatible with $R^{\textrm{\tiny T,L}}_{K^{*},\phi}[1.1,6] \simeq 0.7$ within $1\sigma$ errors.

%%%%%%%%%%%%%%%%%%%%%%%%%%%%%%%%%%%%%%%%%%%%%%%%%%%%%%%%

\section{Conclusions}
\label{sec:conclusion}

In this study we investigated the impact of the very recent $R_{K}$ and $R_{K^*}$ measurements on New Physics (NP) in $b \to s \ell^+ \ell^-$ transitions. 
We focused on the study of NP effects related to Lepton Flavour Universality violation (LFUV). We have explicitly shown that an aggressive or conservative approach to hadronic matrix elements may drastically modify the conclusions drawn from the updated $b \to s \ell^+ \ell^-$ global analysis. A set of key messages can be extracted from our comprehensive study:
%, performed both in the Weak Effective Theory (WET) and in the Standard Model Effective Field Theory (SMEFT):
\begin{itemize}
    \item in the considered ``WET scenarios'', i.e. the cases where NP contributions do not necessarily stem a priori from $SU(2)_{L} \otimes U(1)_{Y}$ gauge-invariant operators at high energies, a preference for NP coupled to both left-handed quark currents with vector muon coupling and to right-handed quark currents with axial muon coupling stands out regardless of the treatment of hadronic uncertainties;
    \item in the instance of ``SMEFT scenarios'', namely when NP effects are explicitly correlated by $SU(2)_{L} \otimes U(1)_{Y}$ gauge invariance in the UV, several distinct cases are able to address present experimental information depending on the treatment of hadronic effects undertaken; aggressive estimates of hadronic uncertainties point to the simultaneous presence of left-handed quark and muon couplings and left-handed quark and right-handed muon couplings; a more conservative analysis leaves room for a broader set of scenarios, including the case of the single purely left-handed operator with muon coupling;
    \item LFUV effects in the electron sector provide a good description of current $R_{K^{(*)}}$ measurements, but an overall satisfactory description of experimental results can be obtained only within a conservative approach to QCD effects; within this framework, these NP scenarios are not favoured over ones featuring muon couplings.
\end{itemize}

We look forward to strengthening and improving our conclusions with the help of forthcoming experimental results: \textit{a)} novel LFUV data from $B_{s} \to \phi \ell^{+} \ell^{-}$, that would corroborate the current ones for $B \to K^* \ell^{+} \ell^{-}$; \textit{b)} possible measurements of LFUV ratios as $R_{K^*,\phi}^{\textrm{\tiny T,L}}$ in the $q^2$-bin [1.1,6] GeV$^2$, that may help to further disentangle the different NP effects highlighted in this work, see tables~\ref{tab:OBS_WET}~-~\ref{tab:OBS_SMEFT}; \textit{c)} new measurements of lepton-flavour dependent angular observables as in ref.~\cite{Wehle:2016yoi}, that would also help to single out heavy new dynamics from standard hadronic physics. 
\begin{minipage}{50cm}\end{minipage}

\begin{acknowledgements}
We wish to acknowledge Jorge de Blas, Julian Heeck and Marco Nardecchia for insightful discussions.
The work of M.F. is supported by the MINECO grant FPA2016-76005-C2-1-P and by Maria de Maetzu program grant MDM-2014-0367 of ICCUB and 2017 SGR 929.
The work of M.V. is supported by the NSF Grant No.~PHY-1620638. This project has received funding from the European Research Council (ERC) under the European Union's Horizon 2020 research and innovation program (grant agreement n$^o$ 772369).
M.C., M.F. and M.V. are grateful to the Mainz Institute for Theoretical
Physics (MITP) for its hospitality and its partial support during the completion of this work. 
\end{acknowledgements}
%%%%%%%%%%%%%%%%%%%%%%%%%%%%%%%%%%%%%%%%%%%%%%%%%%%%%%%%

\appendix
\section{Highlights on current and future measurements}
\label{app:A}
In this appendix we collect a subset of the interesting measurements related to the set of $b \to s \ell^+ \ell^-$ anomalies, namely $R_K$, $R_{K^*}$ (for this one both low-$q^2$ bin, [0.045,1.1]~GeV$^2$, and the central one, [1.1,6]~GeV$^2$), and the two interesting bins of $P_5'$ falling in the region towards the $J/\psi$ resonance. Most importantly, we show the prediction in the $q^2$ bin [1.1,6] relative to $R_\phi$ and to the transverse and longitudinal ratios $R_{K^*}^{\textrm{\tiny T,L}}$ and $R_{\phi}^{\textrm{\tiny T,L}}$.
In the first row of tables~\ref{tab:OBS_WET}-\ref{tab:OBS_SMEFT} we report the experimental measurement (when available), with statistical and systematic errors combined and symmetrized. In particular, concerning both bins of $R_{K^*}$ and the critical bins of\; $P_5'$,\; we\; limit\; ourselves\; to\; report\; the experimental

%% CONTINUED AFTER THE PLOT AND TABLES

\begin{landscape}
\begin{figure}
  \centering
  \includegraphics[height=0.9\textwidth, width=1.2\textwidth]{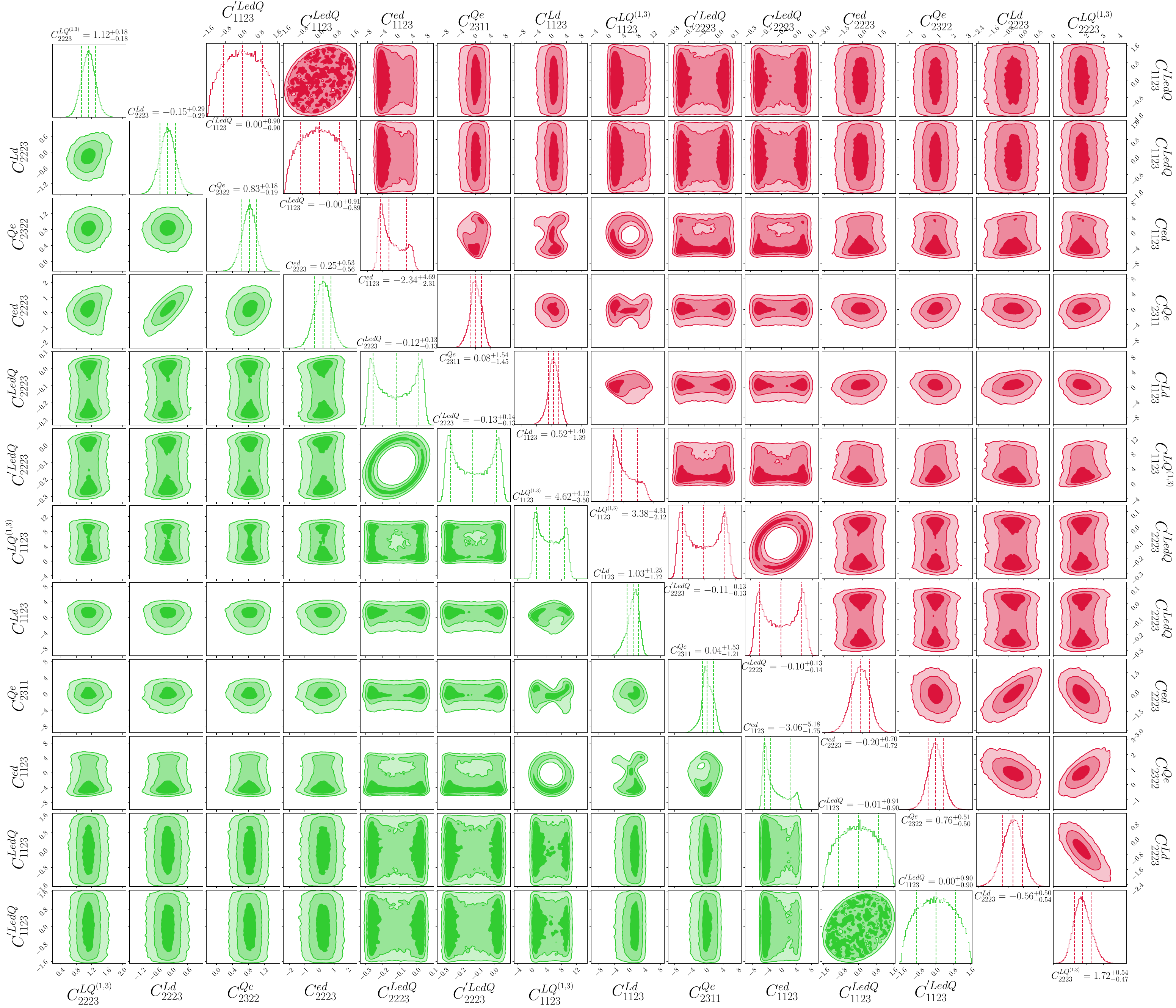}
  \caption{\emph{Global fit of all tree-level SMEFT WCs ($\Lambda = 30$ \emph{TeV}) constrained by $b \to s \ell^+ \ell^-$ measurements within the PMD (green) and PDD (red) approach, including the 2019 $R_K$ measurement. 16-$th$, 50-$th$, 84-$th$ percentile for marginalized distributions is reported.}}
  \label{fig:SMEFT_all}
\end{figure}
\end{landscape}

%%%%%%%%%%%%%%%%%%%%%%%%%%%%%%%%%%%%%%%%%%%%%%%%%%%%%%%%

\begin{landscape}
\begin{table}
\renewcommand{\arraystretch}{1.5}
\vspace{1.2in}
{\footnotesize
\begin{tabular}{|c|ccccc|ccccc|}
\hline
& $ R_K[1.1,6]$ & $ R_{K^*}[0.045,1.1]$ & $ R_{K^*}[1.1,6]$
& $ P_5'[4,6]$ & $ P_5'[6,8]$ & $R_{K^*}^T[1.1,6]$ & $ R_{K^*}^L[1.1,6]$
& $ R_{\phi}[1.1,6]$ & $ R_{\phi}^T[1.1,6]$ & $ R_{\phi}^L[1.1,6]$ \\
\hline
Exp.
& 0.850(59) & 0.680(93) & 0.71(10) & -0.30(16) & -0.51(12) & - & - & - & - & - \\
\hline
\hline
\multirow{2}{*}{$ C_{9,\mu}^{\rm NP} $}
& 0.762(43) & 0.889(12) & 0.845(38) & -0.441(59) & -0.558(76) & 1.032(22) & 0.783(48) & 0.851(39) & 1.039(25) & 0.787(48) \\
& \cg 0.767(25) & \cg 0.889(8) & \cg 0.833(17) & \cg -0.472(46) & \cg -0.616(43) & \cg 1.029(15) & \cg 0.771(25) & \cg 0.838(18) & \cg 1.037(20) & \cg 0.773(25) \\
\hline
$ C_{10,e}^{\rm NP} $
& 0.807(46) & 0.841(23) & 0.733(58) & -0.462(59) & -0.552(77) & 0.726(58) & 0.736(59) & 0.732(59) & 0.729(57) & 0.733(60) \\
\hline
\hline
\multirow{2}{*}{$ (C_{9,\mu}^{\rm NP},C_{9,e}^{\rm NP}) $}
& 0.802(56) & 0.895(13) & 0.869(45) & -0.436(59) & -0.564(76) & 1.036(22) & 0.812(57) & 0.874(45) & 1.042(25) & 0.816(57) \\
& \cg 0.807(56) & \cg 0.896(12) & \cg 0.865(44) & \cg -0.456(50) & \cg -0.606(44) & \cg 1.030(16) & \cg 0.810(55) & \cg 0.869(44) & \cg 1.037(20) & \cg 0.813(55) \\
\hline
\multirow{2}{*}{$ (C_{9,\mu}^{\rm NP},C_{9,\mu}^{\prime, \rm{NP}}) $}
& 0.818(51) & 0.828(34) & 0.708(72) & -0.445(60) & -0.561(79) & 1.042(47) & 0.619(80) & 0.714(74) & 1.045(50) & 0.623(81) \\
& \cg 0.845(46) & \cg 0.860(16) & \cg 0.771(28) & \cg -0.471(47) & \cg -0.627(44) & \cg 1.050(24) & \cg 0.690(38) & \cg 0.779(28) & \cg 1.054(26) & \cg 0.695(37) \\
\hline
\multirow{2}{*}{$ (C_{9,\mu}^{\rm NP},C_{10,\mu}^{\prime, \rm{NP}}) $}
& 0.822(46) & 0.831(25) & 0.703(57) & -0.459(61) & -0.567(79) & 1.016(29) & 0.617(63) & 0.712(57) & 1.029(34) & 0.620(64) \\
& \cg 0.845(35) & \cg 0.845(17) & \cg 0.731(36) & \cg -0.497(48) & \cg -0.646(45) & \cg 1.024(18) & \cg 0.649(43) & \cg 0.739(35) & \cg 1.039(23) & \cg 0.650(44) \\
\hline
$ (C_{10,\mu}^{\rm NP},C_{9,\mu}^{\prime, \rm{NP}}) $
& 0.831(52) & 0.859(22) & 0.774(55) & -0.482(64) & -0.552(83) & 0.755(43) & 0.782(64) & 0.776(54) & 0.764(42) & 0.781(63) \\
\hline
$ (C_{10,\mu}^{\rm NP},C_{10,\mu}^{\prime, \rm{NP}}) $
& 0.826(50) & 0.859(18) & 0.773(46) & -0.483(64) & -0.553(83) & 0.752(42) & 0.781(53) & 0.775(46) & 0.761(40) & 0.780(53) \\
\hline
$ (C_{9,e}^{\rm NP},C_{9,e}^{\prime, \rm{NP}}) $
& 0.853(59) & 0.818(41) & 0.697(88) & -0.453(60) & -0.540(78) & 0.972(56) & 0.622(95) & 0.705(89) & 0.984(59) & 0.628(96) \\
\hline
$ (C_{9,e}^{\rm NP},C_{10,e}^{\prime, \rm{NP}}) $
& 0.851(58) & 0.809(41) & 0.671(86) & -0.456(60) & -0.544(78) & 0.923(59) & 0.601(90) & 0.676(87) & 0.942(59) & 0.602(91) \\
\hline
$ (C_{10,e}^{\rm NP},C_{9,e}^{\prime, \rm{NP}}) $
& 0.840(60) & 0.813(43) & 0.670(85) & -0.464(59) & -0.555(77) & 0.682(68) & 0.652(94) & 0.670(84) & 0.685(68) & 0.650(93) \\
\hline
$ (C_{10,e}^{\rm NP}, C_{10,e}^{\prime, \rm{NP}}) $
& 0.841(60) & 0.806(42) & 0.663(85) & -0.465(59) & -0.557(77) & 0.688(66) & 0.654(95) & 0.663(85) & 0.693(66) & 0.651(96) \\
\hline
\end{tabular}
}
\parbox{21.5cm}{\caption{\emph{Experimental measurements with symmetrized errors (for $R_{K^*}$ and $P_5'$ we report the LHCb ones) and results from the fit for key observables in the WET scenarios considered here. The gray rows highlight the PMD results when experimental data can be well described within the approach. The PDD results are presented for all cases. For the definition of the two approaches, see section~\ref{sec:SDvsLS}.\label{tab:OBS_WET}}}}
\end{table}
\end{landscape}

\begin{landscape}
\begin{table}
\renewcommand{\arraystretch}{1.5}
{\footnotesize
\vspace{1.2in}
\begin{tabular}{|c|ccccc|ccccc|}
\hline
& $ R_K[1.1,6]$ & $ R_{K^*}[0.045,1.1]$ & $ R_{K^*}[1.1,6]$
& $ P_5'[4,6]$ & $ P_5'[6,8]$ & $R_{K^*}^T[1.1,6]$ & $ R_{K^*}^L[1.1,6]$
& $ R_{\phi}[1.1,6]$ & $ R_{\phi}^T[1.1,6]$ & $ R_{\phi}^L[1.1,6]$ \\
\hline
Exp.
& 0.850(59) & 0.680(93) & 0.71(10) & -0.30(16) & -0.51(12) & - & - & - & - & - \\
\hline
\hline
\multirow{2}{*}{$ C^{LQ}_{2223} $}
& 0.764(41) & 0.860(13) & 0.771(38) & -0.471(63) & -0.555(83) & 0.836(30) & 0.748(40) & 0.775(37) & 0.846(29) & 0.749(41) \\
& \cg 0.746(36) & \cg 0.857(12) & \cg 0.754(34) & \cg -0.670(28) & \cg -0.764(30) & \cg 0.799(28) & \cg 0.741(36) & \cg 0.754(34) & \cg 0.803(27) & \cg 0.739(36) \\
\hline
\hline
\multirow{2}{*}{$ (C^{LQ}_{2223},C^{Qe}_{2322}) $}
& 0.760(49) & 0.860(14) & 0.770(39) & -0.469(63) & -0.556(82) & 0.844(64) & 0.744(42) & 0.774(39) & 0.855(64) & 0.745(42) \\
& \cg 0.728(34) & \cg 0.871(13) & \cg 0.781(36) & \cg -0.511(53) & \cg -0.650(47) & \cg 0.947(51) & \cg 0.729(35) & \cg 0.784(36) & \cg 0.957(52) & \cg 0.731(35) \\
\hline
$ (C^{LQ}_{1123},C^{Qe}_{2311}) $
& 0.819(55) & 0.827(32) & 0.705(72) & -0.469(60) & -0.560(77) & 0.68(11) & 0.720(63) & 0.703(74) & 0.68(11) & 0.715(65) \\
\hline
\multirow{2}{*}{$ (C^{LQ}_{2223},C^{ed}_{2223}) $}
& 0.773(46) & 0.858(14) & 0.767(38) & -0.472(63) & -0.552(82) & 0.839(31) & 0.742(42) & 0.771(38) & 0.851(30) & 0.742(42) \\
& \cg 0.738(37) & \cg 0.858(12) & \cg 0.758(34) & \cg -0.653(38) & \cg -0.754(34) & \cg 0.804(28) & \cg 0.744(36) & \cg 0.758(34) & \cg 0.806(27) & \cg 0.743(36) \\
\hline
\multirow{2}{*}{$ (C^{LQ}_{2223},C^{Ld}_{2223}) $}
& 0.805(52) & 0.834(24) & 0.714(56) & -0.478(64) & -0.547(83) & 0.819(35) & 0.680(63) & 0.719(55) & 0.831(34) & 0.681(63) \\
& \cg 0.775(49) & \cg 0.842(20) & \cg 0.724(47) & \cg -0.693(38) & \cg -0.786(39) & \cg 0.792(29) & \cg 0.704(53) & \cg 0.724(47) & \cg 0.796(28) & \cg 0.702(53) \\
\hline
$ (C^{LQ}_{1123},C^{ed}_{1123}) $
& 0.837(59) & 0.819(44) & 0.672(90) & -0.463(59) & -0.553(77) & 0.71(10) & 0.644(90) & 0.670(92) & 0.726(99) & 0.636(93) \\
\hline
$ C^{LQ}_{1123},C^{Ld}_{1123}) $
& 0.844(59) & 0.807(41) & 0.679(84) & -0.457(59) & -0.548(77) & 0.784(57) & 0.629(93) & 0.681(84) & 0.791(56) & 0.630(93) \\
\hline
$ (C^{Qe}_{2311},C^{ed}_{1123}) $
& 0.857(62) & 0.816(44) & 0.692(92) & -0.468(60) & -0.558(77) & 0.81(35) & 0.67(12) & 0.690(94) & 0.82(37) & 0.66(12) \\
\hline
$ (C^{Qe}_{2311},C^{Ld}_{1123}) $
& 0.856(59) & 0.825(41) & 0.716(87) & -0.481(59) & -0.578(76) & 0.590(75) & 0.77(10) & 0.729(87) & 0.590(75) & 0.76(10) \\
\hline
\hline
\multirow{2}{*}{$ (C^{LQ}_{2223},C^{Qe}_{2322},C^{LQ}_{1123},C^{Qe}_{2311}) $}
& 0.806(56) & 0.845(29) & 0.743(69) & -0.477(63) & -0.584(83) & 0.71(16) & 0.768(61) & 0.744(70) & 0.71(17) & 0.765(62) \\
& \cg 0.805(55) & \cg 0.849(26) & \cg 0.763(59) & \cg -0.495(54) & \cg -0.648(48) & \cg 0.71(16) & \cg 0.795(54) & \cg 0.761(61) & \cg 0.72(16) & \cg 0.790(54) \\
\hline
\hline
\multirow{2}{*}{ All }
& 0.838(60) & 0.792(48) & 0.638(89) & -0.475(65) & -0.558(83) & 0.66(32) & 0.70(19) & 0.641(90) & 0.67(33) & 0.70(19) \\
& \cg 0.838(60) & \cg 0.793(49) & \cg 0.650(89) & \cg -0.508(57) & \cg -0.660(53) & \cg 0.60(28) & \cg 0.77(23) & \cg 0.642(90) & \cg 0.61(29) & \cg 0.75(21) \\
\hline
\end{tabular}
}
\caption{\emph{Experimental measurements with symmetrized errors (for $R_{K^*}$ and $P_5'$ we report the LHCb ones) and results from the fit for key observables in the SMEFT scenarios considered here. The gray rows highlight the PMD results when experimental data can be well described within the approach. The PDD results are presented for all cases. For the definition of the two approaches, see section~\ref{sec:SDvsLS}.\label{tab:OBS_SMEFT}}}
\end{table}
\end{landscape}

%\FloatBarrier

\begin{figure*}%[h!]
  \centering
  \includegraphics[width=\textwidth]{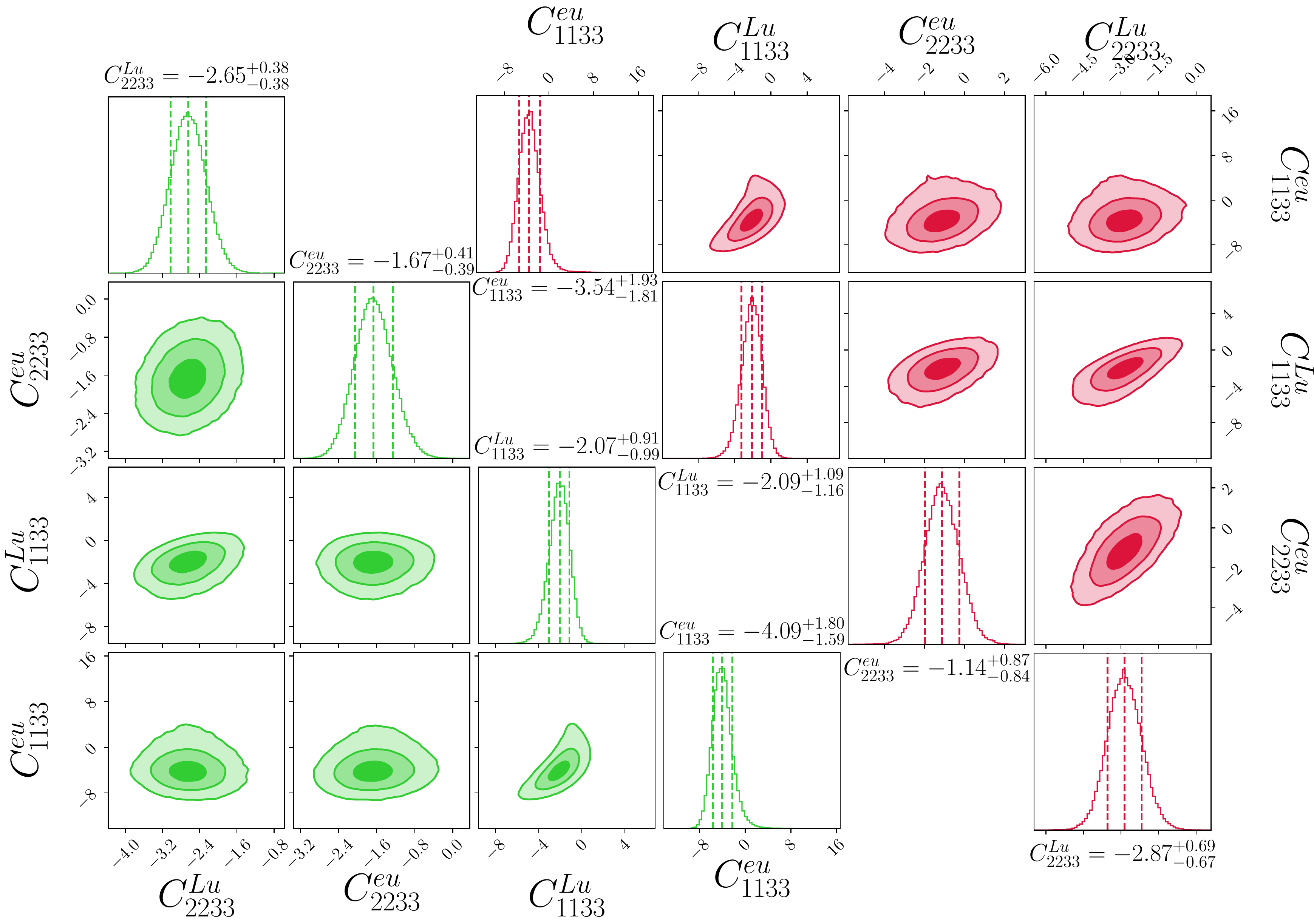}
  \caption{\emph{Global fit of RGE-induced contributions ($\Lambda = 1$ \emph{TeV}) from eq.~\eqref{eq:SMEFT_op_loop_lu} constrained by $b \to s \ell^+ \ell^-$ data within the PMD (green) and PDD (red) approach, including the new $R_K$ measurement. 16-$th$, 50-$th$, 84-$th$ percentile for marginalized distributions is shown.}}
  \label{fig:C_all_RGE}
\end{figure*}

\noindent measurements of LHCb only, the most precise ones currently available for these observables. In the following rows, we give the values obtained for each of the mentioned observables in all the scenario described in section~\ref{sec:results}. Table~\ref{tab:OBS_WET} is relative to the scenarios investigated within the WET framework, while table~\ref{tab:OBS_SMEFT} refers to the ones described within the SMEFT framework.

\section{Top-quark operators for \texorpdfstring{$b \to s \ell^+ \ell^-$}{btosll} anomalies}
\label{app:B}

As anticipated in section~\ref{sec:SMEFT}, in this appendix we detail the fit results for the interesting scenario related to loop-generated effects relevant for $b \to s \ell^{+} \ell^{-}$ anomalies from dimension-six gauge-invariant effective operators involving right-handed top quarks, see eq.~\eqref{eq:SMEFT_op_loop_lu}.

The SMEFT operators of interest enter in our analysis according to the matching condition reported in eq.~\eqref{eq:SMEFT_matching_1loop}: they do not mix into WET operators with right-handed $b \to s$ currents, but contribute only to WCs of $Q_{9V,10A}$. Consequently, a quite important observation can be drawn from what is already illustrated in section~\ref{sec:NPinRK}: the new $R_{K}$ measurement, on general grounds, disfavours scenarios with WET operators involving only left-handed $b \to s$ currents. As we found in section~\ref{sec:results}, it is nevertheless important to note that once hadronic corrections are treated conservatively, such a hint for NP contributions to primed operators in the WET gets weakened. Hence, $C_{9,10}^{\rm NP}$ effects from top-quark loops remain particularly appealing in light of the new $R_{K}$ measurement.

The fit of the full set of top-quark SMEFT operators involved is shown in fig.~\ref{fig:C_all_RGE}. In green we show the result within the PMD approach, while in red we present the PDD one. In the former approach, we find evidence for NP at more than the 6$\sigma$ level in the operator $O^{Lu}_{2233}$, together with the far-reaching evidence for a non-zero contribution from $O^{eu}_{2233}$ too, and a minor role played by the two operators related to the electron mode. This outcome falls within the expectations of what has been repeatedly observed in the present study for the $B \to K^* \mu^{+} \mu^{-}$ angular analysis. In the PDD approach, the interplay between the two muonic operators redistribute the effects amongst the whole set of four operators in adjustment of possible large hadronic effects in the $B \to K^* \ell^{+} \ell^{-}$ amplitude. The WC of $O^{Lu}_{2233}$ is well-determined at more than 4$\sigma$, the other ones are compatible with 0 within 2$\sigma$.

The overall goodness of the fit is exactly the same as for the case of the tree-level SMEFT scenario delineated by the set $(O^{LQ}_{2223,1123},O^{Qe}_{2322,2311})$, inspected in section~\ref{sec:results}. Identical considerations hold in particular for the observables presented in table~\ref{tab:OBS_SMEFT}. Within the PDD approach, such scenario adequately describes the current dataset, as in the right panel of fig.~\ref{fig:CLQ1_CQe}. On the other hand -- if a more restrictive role needs to be assigned eventually to QCD power corrections -- the LHCb update on $R_{K}$ would actually disfavour this scenario in comparison with alternatives featuring NP effects in right-handed $b \to s$ currents too.

Eventually, as first studied in ref.~\cite{deBlas:2015aea}, and recently reviewed in light of $R_{K^{(*)}}$~anomalies in~\cite{Camargo-Molina:2018cwu}, the top-quark operators can be sensitive at the loop level to LEP-I measurements, mainly via the modification of the $Z$-boson decay rate and the corresponding left-right leptonic asymmetries. A convenient language in order to easily capture these modifications of $Z$-boson properties is given by the parameters $[\delta g^{Ze}_{L,R}]_{\ell^+ \ell^-}$ scrutinized in ref.~\cite{Efrati:2015eaa}. 
In the leading-log approximation and at the leading order in the top Yukawa coupling, the contribution from $O^{Lu,eu}_{\ell\ell 33}$ via RGE is~\cite{Jenkins:2013zja}:
\begin{eqnarray}
\label{eq:OLuedRGE}
[\delta g^{Ze}_{L}]_{\ell\ell} & = & 3 \left( \frac{ y_t \, v}{4 \pi \Lambda} \right)^2 \log\left(\Lambda /\mu_{\textrm{\tiny{EW}}}\right) \, C^{Lu}_{\ell\ell33} \ , \\ \nonumber
[\delta g^{Ze}_{R}]_{\ell\ell} & = & 3 \left( \frac{ y_t \, v}{4 \pi \Lambda} \right)^2 \log\left(\Lambda /\mu_{\textrm{\tiny{EW}}}\right) \, C^{eu}_{\ell\ell33} \ .
\end{eqnarray}

\begin{figure*}[!t]
  \centering
  \includegraphics[width=\textwidth]{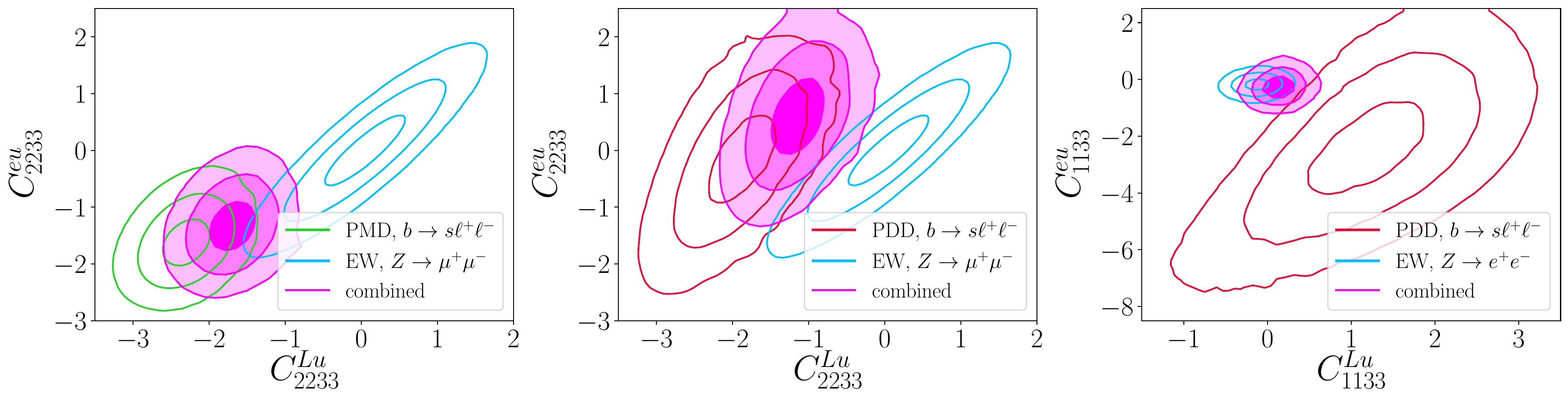}
  \caption{\emph{2D {p.d.f.}
  for the scenario with WCs $(C^{Lu}_{\ell\ell33}, C^{eu}_{\ell\ell33})$
  in both the PMD and the PDD approach, following the colour scheme defined below fig.~\ref{fig:1D_mu_e}. EW constraints on these scenario from $Z \to \ell^+ \ell^-$ measurements~\cite{Efrati:2015eaa} are shown in light blue. The resulting region obtained from the combined information of EW and $b \to s \ell^+ \ell^-$ data is highlighted in fuchsia.}}
  \label{fig:C_Lu_eu}
\end{figure*}

In fig.~\ref{fig:C_Lu_eu} we show the constraints on the WCs of $O^{Lu,eu}_{\ell\ell33}$ coming from EW data. We focus on the muon and electron sector separately, mapping what is given in equation~(4.6) of ref.~\cite{Efrati:2015eaa}, using the correlations given in appendix~B of their work and noting that the two sets $[\delta g^{Ze}_{L,R}]_{ee}$ and $[\delta g^{Ze}_{L,R}]_{\mu \mu}$ are weakly correlated. The resulting bounds are shown as 1,2,3$\sigma$ contours in light blue in the  $C^{Lu}_{\ell\ell33}$~-~$C^{eu}_{\ell\ell33}$ plane. They are obtained assuming $\Lambda = 1$~TeV and evaluated at a matching scale $\mu_{\textrm{\tiny{EW}}}$ close to the top-quark mass to possibly minimize the overall matching-scale dependence~\cite{Bobeth:2017xry,Celis:2017doq}. In the same figure, we also show the joint probability distribution obtained for the same WCs from the analysis of $b \to s \ell^{+} \ell^{-}$ data. As usual, we restrict the analysis of the electron mode to the PDD framework only, in virtue of the information arising from the $B \to K^{*} \mu \mu$ angular observables. Eventually, we also highlight in fuchsia the region obtained combining flavour data of interest with the aforementioned LEP-I results.

The outcome depicted in fig.~\ref{fig:C_Lu_eu} is pretty informative. On the left and central panel, a mild tension stands out between the highest probability region controlled by the $Z \to \mu^{+} \mu^{-}$ data and the one identified by the global $b \to s \ell^{+} \ell^{-}$ analysis. As a result of the interplay between the two experimental handles, a very different outcome for NP is expected depending on the hadronic assumptions involved. In the PDD approach, one may end up looking for NP effects mainly in $O^{Lu}_{2233}$, with a minor role played by the operator $O^{eu}_{2233}$ possibly identified by  $C^{eu}_{2233}>0$; in the PMD approach the presence of the latter is actually ensured by $C^{eu}_{2233}<0$ at the 3$\sigma$ level. Lastly, the right panel of fig.~\ref{fig:C_Lu_eu} shows how LEP-I measurements strongly constrain such a possibility with top-quark operators. 

To sum up, for $O^{Lu,eu}_{\ell \ell 33}$ and $b \to s \ell^+ \ell^-$ anomalies the coupling to muons is allowed by EW data and remains interesting as long as the hint for NP effects in right-handed $b \to s$  currents -- as a mild implication of the present $R_K$ update -- will not be further corroborated by other new measurements. 
An important caveat here is that the assumption regarding the absence of tree-level contributions in the EW fit that may easily relax the $Z \to \ell^+ \ell^-$ constraints. A more refined analysis on this aspect will be presented elsewhere.

%\begin{multicols}{2}
\bibliographystyle{JHEP}
\bibliography{LFUV}

\providecommand{\href}[2]{#2}\begingroup\raggedright\begin{thebibliography}{100}

\bibitem{Gratrex:2015hna}
J.~Gratrex, M.~Hopfer, and R.~Zwicky, {\it {Generalised helicity formalism,
  higher moments and the $B \to K_{J_K}(\to K \pi) \bar{\ell}_1 \ell_2$ angular
  distributions}},  {\em Phys. Rev.} {\bf D93} (2016), no.~5 054008,
  [\href{http://arxiv.org/abs/1506.03970}{{\tt arXiv:1506.03970}}].

\bibitem{Hiller:2003js}
G.~Hiller and F.~Kruger, {\it {More model-independent analysis of $b \to s$
  processes}},  {\em Phys. Rev.} {\bf D69} (2004) 074020,
  [\href{http://arxiv.org/abs/hep-ph/0310219}{{\tt hep-ph/0310219}}].

\bibitem{Bobeth:2007dw}
C.~Bobeth, G.~Hiller, and G.~Piranishvili, {\it {Angular distributions of
  $\bar{B} \to \bar{K} \ell^+\ell^-$ decays}},  {\em JHEP} {\bf 12} (2007) 040,
  [\href{http://arxiv.org/abs/0709.4174}{{\tt arXiv:0709.4174}}].

\bibitem{Bobeth:2008ij}
C.~Bobeth, G.~Hiller, and G.~Piranishvili, {\it {CP Asymmetries in bar $B \to
  \bar{K}^* (\to \bar{K} \pi) \bar{\ell} \ell$ and Untagged $\bar{B}_s$, $B_s
  \to \phi (\to K^{+} K^-) \bar{\ell} \ell$ Decays at NLO}},  {\em JHEP} {\bf
  0807} (2008) 106, [\href{http://arxiv.org/abs/0805.2525}{{\tt
  arXiv:0805.2525}}].

\bibitem{Egede:2008uy}
U.~Egede, T.~Hurth, J.~Matias, M.~Ramon, and W.~Reece, {\it {New observables in
  the decay mode $\bar{B}_d \to \bar{K}^{*0} l^+ l^-$}},  {\em JHEP} {\bf 0811}
  (2008) 032, [\href{http://arxiv.org/abs/0807.2589}{{\tt arXiv:0807.2589}}].

\bibitem{Matias:2012xw}
J.~Matias, F.~Mescia, M.~Ramon, and J.~Virto, {\it {Complete Anatomy of
  $\bar{B}_d \to \bar{K}^{* 0} (\to K \pi)l^+l^-$ and its angular
  distribution}},  {\em JHEP} {\bf 1204} (2012) 104,
  [\href{http://arxiv.org/abs/1202.4266}{{\tt arXiv:1202.4266}}].

\bibitem{Aaij:2013qta}
{\bf LHCb} Collaboration, R.~Aaij et~al., {\it {Measurement of
  Form-Factor-Independent Observables in the Decay $B^{0} \to K^{*0} \mu^+
  \mu^-$}},  {\em Phys.Rev.Lett.} {\bf 111} (2013), no.~19 191801,
  [\href{http://arxiv.org/abs/1308.1707}{{\tt arXiv:1308.1707}}].

\bibitem{Aaij:2014pli}
{\bf LHCb} Collaboration, R.~Aaij et~al., {\it {Differential branching
  fractions and isospin asymmetries of $B \to K^{(*)} \mu^+ \mu^-$ decays}},
  {\em JHEP} {\bf 06} (2014) 133, [\href{http://arxiv.org/abs/1403.8044}{{\tt
  arXiv:1403.8044}}].

\bibitem{Aaij:2014tfa}
{\bf LHCb} Collaboration, R.~Aaij et~al., {\it {Angular analysis of charged and
  neutral $B \to K \mu^+\mu^-$ decays}},  {\em JHEP} {\bf 05} (2014) 082,
  [\href{http://arxiv.org/abs/1403.8045}{{\tt arXiv:1403.8045}}].

\bibitem{Aaij:2015oid}
{\bf LHCb} Collaboration, R.~Aaij et~al., {\it {Angular analysis of the $B^{0}
  \to K^{*0} \mu^{+} \mu^{-}$ decay using 3 fb$^{-1}$ of integrated
  luminosity}},  {\em JHEP} {\bf 02} (2016) 104,
  [\href{http://arxiv.org/abs/1512.04442}{{\tt arXiv:1512.04442}}].

\bibitem{Aaij:2015dea}
{\bf LHCb} Collaboration, R.~Aaij et~al., {\it {Angular analysis of the B$^{0}$
  $\to$ K$^{*0}$ e$^{+}$ e$^{-}$ decay in the low-q$^{2}$ region}},  {\em JHEP}
  {\bf 04} (2015) 064, [\href{http://arxiv.org/abs/1501.03038}{{\tt
  arXiv:1501.03038}}].

\bibitem{Aaij:2015esa}
{\bf LHCb} Collaboration, R.~Aaij et~al., {\it {Angular analysis and
  differential branching fraction of the decay $B^0_s\to\phi\mu^+\mu^-$}},
  {\em JHEP} {\bf 09} (2015) 179, [\href{http://arxiv.org/abs/1506.08777}{{\tt
  arXiv:1506.08777}}].

\bibitem{Aaij:2016flj}
{\bf LHCb} Collaboration, R.~Aaij et~al., {\it {Measurements of the S-wave
  fraction in $B^{0}\rightarrow K^{+}\pi^{-}\mu^{+}\mu^{-}$ decays and the
  $B^{0}\rightarrow K^{\ast}(892)^{0}\mu^{+}\mu^{-}$ differential branching
  fraction}},  {\em JHEP} {\bf 11} (2016) 047,
  [\href{http://arxiv.org/abs/1606.04731}{{\tt arXiv:1606.04731}}].

\bibitem{Wehle:2016yoi}
{\bf Belle} Collaboration, S.~Wehle et~al., {\it {Lepton-Flavor-Dependent
  Angular Analysis of $B\to K^\ast \ell^+\ell^-$}},  {\em Phys. Rev. Lett.}
  {\bf 118} (2017), no.~11 111801, [\href{http://arxiv.org/abs/1612.05014}{{\tt
  arXiv:1612.05014}}].

\bibitem{Aaboud:2018krd}
{\bf ATLAS} Collaboration, M.~Aaboud et~al., {\it {Angular analysis of $B^0_d
  \rightarrow K^{*}\mu^+\mu^-$ decays in $pp$ collisions at $\sqrt{s}= 8$ TeV
  with the ATLAS detector}},  {\em JHEP} {\bf 10} (2018) 047,
  [\href{http://arxiv.org/abs/1805.04000}{{\tt arXiv:1805.04000}}].

\bibitem{Khachatryan:2015isa}
{\bf CMS} Collaboration, V.~Khachatryan et~al., {\it {Angular analysis of the
  decay $B^0 \to K^{*0} \mu^+ \mu^-$ from pp collisions at $\sqrt s = 8$ TeV}},
   {\em Phys. Lett.} {\bf B753} (2016) 424--448,
  [\href{http://arxiv.org/abs/1507.08126}{{\tt arXiv:1507.08126}}].

\bibitem{Sirunyan:2017dhj}
{\bf CMS} Collaboration, A.~M. Sirunyan et~al., {\it {Measurement of angular
  parameters from the decay $\mathrm{B}^0 \to \mathrm{K}^{*0} \mu^+ \mu^-$ in
  proton-proton collisions at $\sqrt{s} = $ 8 TeV}},  {\em Phys. Lett.} {\bf
  B781} (2018) 517--541, [\href{http://arxiv.org/abs/1710.02846}{{\tt
  arXiv:1710.02846}}].

\bibitem{Sirunyan:2018jll}
{\bf CMS} Collaboration, A.~M. Sirunyan et~al., {\it {Angular analysis of the
  decay $B^+\to K^+\mu^+\mu^-$ in proton-proton collisions at $\sqrt{s} =$ 8
  TeV}},  {\em Phys. Rev.} {\bf D98} (2018), no.~11 112011,
  [\href{http://arxiv.org/abs/1806.00636}{{\tt arXiv:1806.00636}}].

\bibitem{Bobeth:2011nj}
C.~Bobeth, G.~Hiller, D.~van Dyk, and C.~Wacker, {\it {The Decay $B \to K l^+
  l^-$ at Low Hadronic Recoil and Model-Independent $\Delta$ B = 1
  Constraints}},  {\em JHEP} {\bf 1201} (2012) 107,
  [\href{http://arxiv.org/abs/1111.2558}{{\tt arXiv:1111.2558}}].

\bibitem{DescotesGenon:2012zf}
S.~Descotes-Genon, J.~Matias, M.~Ramon, and J.~Virto, {\it {Implications from
  clean observables for the binned analysis of $B \to K^*\mu^+\mu^-$ at large
  recoil}},  {\em JHEP} {\bf 1301} (2013) 048,
  [\href{http://arxiv.org/abs/1207.2753}{{\tt arXiv:1207.2753}}].

\bibitem{Descotes-Genon:2013vna}
S.~Descotes-Genon, T.~Hurth, J.~Matias, and J.~Virto, {\it {Optimizing the
  basis of ${B} \to {K}^{*}\ell^+ \ell^-$ observables in the full kinematic
  range}},  {\em JHEP} {\bf 1305} (2013) 137,
  [\href{http://arxiv.org/abs/1303.5794}{{\tt arXiv:1303.5794}}].

\bibitem{Matias:2014jua}
J.~Matias and N.~Serra, {\it {Symmetry relations between angular observables in
  $B^0 \to K^* \mu^+\mu^-$ and the LHCb $P_5^\prime$ anomaly}},  {\em
  Phys.Rev.} {\bf D90} (2014), no.~3 034002,
  [\href{http://arxiv.org/abs/1402.6855}{{\tt arXiv:1402.6855}}].

\bibitem{Khodjamirian:2010vf}
A.~Khodjamirian, T.~Mannel, A.~Pivovarov, and Y.-M. Wang, {\it {Charm-loop
  effect in $B \to K^{(*)} \ell^{+} \ell^{-}$ and $B\to K^*\gamma$}},  {\em
  JHEP} {\bf 1009} (2010) 089, [\href{http://arxiv.org/abs/1006.4945}{{\tt
  arXiv:1006.4945}}].

\bibitem{Khodjamirian:2012rm}
A.~Khodjamirian, T.~Mannel, and Y.~M. Wang, {\it {$B \to K \ell^{+}\ell^{-}$
  decay at large hadronic recoil}},  {\em JHEP} {\bf 02} (2013) 010,
  [\href{http://arxiv.org/abs/1211.0234}{{\tt arXiv:1211.0234}}].

\bibitem{Lyon:2014hpa}
J.~Lyon and R.~Zwicky, {\it {Resonances gone topsy turvy - the charm of QCD or
  new physics in $b \to s \ell^+ \ell^-$?}},
  \href{http://arxiv.org/abs/1406.0566}{{\tt arXiv:1406.0566}}.

\bibitem{Capdevila:2017ert}
B.~Capdevila, S.~Descotes-Genon, L.~Hofer, and J.~Matias, {\it {Hadronic
  uncertainties in B $\to$ K$^{*}$ $\mu^{+}$ $\mu^{-}$: a state-of-the-art
  analysis}},  {\em JHEP} {\bf 04} (2017) 016,
  [\href{http://arxiv.org/abs/1701.08672}{{\tt arXiv:1701.08672}}].

\bibitem{Blake:2017fyh}
T.~Blake, U.~Egede, P.~Owen, K.~A. Petridis, and G.~Pomery, {\it {An empirical
  model to determine the hadronic resonance contributions to $\overline{B}{} ^0
  \!\rightarrow \overline{K}{} ^{*0} \mu ^+ \mu ^- $ transitions}},  {\em Eur.
  Phys. J.} {\bf C78} (2018), no.~6 453,
  [\href{http://arxiv.org/abs/1709.03921}{{\tt arXiv:1709.03921}}].

\bibitem{Bobeth:2017vxj}
C.~Bobeth, M.~Chrzaszcz, D.~van Dyk, and J.~Virto, {\it {Long-distance effects
  in $B\rightarrow K^*\ell \ell $ from analyticity}},  {\em Eur. Phys. J.} {\bf
  C78} (2018), no.~6 451, [\href{http://arxiv.org/abs/1707.07305}{{\tt
  arXiv:1707.07305}}].

\bibitem{Jager:2012uw}
S.~J{\"a}ger and J.~Martin~Camalich, {\it {On $B \to V l l$ at small dilepton
  invariant mass, power corrections, and new physics}},  {\em JHEP} {\bf 1305}
  (2013) 043, [\href{http://arxiv.org/abs/1212.2263}{{\tt arXiv:1212.2263}}].

\bibitem{Jager:2014rwa}
S.~Jäger and J.~Martin~Camalich, {\it {Reassessing the discovery potential of
  the $B \to K^{*} \ell^+\ell^-$ decays in the large-recoil region: SM
  challenges and BSM opportunities}},  {\em Phys. Rev.} {\bf D93} (2016), no.~1
  014028, [\href{http://arxiv.org/abs/1412.3183}{{\tt arXiv:1412.3183}}].

\bibitem{Ciuchini:2015qxb}
M.~Ciuchini, M.~Fedele, E.~Franco, S.~Mishima, A.~Paul, L.~Silvestrini, and
  M.~Valli, {\it {$B\to K^* \ell^+ \ell^-$ decays at large recoil in the
  Standard Model: a theoretical reappraisal}},  {\em JHEP} {\bf 06} (2016) 116,
  [\href{http://arxiv.org/abs/1512.07157}{{\tt arXiv:1512.07157}}].

\bibitem{Ciuchini:2016weo}
M.~Ciuchini, M.~Fedele, E.~Franco, S.~Mishima, A.~Paul, L.~Silvestrini, and
  M.~Valli, {\it {$B\to K^*\ell^+\ell^-$ in the Standard Model: Elaborations
  and Interpretations}},  {\em PoS} {\bf ICHEP2016} (2016) 584,
  [\href{http://arxiv.org/abs/1611.04338}{{\tt arXiv:1611.04338}}].

\bibitem{Ciuchini:2017gva}
M.~Ciuchini, M.~Fedele, E.~Franco, S.~Mishima, A.~Paul, L.~Silvestrini, and
  M.~Valli, {\it {Knowns and Unknowns in the Predictions for $B \to
  K^{*}\mu^{+}\mu^{-}$}},  {\em Nucl. Part. Phys. Proc.} {\bf 285-286} (2017)
  45--49.

\bibitem{Kou:2018nap}
{\bf Belle II} Collaboration, W.~Altmannshofer et~al., {\it {The Belle II
  Physics Book}},  \href{http://arxiv.org/abs/1808.10567}{{\tt
  arXiv:1808.10567}}.

\bibitem{Capdevila:2016ivx}
B.~Capdevila, S.~Descotes-Genon, J.~Matias, and J.~Virto, {\it {Assessing
  lepton-flavour non-universality from $B\to K^*\ell\ell$ angular analyses}},
  {\em JHEP} {\bf 10} (2016) 075, [\href{http://arxiv.org/abs/1605.03156}{{\tt
  arXiv:1605.03156}}].

\bibitem{Serra:2016ivr}
N.~Serra, R.~Silva~Coutinho, and D.~van Dyk, {\it {Measuring the breaking of
  lepton flavor universality in $B\to K^*\ell^+\ell^-$}},  {\em Phys. Rev.}
  {\bf D95} (2017), no.~3 035029, [\href{http://arxiv.org/abs/1610.08761}{{\tt
  arXiv:1610.08761}}].

\bibitem{Albrecht:2017odf}
J.~Albrecht, F.~Bernlochner, M.~Kenzie, S.~Reichert, D.~Straub, and A.~Tully,
  {\it {Future prospects for exploring present day anomalies in flavour physics
  measurements with Belle II and LHCb}},
  \href{http://arxiv.org/abs/1709.10308}{{\tt arXiv:1709.10308}}.

\bibitem{Mauri:2018vbg}
A.~Mauri, N.~Serra, and R.~Silva~Coutinho, {\it {Towards establishing lepton
  flavor universality violation in $\bar{B}\to \bar{K}^*\ell^+\ell^-$ decays}},
   {\em Phys. Rev.} {\bf D99} (2019), no.~1 013007,
  [\href{http://arxiv.org/abs/1805.06401}{{\tt arXiv:1805.06401}}].

\bibitem{Ciuchini:2018anp}
M.~Ciuchini, A.~M. Coutinho, M.~Fedele, E.~Franco, A.~Paul, L.~Silvestrini, and
  M.~Valli, {\it {Hadronic uncertainties in semileptonic $B\to K^*\mu^+\mu^-$
  decays}},  {\em PoS} {\bf BEAUTY2018} (2018) 044,
  [\href{http://arxiv.org/abs/1809.03789}{{\tt arXiv:1809.03789}}].

\bibitem{Bordone:2016gaq}
M.~Bordone, G.~Isidori, and A.~Pattori, {\it {On the Standard Model predictions
  for $R_K$ and $R_{K^*}$}},  {\em Eur. Phys. J.} {\bf C76} (2016), no.~8 440,
  [\href{http://arxiv.org/abs/1605.07633}{{\tt arXiv:1605.07633}}].

\bibitem{Aaij:2014ora}
{\bf LHCb} Collaboration, R.~Aaij et~al., {\it {Test of lepton universality
  using $B^{+}\rightarrow K^{+}\ell^{+}\ell^{-}$ decays}},  {\em Phys. Rev.
  Lett.} {\bf 113} (2014) 151601, [\href{http://arxiv.org/abs/1406.6482}{{\tt
  arXiv:1406.6482}}].

\bibitem{Aaij:2017vbb}
{\bf LHCb} Collaboration, R.~Aaij et~al., {\it {Test of lepton universality
  with $B^{0} \rightarrow K^{*0}\ell^{+}\ell^{-}$ decays}},  {\em JHEP} {\bf
  08} (2017) 055, [\href{http://arxiv.org/abs/1705.05802}{{\tt
  arXiv:1705.05802}}].

\bibitem{Alonso:2014csa}
R.~Alonso, B.~Grinstein, and J.~Martin~Camalich, {\it {$SU(2)\times U(1)$ gauge
  invariance and the shape of new physics in rare $B$ decays}},  {\em Phys.
  Rev. Lett.} {\bf 113} (2014) 241802,
  [\href{http://arxiv.org/abs/1407.7044}{{\tt arXiv:1407.7044}}].

\bibitem{Hiller:2014yaa}
G.~Hiller and M.~Schmaltz, {\it {$R_K$ and future $b \to s \ell \ell$ physics
  beyond the standard model opportunities}},  {\em Phys. Rev.} {\bf D90} (2014)
  054014, [\href{http://arxiv.org/abs/1408.1627}{{\tt arXiv:1408.1627}}].

\bibitem{Ghosh:2014awa}
D.~Ghosh, M.~Nardecchia, and S.~A. Renner, {\it {Hint of Lepton Flavour
  Non-Universality in $B$ Meson Decays}},  {\em JHEP} {\bf 12} (2014) 131,
  [\href{http://arxiv.org/abs/1408.4097}{{\tt arXiv:1408.4097}}].

\bibitem{Glashow:2014iga}
S.~L. Glashow, D.~Guadagnoli, and K.~Lane, {\it {Lepton Flavor Violation in $B$
  Decays?}},  {\em Phys. Rev. Lett.} {\bf 114} (2015) 091801,
  [\href{http://arxiv.org/abs/1411.0565}{{\tt arXiv:1411.0565}}].

\bibitem{Hiller:2014ula}
G.~Hiller and M.~Schmaltz, {\it {Diagnosing lepton-nonuniversality in $b \to s
  \ell \ell$}},  {\em JHEP} {\bf 02} (2015) 055,
  [\href{http://arxiv.org/abs/1411.4773}{{\tt arXiv:1411.4773}}].

\bibitem{Gripaios:2014tna}
B.~Gripaios, M.~Nardecchia, and S.~A. Renner, {\it {Composite leptoquarks and
  anomalies in $B$-meson decays}},  {\em JHEP} {\bf 05} (2015) 006,
  [\href{http://arxiv.org/abs/1412.1791}{{\tt arXiv:1412.1791}}].

\bibitem{Sahoo:2015wya}
S.~Sahoo and R.~Mohanta, {\it {Scalar leptoquarks and the rare $B$ meson
  decays}},  {\em Phys. Rev.} {\bf D91} (2015), no.~9 094019,
  [\href{http://arxiv.org/abs/1501.05193}{{\tt arXiv:1501.05193}}].

\bibitem{Crivellin:2015lwa}
A.~Crivellin, G.~D'Ambrosio, and J.~Heeck, {\it {Addressing the LHC flavor
  anomalies with horizontal gauge symmetries}},  {\em Phys. Rev.} {\bf D91}
  (2015), no.~7 075006, [\href{http://arxiv.org/abs/1503.03477}{{\tt
  arXiv:1503.03477}}].

\bibitem{Crivellin:2015era}
A.~Crivellin, L.~Hofer, J.~Matias, U.~Nierste, S.~Pokorski, and J.~Rosiek, {\it
  {Lepton-flavour violating $B$ decays in generic $Z'$ models}},  {\em Phys.
  Rev.} {\bf D92} (2015), no.~5 054013,
  [\href{http://arxiv.org/abs/1504.07928}{{\tt arXiv:1504.07928}}].

\bibitem{Celis:2015ara}
A.~Celis, J.~Fuentes-Martin, M.~Jung, and H.~Serodio, {\it {Family nonuniversal
  Z' models with protected flavor-changing interactions}},  {\em Phys. Rev.}
  {\bf D92} (2015), no.~1 015007, [\href{http://arxiv.org/abs/1505.03079}{{\tt
  arXiv:1505.03079}}].

\bibitem{Alonso:2015sja}
R.~Alonso, B.~Grinstein, and J.~Martin~Camalich, {\it {Lepton universality
  violation and lepton flavor conservation in $B$-meson decays}},  {\em JHEP}
  {\bf 10} (2015) 184, [\href{http://arxiv.org/abs/1505.05164}{{\tt
  arXiv:1505.05164}}].

\bibitem{Greljo:2015mma}
A.~Greljo, G.~Isidori, and D.~Marzocca, {\it {On the breaking of Lepton Flavor
  Universality in B decays}},  {\em JHEP} {\bf 07} (2015) 142,
  [\href{http://arxiv.org/abs/1506.01705}{{\tt arXiv:1506.01705}}].

\bibitem{Calibbi:2015kma}
L.~Calibbi, A.~Crivellin, and T.~Ota, {\it {Effective Field Theory Approach to
  $b \to sll^{(')}$, $B \to K^{(*)}\nu\overline{\nu}$ and $B \to
  D^{(*)}\tau\nu$ with Third Generation Couplings}},  {\em Phys. Rev. Lett.}
  {\bf 115} (2015) 181801, [\href{http://arxiv.org/abs/1506.02661}{{\tt
  arXiv:1506.02661}}].

\bibitem{Falkowski:2015zwa}
A.~Falkowski, M.~Nardecchia, and R.~Ziegler, {\it {Lepton Flavor
  Non-Universality in B-meson Decays from a U(2) Flavor Model}},  {\em JHEP}
  {\bf 11} (2015) 173, [\href{http://arxiv.org/abs/1509.01249}{{\tt
  arXiv:1509.01249}}].

\bibitem{Carmona:2015ena}
A.~Carmona and F.~Goertz, {\it {Lepton Flavor and Nonuniversality from Minimal
  Composite Higgs Setups}},  {\em Phys. Rev. Lett.} {\bf 116} (2016), no.~25
  251801, [\href{http://arxiv.org/abs/1510.07658}{{\tt arXiv:1510.07658}}].

\bibitem{Allanach:2015gkd}
B.~Allanach, F.~S. Queiroz, A.~Strumia, and S.~Sun, {\it {$Z^{\prime}$ models
  for the LHCb and $g-2$ muon anomalies}},  {\em Phys. Rev.} {\bf D93} (2016),
  no.~5 055045, [\href{http://arxiv.org/abs/1511.07447}{{\tt
  arXiv:1511.07447}}]. [Erratum: Phys. Rev.D95,no.11,119902(2017)].

\bibitem{Chiang:2016qov}
C.-W. Chiang, X.-G. He, and G.~Valencia, {\it {$Z'$ model for $b\to s \ell
  \overline{\ell}$ flavor anomalies}},  {\em Phys. Rev.} {\bf D93} (2016),
  no.~7 074003, [\href{http://arxiv.org/abs/1601.07328}{{\tt
  arXiv:1601.07328}}].

\bibitem{Becirevic:2016zri}
D.~Be{\v c}irevi{\'c}, O.~Sumensari, and R.~Zukanovich~Funchal, {\it {Lepton
  flavor violation in exclusive $b\rightarrow s$ decays}},  {\em Eur. Phys. J.}
  {\bf C76} (2016), no.~3 134, [\href{http://arxiv.org/abs/1602.00881}{{\tt
  arXiv:1602.00881}}].

\bibitem{Feruglio:2016gvd}
F.~Feruglio, P.~Paradisi, and A.~Pattori, {\it {Revisiting Lepton Flavor
  Universality in B Decays}},  {\em Phys. Rev. Lett.} {\bf 118} (2017), no.~1
  011801, [\href{http://arxiv.org/abs/1606.00524}{{\tt arXiv:1606.00524}}].

\bibitem{Megias:2016bde}
E.~Megias, G.~Panico, O.~Pujolas, and M.~Quiros, {\it {A Natural origin for the
  LHCb anomalies}},  {\em JHEP} {\bf 09} (2016) 118,
  [\href{http://arxiv.org/abs/1608.02362}{{\tt arXiv:1608.02362}}].

\bibitem{Becirevic:2016oho}
D.~Be{\v c}irevi{\'c}, N.~Ko{\v s}nik, O.~Sumensari, and R.~Zukanovich~Funchal,
  {\it {Palatable Leptoquark Scenarios for Lepton Flavor Violation in Exclusive
  $b\to s\ell_1\ell_2$ modes}},  {\em JHEP} {\bf 11} (2016) 035,
  [\href{http://arxiv.org/abs/1608.07583}{{\tt arXiv:1608.07583}}].

\bibitem{Arnan:2016cpy}
P.~Arnan, L.~Hofer, F.~Mescia, and A.~Crivellin, {\it {Loop effects of heavy
  new scalars and fermions in $b\to s\mu^+\mu^-$}},  {\em JHEP} {\bf 04} (2017)
  043, [\href{http://arxiv.org/abs/1608.07832}{{\tt arXiv:1608.07832}}].

\bibitem{Altmannshofer:2016jzy}
W.~Altmannshofer, S.~Gori, S.~Profumo, and F.~S. Queiroz, {\it {Explaining dark
  matter and B decay anomalies with an $L_\mu - L_\tau$ model}},  {\em JHEP}
  {\bf 12} (2016) 106, [\href{http://arxiv.org/abs/1609.04026}{{\tt
  arXiv:1609.04026}}].

\bibitem{Sahoo:2016pet}
S.~Sahoo, R.~Mohanta, and A.~K. Giri, {\it {Explaining the $R_{K}$ and
  $R_{D^{(*)}}$ anomalies with vector leptoquarks}},  {\em Phys. Rev.} {\bf
  D95} (2017), no.~3 035027, [\href{http://arxiv.org/abs/1609.04367}{{\tt
  arXiv:1609.04367}}].

\bibitem{Alonso:2016onw}
R.~Alonso, E.~Fernandez~Martinez, M.~B. Gavela, B.~Grinstein, L.~Merlo, and
  P.~Quilez, {\it {Gauged Lepton Flavour}},  {\em JHEP} {\bf 12} (2016) 119,
  [\href{http://arxiv.org/abs/1609.05902}{{\tt arXiv:1609.05902}}].

\bibitem{Hiller:2016kry}
G.~Hiller, D.~Loose, and K.~Schonwald, {\it {Leptoquark Flavor Patterns \& B
  Decay Anomalies}},  {\em JHEP} {\bf 12} (2016) 027,
  [\href{http://arxiv.org/abs/1609.08895}{{\tt arXiv:1609.08895}}].

\bibitem{Galon:2016bka}
I.~Galon, A.~Kwa, and P.~Tanedo, {\it {Lepton-Flavor Violating Mediators}},
  {\em JHEP} {\bf 03} (2017) 064, [\href{http://arxiv.org/abs/1610.08060}{{\tt
  arXiv:1610.08060}}].

\bibitem{Crivellin:2016ejn}
A.~Crivellin, J.~Fuentes-Martin, A.~Greljo, and G.~Isidori, {\it {Lepton Flavor
  Non-Universality in B decays from Dynamical Yukawas}},  {\em Phys. Lett.}
  {\bf B766} (2017) 77--85, [\href{http://arxiv.org/abs/1611.02703}{{\tt
  arXiv:1611.02703}}].

\bibitem{GarciaGarcia:2016nvr}
I.~Garc{\'\i}a~Garc{\'\i}a, {\it {LHCb anomalies from a natural perspective}},
  {\em JHEP} {\bf 03} (2017) 040, [\href{http://arxiv.org/abs/1611.03507}{{\tt
  arXiv:1611.03507}}].

\bibitem{Cox:2016epl}
P.~Cox, A.~Kusenko, O.~Sumensari, and T.~T. Yanagida, {\it {SU(5) Unification
  with TeV-scale Leptoquarks}},  {\em JHEP} {\bf 03} (2017) 035,
  [\href{http://arxiv.org/abs/1612.03923}{{\tt arXiv:1612.03923}}].

\bibitem{Jager:2017gal}
S.~Jäger, M.~Kirk, A.~Lenz, and K.~Leslie, {\it {Charming new physics in rare
  B-decays and mixing?}},  {\em Phys. Rev.} {\bf D97} (2018), no.~1 015021,
  [\href{http://arxiv.org/abs/1701.09183}{{\tt arXiv:1701.09183}}].

\bibitem{Megias:2017ove}
E.~Megias, M.~Quiros, and L.~Salas, {\it {Lepton-flavor universality violation
  in $R_{K}$ and $ {R}_{D^{{\left(\ast \right)}}} $ from warped space}},  {\em
  JHEP} {\bf 07} (2017) 102, [\href{http://arxiv.org/abs/1703.06019}{{\tt
  arXiv:1703.06019}}].

\bibitem{Crivellin:2017zlb}
A.~Crivellin, D.~M{\"u}ller, and T.~Ota, {\it {Simultaneous explanation of
  $R(D^{(*)})$ and $b \to s \mu^{+} \mu^{-}$: the last scalar leptoquarks
  standing}},  {\em JHEP} {\bf 09} (2017) 040,
  [\href{http://arxiv.org/abs/1703.09226}{{\tt arXiv:1703.09226}}].

\bibitem{Celis:2017doq}
A.~Celis, J.~Fuentes-Martin, A.~Vicente, and J.~Virto, {\it {Gauge-invariant
  implications of the LHCb measurements on lepton-flavor nonuniversality}},
  {\em Phys. Rev.} {\bf D96} (2017), no.~3 035026,
  [\href{http://arxiv.org/abs/1704.05672}{{\tt arXiv:1704.05672}}].

\bibitem{Becirevic:2017jtw}
D.~Be{\v c}irevi{\'c} and O.~Sumensari, {\it {A leptoquark model to accommodate
  $R_K^\mathrm{exp} < R_K^\mathrm{SM}$ and $R_{K^\ast}^\mathrm{exp} <
  R_{K^\ast}^\mathrm{SM}$}},  {\em JHEP} {\bf 08} (2017) 104,
  [\href{http://arxiv.org/abs/1704.05835}{{\tt arXiv:1704.05835}}].

\bibitem{Cai:2017wry}
Y.~Cai, J.~Gargalionis, M.~A. Schmidt, and R.~R. Volkas, {\it {Reconsidering
  the One Leptoquark solution: flavor anomalies and neutrino mass}},  {\em
  JHEP} {\bf 10} (2017) 047, [\href{http://arxiv.org/abs/1704.05849}{{\tt
  arXiv:1704.05849}}].

\bibitem{Kamenik:2017tnu}
J.~F. Kamenik, Y.~Soreq, and J.~Zupan, {\it {Lepton flavor universality
  violation without new sources of quark flavor violation}},  {\em Phys. Rev.}
  {\bf D97} (2018), no.~3 035002, [\href{http://arxiv.org/abs/1704.06005}{{\tt
  arXiv:1704.06005}}].

\bibitem{Sala:2017ihs}
F.~Sala and D.~M. Straub, {\it {A New Light Particle in B Decays?}},  {\em
  Phys. Lett.} {\bf B774} (2017) 205--209,
  [\href{http://arxiv.org/abs/1704.06188}{{\tt arXiv:1704.06188}}].

\bibitem{DiChiara:2017cjq}
S.~Di~Chiara, A.~Fowlie, S.~Fraser, C.~Marzo, L.~Marzola, M.~Raidal, and
  C.~Spethmann, {\it {Minimal flavor-changing $Z'$ models and muon $g-2$ after
  the $R_{K^*}$ measurement}},  {\em Nucl. Phys.} {\bf B923} (2017) 245--257,
  [\href{http://arxiv.org/abs/1704.06200}{{\tt arXiv:1704.06200}}].

\bibitem{Ghosh:2017ber}
D.~Ghosh, {\it {Explaining the $R_K$ and $R_{K^*}$ anomalies}},  {\em Eur.
  Phys. J.} {\bf C77} (2017), no.~10 694,
  [\href{http://arxiv.org/abs/1704.06240}{{\tt arXiv:1704.06240}}].

\bibitem{Alonso:2017bff}
R.~Alonso, P.~Cox, C.~Han, and T.~T. Yanagida, {\it {Anomaly-free local
  horizontal symmetry and anomaly-full rare B-decays}},  {\em Phys. Rev.} {\bf
  D96} (2017), no.~7 071701, [\href{http://arxiv.org/abs/1704.08158}{{\tt
  arXiv:1704.08158}}].

\bibitem{Greljo:2017vvb}
A.~Greljo and D.~Marzocca, {\it {High-$p_T$ dilepton tails and flavor
  physics}},  {\em Eur. Phys. J.} {\bf C77} (2017), no.~8 548,
  [\href{http://arxiv.org/abs/1704.09015}{{\tt arXiv:1704.09015}}].

\bibitem{Bonilla:2017lsq}
C.~Bonilla, T.~Modak, R.~Srivastava, and J.~W.~F. Valle, {\it
  {$U(1)_{B_3-3L_\mu}$ gauge symmetry as a simple description of $b\to s$
  anomalies}},  {\em Phys. Rev.} {\bf D98} (2018), no.~9 095002,
  [\href{http://arxiv.org/abs/1705.00915}{{\tt arXiv:1705.00915}}].

\bibitem{Feruglio:2017rjo}
F.~Feruglio, P.~Paradisi, and A.~Pattori, {\it {On the Importance of
  Electroweak Corrections for B Anomalies}},  {\em JHEP} {\bf 09} (2017) 061,
  [\href{http://arxiv.org/abs/1705.00929}{{\tt arXiv:1705.00929}}].

\bibitem{Ellis:2017nrp}
J.~Ellis, M.~Fairbairn, and P.~Tunney, {\it {Anomaly-Free Models for Flavour
  Anomalies}},  {\em Eur. Phys. J.} {\bf C78} (2018), no.~3 238,
  [\href{http://arxiv.org/abs/1705.03447}{{\tt arXiv:1705.03447}}].

\bibitem{Bishara:2017pje}
F.~Bishara, U.~Haisch, and P.~F. Monni, {\it {Regarding light resonance
  interpretations of the B decay anomalies}},  {\em Phys. Rev.} {\bf D96}
  (2017), no.~5 055002, [\href{http://arxiv.org/abs/1705.03465}{{\tt
  arXiv:1705.03465}}].

\bibitem{Alonso:2017uky}
R.~Alonso, P.~Cox, C.~Han, and T.~T. Yanagida, {\it {Flavoured $B-L$ local
  symmetry and anomalous rare $B$ decays}},  {\em Phys. Lett.} {\bf B774}
  (2017) 643--648, [\href{http://arxiv.org/abs/1705.03858}{{\tt
  arXiv:1705.03858}}].

\bibitem{Tang:2017gkz}
Y.~Tang and Y.-L. Wu, {\it {Flavor non-universal gauge interactions and
  anomalies in B-meson decays}},  {\em Chin. Phys.} {\bf C42} (2018), no.~3
  033104, [\href{http://arxiv.org/abs/1705.05643}{{\tt arXiv:1705.05643}}].

\bibitem{Datta:2017ezo}
A.~Datta, J.~Kumar, J.~Liao, and D.~Marfatia, {\it {New light mediators for the
  $R_K$ and $R_{K^*}$ puzzles}},  {\em Phys. Rev.} {\bf D97} (2018), no.~11
  115038, [\href{http://arxiv.org/abs/1705.08423}{{\tt arXiv:1705.08423}}].

\bibitem{Das:2017kfo}
D.~Das, C.~Hati, G.~Kumar, and N.~Mahajan, {\it {Scrutinizing $R$-parity
  violating interactions in light of $R_{K^{(\ast)}}$ data}},  {\em Phys. Rev.}
  {\bf D96} (2017), no.~9 095033, [\href{http://arxiv.org/abs/1705.09188}{{\tt
  arXiv:1705.09188}}].

\bibitem{Dinh:2017smk}
D.~N. Dinh, L.~Merlo, S.~T. Petcov, and R.~Vega-{\'A}lvarez, {\it {Revisiting
  Minimal Lepton Flavour Violation in the Light of Leptonic CP Violation}},
  {\em JHEP} {\bf 07} (2017) 089, [\href{http://arxiv.org/abs/1705.09284}{{\tt
  arXiv:1705.09284}}].

\bibitem{Bardhan:2017xcc}
D.~Bardhan, P.~Byakti, and D.~Ghosh, {\it {Role of Tensor operators in $R_K$
  and $R_{K^*}$}},  {\em Phys. Lett.} {\bf B773} (2017) 505--512,
  [\href{http://arxiv.org/abs/1705.09305}{{\tt arXiv:1705.09305}}].

\bibitem{DiLuzio:2017chi}
L.~Di~Luzio and M.~Nardecchia, {\it {What is the scale of new physics behind
  the $B$-flavour anomalies?}},  {\em Eur. Phys. J.} {\bf C77} (2017), no.~8
  536, [\href{http://arxiv.org/abs/1706.01868}{{\tt arXiv:1706.01868}}].

\bibitem{Chiang:2017hlj}
C.-W. Chiang, X.-G. He, J.~Tandean, and X.-B. Yuan, {\it {$R_{K^{(*)}}$ and
  related $b\to s\ell\bar\ell$ anomalies in minimal flavor violation framework
  with $Z'$ boson}},  {\em Phys. Rev.} {\bf D96} (2017), no.~11 115022,
  [\href{http://arxiv.org/abs/1706.02696}{{\tt arXiv:1706.02696}}].

\bibitem{Chauhan:2017ndd}
B.~Chauhan, B.~Kindra, and A.~Narang, {\it {Discrepancies in simultaneous
  explanation of flavor anomalies and IceCube PeV events using leptoquarks}},
  {\em Phys. Rev.} {\bf D97} (2018), no.~9 095007,
  [\href{http://arxiv.org/abs/1706.04598}{{\tt arXiv:1706.04598}}].

\bibitem{King:2017anf}
S.~F. King, {\it {Flavourful Z$^{'}$ models for $ {R}_{K^{\left(\ast \right)}}
  $}},  {\em JHEP} {\bf 08} (2017) 019,
  [\href{http://arxiv.org/abs/1706.06100}{{\tt arXiv:1706.06100}}].

\bibitem{Chivukula:2017qsi}
R.~S. Chivukula, J.~Isaacson, K.~A. Mohan, D.~Sengupta, and E.~H. Simmons, {\it
  {$R_K$ anomalies and simplified limits on $Z'$ models at the LHC}},  {\em
  Phys. Rev.} {\bf D96} (2017), no.~7 075012,
  [\href{http://arxiv.org/abs/1706.06575}{{\tt arXiv:1706.06575}}].

\bibitem{Dorsner:2017ufx}
I.~Doršner, S.~Fajfer, D.~A. Faroughy, and N.~Košnik, {\it {The role of the
  $S_3$ GUT leptoquark in flavor universality and collider searches}},  {\em
  JHEP} {\bf 10} (2017) 188, [\href{http://arxiv.org/abs/1706.07779}{{\tt
  arXiv:1706.07779}}].

\bibitem{Buttazzo:2017ixm}
D.~Buttazzo, A.~Greljo, G.~Isidori, and D.~Marzocca, {\it {B-physics anomalies:
  a guide to combined explanations}},  {\em JHEP} {\bf 11} (2017) 044,
  [\href{http://arxiv.org/abs/1706.07808}{{\tt arXiv:1706.07808}}].

\bibitem{Choudhury:2017qyt}
D.~Choudhury, A.~Kundu, R.~Mandal, and R.~Sinha, {\it {Minimal unified
  resolution to $R_{K^{(*)}}$ and $R(D^{(*)})$ anomalies with lepton mixing}},
  {\em Phys. Rev. Lett.} {\bf 119} (2017), no.~15 151801,
  [\href{http://arxiv.org/abs/1706.08437}{{\tt arXiv:1706.08437}}].

\bibitem{Cline:2017ihf}
J.~M. Cline and J.~Martin~Camalich, {\it {$B$ decay anomalies from nonabelian
  local horizontal symmetry}},  {\em Phys. Rev.} {\bf D96} (2017), no.~5
  055036, [\href{http://arxiv.org/abs/1706.08510}{{\tt arXiv:1706.08510}}].

\bibitem{Crivellin:2017dsk}
A.~Crivellin, D.~Müller, A.~Signer, and Y.~Ulrich, {\it {Correlating lepton
  flavor universality violation in $B$ decays with $\mu\to e\gamma$ using
  leptoquarks}},  {\em Phys. Rev.} {\bf D97} (2018), no.~1 015019,
  [\href{http://arxiv.org/abs/1706.08511}{{\tt arXiv:1706.08511}}].

\bibitem{Guo:2017gxp}
S.-Y. Guo, Z.-L. Han, B.~Li, Y.~Liao, and X.-D. Ma, {\it {Interpreting the
  $R_{K^{(*)}}$ anomaly in the colored Zee–Babu model}},  {\em Nucl. Phys.}
  {\bf B928} (2018) 435--447, [\href{http://arxiv.org/abs/1707.00522}{{\tt
  arXiv:1707.00522}}].

\bibitem{Chen:2017usq}
C.-H. Chen and T.~Nomura, {\it {Penguin $b \to s\ell'^+ \ell'^-$ and $B$-meson
  anomalies in a gauged ${L_\mu -L_\tau}$}},  {\em Phys. Lett.} {\bf B777}
  (2018) 420--427, [\href{http://arxiv.org/abs/1707.03249}{{\tt
  arXiv:1707.03249}}].

\bibitem{Baek:2017sew}
S.~Baek, {\it {Dark matter contribution to $b\to s \mu^+ \mu^-$ anomaly in
  local $U(1)_{L_\mu-L_\tau}$ model}},  {\em Phys. Lett.} {\bf B781} (2018)
  376--382, [\href{http://arxiv.org/abs/1707.04573}{{\tt arXiv:1707.04573}}].

\bibitem{Bian:2017rpg}
L.~Bian, S.-M. Choi, Y.-J. Kang, and H.~M. Lee, {\it {A minimal flavored
  $U(1)'$ for $B$-meson anomalies}},  {\em Phys. Rev.} {\bf D96} (2017), no.~7
  075038, [\href{http://arxiv.org/abs/1707.04811}{{\tt arXiv:1707.04811}}].

\bibitem{Megias:2017vdg}
E.~Megias, M.~Quiros, and L.~Salas, {\it {Lepton-flavor universality limits in
  warped space}},  {\em Phys. Rev.} {\bf D96} (2017), no.~7 075030,
  [\href{http://arxiv.org/abs/1707.08014}{{\tt arXiv:1707.08014}}].

\bibitem{Lee:2017fin}
H.~M. Lee, {\it {Gauged $U(1)$ clockwork theory}},  {\em Phys. Lett.} {\bf
  B778} (2018) 79--87, [\href{http://arxiv.org/abs/1708.03564}{{\tt
  arXiv:1708.03564}}].

\bibitem{Assad:2017iib}
N.~Assad, B.~Fornal, and B.~Grinstein, {\it {Baryon Number and Lepton
  Universality Violation in Leptoquark and Diquark Models}},  {\em Phys. Lett.}
  {\bf B777} (2018) 324--331, [\href{http://arxiv.org/abs/1708.06350}{{\tt
  arXiv:1708.06350}}].

\bibitem{DiLuzio:2017vat}
L.~Di~Luzio, A.~Greljo, and M.~Nardecchia, {\it {Gauge leptoquark as the origin
  of B-physics anomalies}},  {\em Phys. Rev.} {\bf D96} (2017), no.~11 115011,
  [\href{http://arxiv.org/abs/1708.08450}{{\tt arXiv:1708.08450}}].

\bibitem{Calibbi:2017qbu}
L.~Calibbi, A.~Crivellin, and T.~Li, {\it {Model of vector leptoquarks in view
  of the $B$-physics anomalies}},  {\em Phys. Rev.} {\bf D98} (2018), no.~11
  115002, [\href{http://arxiv.org/abs/1709.00692}{{\tt arXiv:1709.00692}}].

\bibitem{Cline:2017aed}
J.~M. Cline, {\it {$B$ decay anomalies and dark matter from vectorlike
  confinement}},  {\em Phys. Rev.} {\bf D97} (2018), no.~1 015013,
  [\href{http://arxiv.org/abs/1710.02140}{{\tt arXiv:1710.02140}}].

\bibitem{Romao:2017qnu}
M.~Crispim~Romao, S.~F. King, and G.~K. Leontaris, {\it {Non-universal $Z'$
  from fluxed GUTs}},  {\em Phys. Lett.} {\bf B782} (2018) 353--361,
  [\href{http://arxiv.org/abs/1710.02349}{{\tt arXiv:1710.02349}}].

\bibitem{Descotes-Genon:2017ptp}
S.~Descotes-Genon, M.~Moscati, and G.~Ricciardi, {\it {Nonminimal 331 model for
  lepton flavor universality violation in $b{\rightarrow}s{\ell}{\ell}$
  decays}},  {\em Phys. Rev.} {\bf D98} (2018), no.~11 115030,
  [\href{http://arxiv.org/abs/1711.03101}{{\tt arXiv:1711.03101}}].

\bibitem{Altmannshofer:2017bsz}
W.~Altmannshofer, M.~J. Baker, S.~Gori, R.~Harnik, M.~Pospelov, E.~Stamou, and
  A.~Thamm, {\it {Light resonances and the low-q$^{2}$ bin of $ {R}_{K^{*}}
  $}},  {\em JHEP} {\bf 03} (2018) 188,
  [\href{http://arxiv.org/abs/1711.07494}{{\tt arXiv:1711.07494}}].

\bibitem{Bian:2017xzg}
L.~Bian, H.~M. Lee, and C.~B. Park, {\it {$B$-meson anomalies and Higgs physics
  in flavored $U(1)'$ model}},  {\em Eur. Phys. J.} {\bf C78} (2018), no.~4
  306, [\href{http://arxiv.org/abs/1711.08930}{{\tt arXiv:1711.08930}}].

\bibitem{Cline:2017qqu}
J.~M. Cline and J.~M. Cornell, {\it {$R({K^{(*)}})$ from dark matter
  exchange}},  {\em Phys. Lett.} {\bf B782} (2018) 232--237,
  [\href{http://arxiv.org/abs/1711.10770}{{\tt arXiv:1711.10770}}].

\bibitem{Botella:2017caf}
F.~J. Botella, G.~C. Branco, and M.~Nebot, {\it {Singlet Heavy Fermions as the
  Origin of B Anomalies in Flavour Changing Neutral Currents}},
  \href{http://arxiv.org/abs/1712.04470}{{\tt arXiv:1712.04470}}.

\bibitem{DiLuzio:2017fdq}
L.~Di~Luzio, M.~Kirk, and A.~Lenz, {\it {Updated $B_s$-mixing constraints on
  new physics models for $b\to s\ell^+\ell^-$ anomalies}},  {\em Phys. Rev.}
  {\bf D97} (2018), no.~9 095035, [\href{http://arxiv.org/abs/1712.06572}{{\tt
  arXiv:1712.06572}}].

\bibitem{Barbieri:2017tuq}
R.~Barbieri and A.~Tesi, {\it {$B$-decay anomalies in Pati-Salam SU(4)}},  {\em
  Eur. Phys. J.} {\bf C78} (2018), no.~3 193,
  [\href{http://arxiv.org/abs/1712.06844}{{\tt arXiv:1712.06844}}].

\bibitem{Sannino:2017utc}
F.~Sannino, P.~Stangl, D.~M. Straub, and A.~E. Thomsen, {\it {Flavor Physics
  and Flavor Anomalies in Minimal Fundamental Partial Compositeness}},  {\em
  Phys. Rev.} {\bf D97} (2018), no.~11 115046,
  [\href{http://arxiv.org/abs/1712.07646}{{\tt arXiv:1712.07646}}].

\bibitem{DAmbrosio:2017wis}
G.~D'Ambrosio and A.~M. Iyer, {\it {Flavour issues in warped custodial models:
  $B$ anomalies and rare $K$ decays}},  {\em Eur. Phys. J.} {\bf C78} (2018),
  no.~6 448, [\href{http://arxiv.org/abs/1712.08122}{{\tt arXiv:1712.08122}}].

\bibitem{Raby:2017igl}
S.~Raby and A.~Trautner, {\it {Vectorlike chiral fourth family to explain muon
  anomalies}},  {\em Phys. Rev.} {\bf D97} (2018), no.~9 095006,
  [\href{http://arxiv.org/abs/1712.09360}{{\tt arXiv:1712.09360}}].

\bibitem{Blanke:2018sro}
M.~Blanke and A.~Crivellin, {\it {$B$ Meson Anomalies in a Pati-Salam Model
  within the Randall-Sundrum Background}},  {\em Phys. Rev. Lett.} {\bf 121}
  (2018), no.~1 011801, [\href{http://arxiv.org/abs/1801.07256}{{\tt
  arXiv:1801.07256}}].

\bibitem{Falkowski:2018dsl}
A.~Falkowski, S.~F. King, E.~Perdomo, and M.~Pierre, {\it {Flavourful $Z'$
  portal for vector-like neutrino Dark Matter and $R_{K^{(*)}}$}},  {\em JHEP}
  {\bf 08} (2018) 061, [\href{http://arxiv.org/abs/1803.04430}{{\tt
  arXiv:1803.04430}}].

\bibitem{Arcadi:2018tly}
G.~Arcadi, T.~Hugle, and F.~S. Queiroz, {\it {The Dark $L_\mu - L_\tau$ Rises
  via Kinetic Mixing}},  {\em Phys. Lett.} {\bf B784} (2018) 151--158,
  [\href{http://arxiv.org/abs/1803.05723}{{\tt arXiv:1803.05723}}].

\bibitem{CarcamoHernandez:2018aon}
A.~E. Cárcamo~Hernández and S.~F. King, {\it {Muon anomalies and the $SU(5)$
  Yukawa relations}},  \href{http://arxiv.org/abs/1803.07367}{{\tt
  arXiv:1803.07367}}.

\bibitem{Marzocca:2018wcf}
D.~Marzocca, {\it {Addressing the B-physics anomalies in a fundamental
  Composite Higgs Model}},  {\em JHEP} {\bf 07} (2018) 121,
  [\href{http://arxiv.org/abs/1803.10972}{{\tt arXiv:1803.10972}}].

\bibitem{Camargo-Molina:2018cwu}
J.~E. Camargo-Molina, A.~Celis, and D.~A. Faroughy, {\it {Anomalies in Bottom
  from new physics in Top}},  {\em Phys. Lett.} {\bf B784} (2018) 284--293,
  [\href{http://arxiv.org/abs/1805.04917}{{\tt arXiv:1805.04917}}].

\bibitem{Bordone:2018nbg}
M.~Bordone, C.~Cornella, J.~Fuentes-Martín, and G.~Isidori, {\it {Low-energy
  signatures of the $\mathrm{PS}^3$ model: from $B$-physics anomalies to LFV}},
   {\em JHEP} {\bf 10} (2018) 148, [\href{http://arxiv.org/abs/1805.09328}{{\tt
  arXiv:1805.09328}}].

\bibitem{Earl:2018snx}
K.~Earl and T.~Grégoire, {\it {Contributions to ${b \rightarrow s \ell \ell}$
  Anomalies from ${R}$-Parity Violating Interactions}},  {\em JHEP} {\bf 08}
  (2018) 201, [\href{http://arxiv.org/abs/1806.01343}{{\tt arXiv:1806.01343}}].

\bibitem{Matsuzaki:2018jui}
S.~Matsuzaki, K.~Nishiwaki, and K.~Yamamoto, {\it {Simultaneous interpretation
  of $K$ and $B$ anomalies in terms of chiral-flavorful vectors}},  {\em JHEP}
  {\bf 11} (2018) 164, [\href{http://arxiv.org/abs/1806.02312}{{\tt
  arXiv:1806.02312}}].

\bibitem{Becirevic:2018afm}
D.~Bečirević, I.~Doršner, S.~Fajfer, N.~Košnik, D.~A. Faroughy, and
  O.~Sumensari, {\it {Scalar leptoquarks from grand unified theories to
  accommodate the $B$-physics anomalies}},  {\em Phys. Rev.} {\bf D98} (2018),
  no.~5 055003, [\href{http://arxiv.org/abs/1806.05689}{{\tt
  arXiv:1806.05689}}].

\bibitem{Baek:2018aru}
S.~Baek and C.~Yu, {\it {Dark matter for $b\to s \mu^+ \mu^-$ anomaly in a
  gauged $U(1)_X$ model}},  {\em JHEP} {\bf 11} (2018) 054,
  [\href{http://arxiv.org/abs/1806.05967}{{\tt arXiv:1806.05967}}].

\bibitem{King:2018fcg}
S.~F. King, {\it {$ {R}_{K^{\left(*\right)}} $ and the origin of Yukawa
  couplings}},  {\em JHEP} {\bf 09} (2018) 069,
  [\href{http://arxiv.org/abs/1806.06780}{{\tt arXiv:1806.06780}}].

\bibitem{Kumar:2018kmr}
J.~Kumar, D.~London, and R.~Watanabe, {\it {Combined Explanations of the $b \to
  s \mu^+ \mu^-$ and $b \to c \tau^- {\bar\nu}$ Anomalies: a General Model
  Analysis}},  {\em Phys. Rev.} {\bf D99} (2019), no.~1 015007,
  [\href{http://arxiv.org/abs/1806.07403}{{\tt arXiv:1806.07403}}].

\bibitem{Hati:2018fzc}
C.~Hati, G.~Kumar, J.~Orloff, and A.~M. Teixeira, {\it {Reconciling $B$-meson
  decay anomalies with neutrino masses, dark matter and constraints from
  flavour violation}},  {\em JHEP} {\bf 11} (2018) 011,
  [\href{http://arxiv.org/abs/1806.10146}{{\tt arXiv:1806.10146}}].

\bibitem{Guadagnoli:2018ojc}
D.~Guadagnoli, M.~Reboud, and O.~Sumensari, {\it {A gauged horizontal $SU(2)$
  symmetry and $R_{K^{(\ast)}}$}},  {\em JHEP} {\bf 11} (2018) 163,
  [\href{http://arxiv.org/abs/1807.03285}{{\tt arXiv:1807.03285}}].

\bibitem{deMedeirosVarzielas:2018bcy}
I.~de~Medeiros~Varzielas and S.~F. King, {\it {$ {R}_{K^{\left(*\right)}} $
  with leptoquarks and the origin of Yukawa couplings}},  {\em JHEP} {\bf 11}
  (2018) 100, [\href{http://arxiv.org/abs/1807.06023}{{\tt arXiv:1807.06023}}].

\bibitem{Li:2018rax}
S.-P. Li, X.-Q. Li, Y.-D. Yang, and X.~Zhang, {\it {$
  {R}_{D^{\left(*\right)}},{R}_{K^{\left(*\right)}} $ and neutrino mass in the
  2HDM-III with right-handed neutrinos}},  {\em JHEP} {\bf 09} (2018) 149,
  [\href{http://arxiv.org/abs/1807.08530}{{\tt arXiv:1807.08530}}].

\bibitem{Alonso:2018bcg}
R.~Alonso, A.~Carmona, B.~M. Dillon, J.~F. Kamenik, J.~Martin~Camalich, and
  J.~Zupan, {\it {A clockwork solution to the flavor puzzle}},  {\em JHEP} {\bf
  10} (2018) 099, [\href{http://arxiv.org/abs/1807.09792}{{\tt
  arXiv:1807.09792}}].

\bibitem{Azatov:2018kzb}
A.~Azatov, D.~Barducci, D.~Ghosh, D.~Marzocca, and L.~Ubaldi, {\it {Combined
  explanations of B-physics anomalies: the sterile neutrino solution}},  {\em
  JHEP} {\bf 10} (2018) 092, [\href{http://arxiv.org/abs/1807.10745}{{\tt
  arXiv:1807.10745}}].

\bibitem{DiLuzio:2018zxy}
L.~Di~Luzio, J.~Fuentes-Martin, A.~Greljo, M.~Nardecchia, and S.~Renner, {\it
  {Maximal Flavour Violation: a Cabibbo mechanism for leptoquarks}},  {\em
  JHEP} {\bf 11} (2018) 081, [\href{http://arxiv.org/abs/1808.00942}{{\tt
  arXiv:1808.00942}}].

\bibitem{Duan:2018akc}
G.~H. Duan, X.~Fan, M.~Frank, C.~Han, and J.~M. Yang, {\it {A minimal
  $U(1)^\prime$ extension of MSSM in light of the B decay anomaly}},  {\em
  Phys. Lett.} {\bf B789} (2019) 54--58,
  [\href{http://arxiv.org/abs/1808.04116}{{\tt arXiv:1808.04116}}].

\bibitem{Heeck:2018ntp}
J.~Heeck and D.~Teresi, {\it {Pati-Salam explanations of the B-meson
  anomalies}},  {\em JHEP} {\bf 12} (2018) 103,
  [\href{http://arxiv.org/abs/1808.07492}{{\tt arXiv:1808.07492}}].

\bibitem{Angelescu:2018tyl}
A.~Angelescu, D.~Bečirević, D.~A. Faroughy, and O.~Sumensari, {\it {Closing
  the window on single leptoquark solutions to the $B$-physics anomalies}},
  {\em JHEP} {\bf 10} (2018) 183, [\href{http://arxiv.org/abs/1808.08179}{{\tt
  arXiv:1808.08179}}].

\bibitem{Grinstein:2018fgb}
B.~Grinstein, S.~Pokorski, and G.~G. Ross, {\it {Lepton non-universality in $B$
  decays and fermion mass structure}},  {\em JHEP} {\bf 12} (2018) 079,
  [\href{http://arxiv.org/abs/1809.01766}{{\tt arXiv:1809.01766}}].

\bibitem{Singirala:2018mio}
S.~Singirala, S.~Sahoo, and R.~Mohanta, {\it {Exploring dark matter, neutrino
  mass and $R_{K^{(*)},\phi}$ anomalies in $L_{\mu}-L_{\tau}$ model}},  {\em
  Phys. Rev.} {\bf D99} (2019), no.~3 035042,
  [\href{http://arxiv.org/abs/1809.03213}{{\tt arXiv:1809.03213}}].

\bibitem{Balaji:2018zna}
S.~Balaji, R.~Foot, and M.~A. Schmidt, {\it {Chiral SU(4) explanation of the
  $b\to s$ anomalies}},  {\em Phys. Rev.} {\bf D99} (2019), no.~1 015029,
  [\href{http://arxiv.org/abs/1809.07562}{{\tt arXiv:1809.07562}}].

\bibitem{Rocha-Moran:2018jzu}
P.~Rocha-Moran and A.~Vicente, {\it {Lepton Flavor Violation in a $Z^\prime$
  model for the $b \to s$ anomalies}},  {\em Phys. Rev.} {\bf D99} (2019),
  no.~3 035016, [\href{http://arxiv.org/abs/1810.02135}{{\tt
  arXiv:1810.02135}}].

\bibitem{Kamada:2018kmi}
A.~Kamada, M.~Yamada, and T.~T. Yanagida, {\it {Self-interacting dark matter
  with a vector mediator: kinetic mixing with the $
  \mathrm{U}{(1)}_{{\left(B-L\right)}_3} $ gauge boson}},  {\em JHEP} {\bf 03}
  (2019) 021, [\href{http://arxiv.org/abs/1811.02567}{{\tt arXiv:1811.02567}}].

\bibitem{Fornal:2018dqn}
B.~Fornal, S.~A. Gadam, and B.~Grinstein, {\it {Left-Right SU(4) Vector
  Leptoquark Model for Flavor Anomalies}},
  \href{http://arxiv.org/abs/1812.01603}{{\tt arXiv:1812.01603}}.

\bibitem{Geng:2018xzd}
C.-Q. Geng and H.~Okada, {\it {Resolving $B$-meson anomalies by
  flavor-dependent gauged symmetries $\displaystyle
  \prod_{i=1}^3U(1)_{B_i-L_i}$}},  \href{http://arxiv.org/abs/1812.07918}{{\tt
  arXiv:1812.07918}}.

\bibitem{Kumar:2019qbv}
J.~Kumar and D.~London, {\it {New physics in $b \to s e^+ e^-$?}},
  \href{http://arxiv.org/abs/1901.04516}{{\tt arXiv:1901.04516}}.

\bibitem{Baek:2019qte}
S.~Baek, {\it {Scalar dark matter behind $b \to s \mu \mu$ anomaly}},
  \href{http://arxiv.org/abs/1901.04761}{{\tt arXiv:1901.04761}}.

\bibitem{Marzo:2019ldg}
C.~Marzo, L.~Marzola, and M.~Raidal, {\it {Common explanation to the
  $R_{K^{(*)}}$, $R_{D^{(*)}}$ and $\epsilon^\prime/\epsilon$ anomalies in a
  3HDM+$\nu_R$ and connections to neutrino physics}},
  \href{http://arxiv.org/abs/1901.08290}{{\tt arXiv:1901.08290}}.

\bibitem{Cerdeno:2019vpd}
D.~G. Cerdeño, A.~Cheek, P.~Martín-Ramiro, and J.~M. Moreno, {\it {B
  anomalies and dark matter: a complex connection}},
  \href{http://arxiv.org/abs/1902.01789}{{\tt arXiv:1902.01789}}.

\bibitem{Bhattacharya:2019eji}
S.~Bhattacharya, A.~Biswas, Z.~Calcuttawala, and S.~K. Patra, {\it {An in-depth
  analysis of $b\to c(s)$ semileptonic observables with possible $\mu - \tau$
  mixing}},  \href{http://arxiv.org/abs/1902.02796}{{\tt arXiv:1902.02796}}.

\bibitem{Ko:2019tts}
P.~Ko, T.~Nomura, and C.~Yu, {\it {$b\rightarrow s \mu^+ \mu^-$ anomalies and
  related phenomenology in $U(1)_{B_3 - x_\mu L_\mu - x_\tau L_\tau}$ flavor
  gauge models}},  \href{http://arxiv.org/abs/1902.06107}{{\tt
  arXiv:1902.06107}}.

\bibitem{DAmico:2017mtc}
G.~D'Amico, M.~Nardecchia, P.~Panci, F.~Sannino, A.~Strumia, R.~Torre, and
  A.~Urbano, {\it {Flavour anomalies after the $R_{K^*}$ measurement}},  {\em
  JHEP} {\bf 09} (2017) 010, [\href{http://arxiv.org/abs/1704.05438}{{\tt
  arXiv:1704.05438}}].

\bibitem{Geng:2017svp}
L.-S. Geng, B.~Grinstein, S.~Jäger, J.~Martin~Camalich, X.-L. Ren, and R.-X.
  Shi, {\it {Towards the discovery of new physics with lepton-universality
  ratios of $b\to s\ell\ell$ decays}},  {\em Phys. Rev.} {\bf D96} (2017),
  no.~9 093006, [\href{http://arxiv.org/abs/1704.05446}{{\tt
  arXiv:1704.05446}}].

\bibitem{Capdevila:2017bsm}
B.~Capdevila, A.~Crivellin, S.~Descotes-Genon, J.~Matias, and J.~Virto, {\it
  {Patterns of New Physics in $b\to s\ell^+\ell^-$ transitions in the light of
  recent data}},  {\em JHEP} {\bf 01} (2018) 093,
  [\href{http://arxiv.org/abs/1704.05340}{{\tt arXiv:1704.05340}}].

\bibitem{Altmannshofer:2017yso}
W.~Altmannshofer, P.~Stangl, and D.~M. Straub, {\it {Interpreting Hints for
  Lepton Flavor Universality Violation}},  {\em Phys. Rev.} {\bf D96} (2017),
  no.~5 055008, [\href{http://arxiv.org/abs/1704.05435}{{\tt
  arXiv:1704.05435}}].

\bibitem{Ciuchini:2017mik}
M.~Ciuchini, A.~M. Coutinho, M.~Fedele, E.~Franco, A.~Paul, L.~Silvestrini, and
  M.~Valli, {\it {On Flavourful Easter eggs for New Physics hunger and Lepton
  Flavour Universality violation}},  {\em Eur. Phys. J.} {\bf C77} (2017),
  no.~10 688, [\href{http://arxiv.org/abs/1704.05447}{{\tt arXiv:1704.05447}}].

\bibitem{Hiller:2017bzc}
G.~Hiller and I.~Nisandzic, {\it {$R_K$ and $R_{K^{\ast}}$ beyond the standard
  model}},  {\em Phys. Rev.} {\bf D96} (2017), no.~3 035003,
  [\href{http://arxiv.org/abs/1704.05444}{{\tt arXiv:1704.05444}}].

\bibitem{Aaij:2019wad}
{\bf LHCb} Collaboration, R.~Aaij et~al., {\it {Search for lepton-universality
  violation in $B^+\to K^+\ell^+\ell^-$ decays}},
  \href{http://arxiv.org/abs/1903.09252}{{\tt arXiv:1903.09252}}.

\bibitem{Abdesselam:2019wac}
{\bf Belle} Collaboration, A.~Abdesselam et~al., {\it {Test of lepton flavor
  universality in ${B\to K^\ast\ell^+\ell^-}$ decays at Belle}},
  \href{http://arxiv.org/abs/1904.02440}{{\tt arXiv:1904.02440}}.

\bibitem{Chetyrkin:1996vx}
K.~G. Chetyrkin, M.~Misiak, and M.~Munz, {\it {Weak radiative B meson decay
  beyond leading logarithms}},  {\em Phys.Lett.} {\bf B400} (1997) 206--219,
  [\href{http://arxiv.org/abs/hep-ph/9612313}{{\tt hep-ph/9612313}}].

\bibitem{Buras:1994dj}
A.~J. Buras and M.~Munz, {\it {Effective Hamiltonian for $B \to X_s e^+ e^-$
  beyond leading logarithms in the NDR and HV schemes}},  {\em Phys.Rev.} {\bf
  D52} (1995) 186--195, [\href{http://arxiv.org/abs/hep-ph/9501281}{{\tt
  hep-ph/9501281}}].

\bibitem{Bobeth:1999mk}
C.~Bobeth, M.~Misiak, and J.~Urban, {\it {Photonic penguins at two loops and
  m(t) dependence of $BR[B \to X(s) l^+ l^-]$}},  {\em Nucl.Phys.} {\bf B574}
  (2000) 291--330, [\href{http://arxiv.org/abs/hep-ph/9910220}{{\tt
  hep-ph/9910220}}].

\bibitem{Gambino:2001au}
P.~Gambino and U.~Haisch, {\it {Complete electroweak matching for radiative B
  decays}},  {\em JHEP} {\bf 10} (2001) 020,
  [\href{http://arxiv.org/abs/hep-ph/0109058}{{\tt hep-ph/0109058}}].

\bibitem{Misiak:2004ew}
M.~Misiak and M.~Steinhauser, {\it {Three loop matching of the dipole operators
  for b $\to$ s gamma and b $\to$ s g}},  {\em Nucl.Phys.} {\bf B683} (2004)
  277--305, [\href{http://arxiv.org/abs/hep-ph/0401041}{{\tt hep-ph/0401041}}].

\bibitem{Bobeth:2003at}
C.~Bobeth, P.~Gambino, M.~Gorbahn, and U.~Haisch, {\it {Complete NNLO QCD
  analysis of $\bar{B} \to X_s \ell^+ \ell^-$ and higher order electroweak
  effects}},  {\em JHEP} {\bf 0404} (2004) 071,
  [\href{http://arxiv.org/abs/hep-ph/0312090}{{\tt hep-ph/0312090}}].

\bibitem{Gambino:2003zm}
P.~Gambino, M.~Gorbahn, and U.~Haisch, {\it {Anomalous dimension matrix for
  radiative and rare semileptonic B decays up to three loops}},  {\em
  Nucl.Phys.} {\bf B673} (2003) 238--262,
  [\href{http://arxiv.org/abs/hep-ph/0306079}{{\tt hep-ph/0306079}}].

\bibitem{Huber:2005ig}
T.~Huber, E.~Lunghi, M.~Misiak, and D.~Wyler, {\it {Electromagnetic logarithms
  in $\bar{B} \to X_s \ell^+ \ell^-$}},  {\em Nucl.Phys.} {\bf B740} (2006)
  105--137, [\href{http://arxiv.org/abs/hep-ph/0512066}{{\tt hep-ph/0512066}}].

\bibitem{Straub:2015ica}
A.~Bharucha, D.~M. Straub, and R.~Zwicky, {\it {$B\to V\ell^+\ell^-$ in the
  Standard Model from light-cone sum rules}},  {\em JHEP} {\bf 08} (2016) 098,
  [\href{http://arxiv.org/abs/1503.05534}{{\tt arXiv:1503.05534}}].

\bibitem{Altmannshofer:2014rta}
W.~Altmannshofer and D.~M. Straub, {\it {New physics in $b\rightarrow s$
  transitions after LHC run 1}},  {\em Eur. Phys. J.} {\bf C75} (2015), no.~8
  382, [\href{http://arxiv.org/abs/1411.3161}{{\tt arXiv:1411.3161}}].

\bibitem{Descotes-Genon:2015uva}
S.~Descotes-Genon, L.~Hofer, J.~Matias, and J.~Virto, {\it {Global analysis of
  $b\to s\ell\ell$ anomalies}},  {\em JHEP} {\bf 06} (2016) 092,
  [\href{http://arxiv.org/abs/1510.04239}{{\tt arXiv:1510.04239}}].

\bibitem{Chobanova:2017ghn}
V.~G. Chobanova, T.~Hurth, F.~Mahmoudi, D.~Martinez~Santos, and S.~Neshatpour,
  {\it {Large hadronic power corrections or new physics in the rare decay $B\to
  K^{*}\mu^{+}\mu^{-}$?}},  {\em JHEP} {\bf 07} (2017) 025,
  [\href{http://arxiv.org/abs/1702.02234}{{\tt arXiv:1702.02234}}].

\bibitem{Alok:2017sui}
A.~K. Alok, B.~Bhattacharya, A.~Datta, D.~Kumar, J.~Kumar, and D.~London, {\it
  {New Physics in $b \to s \mu^+ \mu^-$ after the Measurement of $R_{K^*}$}},
  {\em Phys. Rev.} {\bf D96} (2017), no.~9 095009,
  [\href{http://arxiv.org/abs/1704.07397}{{\tt arXiv:1704.07397}}].

\bibitem{Hurth:2017hxg}
T.~Hurth, F.~Mahmoudi, D.~Martinez~Santos, and S.~Neshatpour, {\it {Lepton
  nonuniversality in exclusive $b{\rightarrow}s{\ell}{\ell}$ decays}},  {\em
  Phys. Rev.} {\bf D96} (2017), no.~9 095034,
  [\href{http://arxiv.org/abs/1705.06274}{{\tt arXiv:1705.06274}}].

\bibitem{Paul:2016urs}
A.~Paul and D.~M. Straub, {\it {Constraints on new physics from radiative $B$
  decays}},  {\em JHEP} {\bf 04} (2017) 027,
  [\href{http://arxiv.org/abs/1608.02556}{{\tt arXiv:1608.02556}}].

\bibitem{Alguero:2018nvb}
M.~Algueró, B.~Capdevila, S.~Descotes-Genon, P.~Masjuan, and J.~Matias, {\it
  {Are we overlooking Lepton Flavour Universal New Physics in $b\to s\ell\ell$
  ?}},  \href{http://arxiv.org/abs/1809.08447}{{\tt arXiv:1809.08447}}.

\bibitem{Alguero:2019pjc}
M.~Algueró, B.~Capdevila, S.~Descotes-Genon, P.~Masjuan, and J.~Matias, {\it
  {What $R_K$ and $Q_5$ can tell us about New Physics in $b\to s\ell\ell$
  transitions?}},  \href{http://arxiv.org/abs/1902.04900}{{\tt
  arXiv:1902.04900}}.

\bibitem{Arbey:2018ics}
A.~Arbey, T.~Hurth, F.~Mahmoudi, and S.~Neshatpour, {\it {Hadronic and New
  Physics Contributions to $b \to s$ Transitions}},  {\em Phys. Rev.} {\bf D98}
  (2018), no.~9 095027, [\href{http://arxiv.org/abs/1806.02791}{{\tt
  arXiv:1806.02791}}].

\bibitem{Gubernari:2018wyi}
N.~Gubernari, A.~Kokulu, and D.~van Dyk, {\it {$B\to P$ and $B\to V$ Form
  Factors from $B$-Meson Light-Cone Sum Rules beyond Leading Twist}},  {\em
  JHEP} {\bf 01} (2019) 150, [\href{http://arxiv.org/abs/1811.00983}{{\tt
  arXiv:1811.00983}}].

\bibitem{Horgan:2013hoa}
R.~R. Horgan, Z.~Liu, S.~Meinel, and M.~Wingate, {\it {Lattice QCD calculation
  of form factors describing the rare decays $B \to K^* \ell^+ \ell^-$ and $B_s
  \to \phi \ell^+ \ell^-$}},  {\em Phys.Rev.} {\bf D89} (2014) 094501,
  [\href{http://arxiv.org/abs/1310.3722}{{\tt arXiv:1310.3722}}].

\bibitem{Beneke:2001at}
M.~Beneke, T.~Feldmann, and D.~Seidel, {\it {Systematic approach to exclusive
  $B \to V l^+ l^-, V \gamma$ decays}},  {\em Nucl.Phys.} {\bf B612} (2001)
  25--58, [\href{http://arxiv.org/abs/hep-ph/0106067}{{\tt hep-ph/0106067}}].

\bibitem{Kozachuk:2018yxf}
A.~Kozachuk and D.~Melikhov, {\it {Revisiting nonfactorizable charm-loop
  effects in exclusive FCNC $B$-decays}},  {\em Phys. Lett.} {\bf B786} (2018)
  378--381, [\href{http://arxiv.org/abs/1805.05720}{{\tt arXiv:1805.05720}}].

\bibitem{Beneke:2009az}
M.~Beneke, G.~Buchalla, M.~Neubert, and C.~Sachrajda, {\it {Penguins with Charm
  and Quark-Hadron Duality}},  {\em Eur.Phys.J.} {\bf C61} (2009) 439--449,
  [\href{http://arxiv.org/abs/0902.4446}{{\tt arXiv:0902.4446}}].

\bibitem{Bobeth:2017xry}
C.~Bobeth, A.~J. Buras, A.~Celis, and M.~Jung, {\it {Yukawa enhancement of
  $Z$-mediated new physics in $\Delta S = 2$ and $\Delta B = 2$ processes}},
  {\em JHEP} {\bf 07} (2017) 124, [\href{http://arxiv.org/abs/1703.04753}{{\tt
  arXiv:1703.04753}}].

\bibitem{Buchmuller:1985jz}
W.~Buchmuller and D.~Wyler, {\it {Effective Lagrangian Analysis of New
  Interactions and Flavor Conservation}},  {\em Nucl. Phys.} {\bf B268} (1986)
  621--653.

\bibitem{Grzadkowski:2010es}
B.~Grzadkowski, M.~Iskrzynski, M.~Misiak, and J.~Rosiek, {\it {Dimension-Six
  Terms in the Standard Model Lagrangian}},  {\em JHEP} {\bf 10} (2010) 085,
  [\href{http://arxiv.org/abs/1008.4884}{{\tt arXiv:1008.4884}}].

\bibitem{Aebischer:2015fzz}
J.~Aebischer, A.~Crivellin, M.~Fael, and C.~Greub, {\it {Matching of gauge
  invariant dimension-six operators for $b\to s$ and $b\to c$ transitions}},
  {\em JHEP} {\bf 05} (2016) 037, [\href{http://arxiv.org/abs/1512.02830}{{\tt
  arXiv:1512.02830}}].

\bibitem{Silvestrini:2018dos}
L.~Silvestrini and M.~Valli, {\it {Model-independent Bounds on the Standard
  Model Effective Theory from Flavour Physics}},
  \href{http://arxiv.org/abs/1812.10913}{{\tt arXiv:1812.10913}}.

\bibitem{Descotes-Genon:2018foz}
S.~Descotes-Genon, A.~Falkowski, M.~Fedele, M.~González-Alonso, and J.~Virto,
  {\it {The CKM parameters in the SMEFT}},
  \href{http://arxiv.org/abs/1812.08163}{{\tt arXiv:1812.08163}}.

\bibitem{deBlas:2016ojx}
J.~de~Blas, M.~Ciuchini, E.~Franco, S.~Mishima, M.~Pierini, L.~Reina, and
  L.~Silvestrini, {\it {Electroweak precision observables and Higgs-boson
  signal strengths in the Standard Model and beyond: present and future}},
  {\em JHEP} {\bf 12} (2016) 135, [\href{http://arxiv.org/abs/1608.01509}{{\tt
  arXiv:1608.01509}}].

\bibitem{Ellis:2018gqa}
J.~Ellis, C.~W. Murphy, V.~Sanz, and T.~You, {\it {Updated Global SMEFT Fit to
  Higgs, Diboson and Electroweak Data}},  {\em JHEP} {\bf 06} (2018) 146,
  [\href{http://arxiv.org/abs/1803.03252}{{\tt arXiv:1803.03252}}].

\bibitem{Buras:2014fpa}
A.~J. Buras, J.~Girrbach-Noe, C.~Niehoff, and D.~M. Straub, {\it {$ B\to
  {K}^{\left(\ast \right)}\nu \overline{\nu} $ decays in the Standard Model and
  beyond}},  {\em JHEP} {\bf 02} (2015) 184,
  [\href{http://arxiv.org/abs/1409.4557}{{\tt arXiv:1409.4557}}].

\bibitem{Lees:2013kla}
{\bf BaBar} Collaboration, J.~P. Lees et~al., {\it {Search for $B \to K^{(*)}
  \nu \overline \nu$ and invisible quarkonium decays}},  {\em Phys. Rev.} {\bf
  D87} (2013), no.~11 112005, [\href{http://arxiv.org/abs/1303.7465}{{\tt
  arXiv:1303.7465}}].

\bibitem{Jenkins:2013zja}
E.~E. Jenkins, A.~V. Manohar, and M.~Trott, {\it {Renormalization Group
  Evolution of the Standard Model Dimension Six Operators I: Formalism and
  lambda Dependence}},  {\em JHEP} {\bf 10} (2013) 087,
  [\href{http://arxiv.org/abs/1308.2627}{{\tt arXiv:1308.2627}}].

\bibitem{Jenkins:2013wua}
E.~E. Jenkins, A.~V. Manohar, and M.~Trott, {\it {Renormalization Group
  Evolution of the Standard Model Dimension Six Operators II: Yukawa
  Dependence}},  {\em JHEP} {\bf 01} (2014) 035,
  [\href{http://arxiv.org/abs/1310.4838}{{\tt arXiv:1310.4838}}].

\bibitem{Hurth:2019ula}
T.~Hurth, S.~Renner, and W.~Shepherd, {\it {Matching for FCNC effects in the
  flavour-symmetric SMEFT}},  \href{http://arxiv.org/abs/1903.00500}{{\tt
  arXiv:1903.00500}}.

\bibitem{Aebischer:2018iyb}
J.~Aebischer, J.~Kumar, P.~Stangl, and D.~M. Straub, {\it {A Global Likelihood
  for Precision Constraints and Flavour Anomalies}},
  \href{http://arxiv.org/abs/1810.07698}{{\tt arXiv:1810.07698}}.

\bibitem{Efrati:2015eaa}
A.~Efrati, A.~Falkowski, and Y.~Soreq, {\it {Electroweak constraints on
  flavorful effective theories}},  {\em JHEP} {\bf 07} (2015) 018,
  [\href{http://arxiv.org/abs/1503.07872}{{\tt arXiv:1503.07872}}].

\bibitem{Bordone:2017lsy}
M.~Bordone, D.~Buttazzo, G.~Isidori, and J.~Monnard, {\it {Probing Lepton
  Flavour Universality with $K \to \pi \nu \bar\nu$ decays}},  {\em Eur. Phys.
  J.} {\bf C77} (2017), no.~9 618, [\href{http://arxiv.org/abs/1705.10729}{{\tt
  arXiv:1705.10729}}].

\bibitem{Descotes-Genon:2013wba}
S.~Descotes-Genon, J.~Matias, and J.~Virto, {\it {Understanding the $B \to
  K^*\mu^+\mu^-$ Anomaly}},  {\em Phys.Rev.} {\bf D88} (2013), no.~7 074002,
  [\href{http://arxiv.org/abs/1307.5683}{{\tt arXiv:1307.5683}}].

\bibitem{Hiller:2013cza}
G.~Hiller and R.~Zwicky, {\it {(A)symmetries of weak decays at and near the
  kinematic endpoint}},  {\em JHEP} {\bf 03} (2014) 042,
  [\href{http://arxiv.org/abs/1312.1923}{{\tt arXiv:1312.1923}}].

\bibitem{HEPfit}
``Hepfit, a tool to combine indirect and direct constraints on high energy
  physics.'' \url{http://hepfit.roma1.infn.it/}.

\bibitem{Caldwell:2008fw}
A.~Caldwell, D.~Kollar, and K.~Kroninger, {\it {BAT: The Bayesian Analysis
  Toolkit}},  {\em Comput. Phys. Commun.} {\bf 180} (2009) 2197--2209,
  [\href{http://arxiv.org/abs/0808.2552}{{\tt arXiv:0808.2552}}].

\bibitem{2013arXiv1307.5928G}
A.~{Gelman}, J.~{Hwang}, and A.~{Vehtari}, {\it {Understanding predictive
  information criteria for Bayesian models}},  {\em arXiv e-prints} (Jul, 2013)
  arXiv:1307.5928, [\href{http://arxiv.org/abs/1307.5928}{{\tt
  arXiv:1307.5928}}].

\bibitem{IC}
T.~Ando, {\it Predictive bayesian model selection},  {\em American Journal of
  Mathematical and Management Sciences} {\bf 31} (2011), no.~1-2 13--38.
  \url{http://dx.doi.org/10.1080/01966324.2011.10737798}.

\bibitem{BayesFactors}
R.~E. Kass and A.~E. Raftery, {\it Bayes factors},  {\em Journal of the
  American Statistical Association} {\bf 90} (1995), no.~430 773--795.
  \url{http://dx.doi.org/10.1080/01621459.1995.10476572}.

\bibitem{Aaij:2017vad}
{\bf LHCb} Collaboration, R.~Aaij et~al., {\it {Measurement of the
  $B^0_s\to\mu^+\mu^-$ branching fraction and effective lifetime and search for
  $B^0\to\mu^+\mu^-$ decays}},  {\em Phys. Rev. Lett.} {\bf 118} (2017), no.~19
  191801, [\href{http://arxiv.org/abs/1703.05747}{{\tt arXiv:1703.05747}}].

\bibitem{Chatrchyan:2013bka}
{\bf CMS} Collaboration, S.~Chatrchyan et~al., {\it {Measurement of the
  $B_{(s)} \to \mu^+ \mu^-$ branching fraction and search for $B^0 \to \mu^+
  \mu^-$ with the CMS Experiment}},  {\em Phys. Rev. Lett.} {\bf 111} (2013)
  101804, [\href{http://arxiv.org/abs/1307.5025}{{\tt arXiv:1307.5025}}].

\bibitem{Aaboud:2018mst}
{\bf ATLAS} Collaboration, M.~Aaboud et~al., {\it {Study of the rare decays of
  $B^0_s$ and $B^0$ mesons into muon pairs using data collected during 2015 and
  2016 with the ATLAS detector}},  {\em Submitted to: JHEP} (2018)
  [\href{http://arxiv.org/abs/1812.03017}{{\tt arXiv:1812.03017}}].

\bibitem{Amhis:2016xyh}
{\bf HFLAV} Collaboration, Y.~Amhis et~al., {\it {Averages of $b$-hadron,
  $c$-hadron, and $\tau$-lepton properties as of summer 2016}},  {\em Eur.
  Phys. J.} {\bf C77} (2017), no.~12 895,
  [\href{http://arxiv.org/abs/1612.07233}{{\tt arXiv:1612.07233}}].

\bibitem{Aaij:2012ita}
{\bf LHCb} Collaboration, R.~Aaij et~al., {\it {Measurement of the ratio of
  branching fractions $BR(B_0 \to K^{\star 0} \gamma)/BR(B_{s0} \to \phi
  \gamma)$ and the direct CP asymmetry in $B_0 \to K^{\star 0} \gamma$}},  {\em
  Nucl. Phys.} {\bf B867} (2013) 1--18,
  [\href{http://arxiv.org/abs/1209.0313}{{\tt arXiv:1209.0313}}].

\bibitem{Bailey:2015dka}
J.~A. Bailey et~al., {\it {$B\to Kl^+l^-$ decay form factors from three-flavor
  lattice QCD}},  {\em Phys. Rev.} {\bf D93} (2016), no.~2 025026,
  [\href{http://arxiv.org/abs/1509.06235}{{\tt arXiv:1509.06235}}].

\bibitem{DAmbrosio:2019tph}
G.~D'Ambrosio, A.~M. Iyer, F.~Piccinini, and A.~D. Polosa, {\it {Confronting
  $B$ anomalies with atomic physics}},
  \href{http://arxiv.org/abs/1902.00893}{{\tt arXiv:1902.00893}}.

\bibitem{Chrzaszcz:2018yza}
M.~Chrzaszcz, A.~Mauri, N.~Serra, R.~Silva~Coutinho, and D.~van Dyk, {\it
  {Prospects for disentangling long- and short-distance effects in the decays
  $B\to K^* \mu^+\mu^-$}},  \href{http://arxiv.org/abs/1805.06378}{{\tt
  arXiv:1805.06378}}.

\bibitem{deBlas:2015aea}
J.~de~Blas, M.~Chala, and J.~Santiago, {\it {Renormalization Group Constraints
  on New Top Interactions from Electroweak Precision Data}},  {\em JHEP} {\bf
  09} (2015) 189, [\href{http://arxiv.org/abs/1507.00757}{{\tt
  arXiv:1507.00757}}].

\end{thebibliography}\endgroup
%\end{multicols}
\end{document}